\pdfoutput=1

\documentclass[graybox]{svmult}

\usepackage{type1cm}        
\usepackage{makeidx}         
\usepackage{graphicx}        
\usepackage{multicol}        
\usepackage[bottom]{footmisc}

\usepackage{newtxtext}       %
\usepackage{newtxmath}       
\usepackage{natbib}
\usepackage{bm,color}

\graphicspath{{./fig/}{./png/}}

\newcommand{\EQ}{\begin{equation}}
\newcommand{\EN}{\end{equation}}
\newcommand{\EQA}{\begin{eqnarray}}
\newcommand{\ENA}{\end{eqnarray}}
\newcommand{\eq}[1]{(\ref{#1})}

\newcommand{\EEq}[1]{Equation~(\ref{#1})}
\newcommand{\Eq}[1]{Eq.~(\ref{#1})}
\newcommand{\Eqs}[2]{Eqs.~(\ref{#1}) and~(\ref{#2})}

\newcommand{\Sec}[1]{Sect.~\ref{#1}}
\newcommand{\Secs}[2]{Sects.~\ref{#1} and \ref{#2}}

\newcommand{\Fig}[1]{Fig.~\ref{#1}}
\newcommand{\FFig}[1]{Figure~\ref{#1}}

\newcommand{\Tab}[1]{Table~\ref{#1}}

\newcommand{\bra}[1]{\langle #1\rangle}

\newcommand{\meanrho}{\overline{\rho}}

\DeclareMathAlphabet\mathbfcal{OMS}{cmsy}{b}{n}
{}
{}

\newcommand{\meanFFFFf}{\overline{\mathbfcal{F}}_{\rm f}}
\newcommand{\meanemf}{\overline{\cal E} {}}

{}
{}
\newcommand{\meanEMF}{\overline{\mathbfcal{E}}}
{}
{}
{}
{}
{}
{}
{}
{}
{}
{}
{}
{}
{}
{}
\newcommand{\meanUU}{\overline{\bm{U}}}

{}

\newcommand{\meanA}{\overline{A}}
\newcommand{\meanB}{\overline{B}}

\newcommand{\meanJ}{\overline{J}}

\newcommand{\meanS}{\overline{S}}

{}

{}
{}
{}

\newcommand{\pphi}{\hat{\bm{\phi}}}

\newcommand{\meanAA}{{\overline{\bm{A}}}}
\newcommand{\meanBB}{{\overline{\bm{B}}}}
\newcommand{\meanJJ}{{\overline{\bm{J}}}}

\newcommand{\kk}{\bm{k}}

\newcommand{\aaaa}{\bm{a}}
\newcommand{\jj}{\bm{j}}

\newcommand{\bb}{\bm{b}}
\newcommand{\BB}{\bm{B}}

\newcommand{\EE}{\bm{E}}
\newcommand{\JJ}{\bm{J}}
\newcommand{\oo}{\bm{\omega}}

\newcommand{\AAA}{\bm{A}}

\newcommand{\UU}{\bm{U}}

\newcommand{\uu}{\bm{u}}

\newcommand{\nab}{{\bm{\nabla}}}

\newcommand{\OO}{\bm{\Omega}}

\newcommand{\ggamma}{\mbox{\boldmath $\gamma$} {}}

\newcommand{\ii}{{\rm i}}

\newcommand{\dd}{{\rm d} {}}

\def\la{\mathrel{\mathchoice {\vcenter{\offinterlineskip\halign{\hfil
$\displaystyle##$\hfil\cr<\cr\sim\cr}}}
{\vcenter{\offinterlineskip\halign{\hfil$\textstyle##$\hfil\cr<\cr\sim\cr}}}
{\vcenter{\offinterlineskip\halign{\hfil$\scriptstyle##$\hfil\cr<\cr\sim\cr}}}
{\vcenter{\offinterlineskip\halign{\hfil$\scriptscriptstyle##$\hfil\cr<\cr\sim\cr}}}}}

\def\Ra{\mbox{\rm Ra}}

\def\Co{\mbox{\rm Co}}
\def\Ro{\mbox{\rm Ro}}

\def\St{\mbox{\rm St}}

\def\PrT{\mbox{\rm Pr}_{\rm T}}
\def\Pm{\mbox{\rm Pr}_{\rm M}}
\def\PmT{\mbox{\rm Pr}_{\rm M,T}}
\def\Rm{\mbox{\rm Re}_{\rm M}}

\def\Rey{\mbox{\rm Re}}
\def\Imag{\mbox{\rm Im}}

\def\Co{\mbox{\rm Co}}

\def\alpK{\alpha_{\rm K}}
\def\alpM{\alpha_{\rm M}}

\def\kf{k_{\rm f}}

\def\kB{k_{\rm B}}

\def\kB{k_{\rm B}}

\def\urms{u_{\rm rms}}

\def\nuT{\nu_{\rm T}}

\def\etat{\eta_{\rm t}}
\def\etatz{\eta_{\rm t0}}
\def\etatz{\eta_{\rm t0}}

\def\etaT{\eta_{\rm T}}

\def\Beq{B_{\rm eq}}

\def\half{\textstyle{\frac{1}{2}}}

\newcommand{\T}{\,{\rm T}}
\newcommand{\nT}{\,{\rm nT}}
\newcommand{\G}{\,{\rm G}}

\newcommand{\nHz}{\,{\rm nHz}}

\newcommand{\kG}{\,{\rm kG}}
\newcommand{\K}{\,{\rm K}}

\newcommand{\s}{\,{\rm s}}

\newcommand{\cm}{\,{\rm cm}}

\newcommand{\m}{\,{\rm m}}

\newcommand{\Mm}{\,{\rm Mm}}

\newcommand{\days}{\,{\rm d}}

\newcommand{\Gyr}{\,{\rm Gyr}}
\newcommand{\erg}{\,{\rm erg}}

\newcommand{\AU}{\,{\rm AU}}

\def\blue{\textcolor{blue}}

\def\red{\textcolor{red}}
\def\blue{\textcolor{blue}}

\makeindex             

\begin{document}

\title*{Turbulent processes and mean-field dynamo}
\author{Axel Brandenburg, Detlef Elstner, Youhei Masada, and Valery Pipin}
\authorrunning{A. Brandenburg et al.}
\institute{
Axel Brandenburg \at
Nordita, KTH Royal Institute of Technology and Stockholm University, Hannes Alfv\'ens v\"ag 12, 10691 Stockholm, Sweden;
The Oskar Klein Centre, Department of Astronomy, Stockholm University, AlbaNova, 10691 Stockholm, Sweden;
School of Natural Sciences and Medicine, Ilia State University, 0194 Tbilisi, Georgia;
McWilliams Center for Cosmology and Department of Physics, Carnegie Mellon University, Pittsburgh, Pennsylvania 15213, USA,
\email{brandenb@nordita.org}
\and Detlef Elstner
\at Leibniz-Institut f\"ur Astrophysik Potsdam (AIP), An der Sternwarte 16, 14482 Potsdam, Germany,
\email{delstner@aip.de}
\and Youhei Masada
\at Department of Applied Physics, Faculty of Science, Fukuoka University, Fukuoka 814-0180, Japan
\email{ymasada@fukuoka-u.ac.jp}
\and Valery Pipin
\at Institute of Solar-Terrestrial Physics, Russian Academy of Sciences, Irkutsk, 664033, Russia,
\email{pip@iszf.irk.ru}
}

\maketitle

\abstract{
Mean-field dynamo theory has important applications in solar physics
and galactic magnetism.
We discuss some of the many turbulence effects relevant to the generation
of large-scale magnetic fields in the solar convection zone.
The mean-field description is then used to illustrate the physics of the
$\alpha$ effect, turbulent pumping, turbulent magnetic diffusivity,
and other effects on a modern solar dynamo model.
We also discuss how turbulence transport coefficients are derived from
local simulations of convection and then used in mean-field models.
}


\section{Introduction}

The problem of solar and stellar dynamos is still an open one.
In spite of tremendous progress over recent decades, we still do not
understand with any degree of certainty the reason behind the equatorward
migration of solar activity belts, the dependence of cycle frequency on
rotation frequency, or the level of magnetic activity.\footnote{
The reader is referred to the review of \cite{Hazra2023} for a discussion
of flux transport dynamos to explain some of the outstanding questions
of large-scale dynamos in the Sun and stars.
We comment on the main differences between the proposed models
in \Secs{DynamoFluxBudget}{StellarCyclePeriods} below.}
All models of solar and stellar magnetism rely on some assumptions.
Even the most realistic simulations suffer from finite resolution and
the compromises in the physics that are made.
The crucial question is then, when and where we are allowed to make
compromises and when not.
Among those approximations is the second-order correlation approximation
(SOCA), also known as the first-order smoothing approximation.
These are nowadays either replaced by other approximations or by numerical
techniques such as the test-field method (TFM), as will be explained
later in this review.

The Sun's magnetic field exhibits a clear mean field with spatio-temporal
order: antisymmetry of radial and toroidal fields about the equator and
the 11-yr cycle.
This mean field can well be described by an azimuthal average.
The radial component of such an azimuthally averaged mean field has a
typical strength of $\pm10\G$.
This is not much compared with the peak strength of $\pm2\kG$ in sunspots,
but much of this is ``lost'' in the process of averaging.
Of course, whatever is lost corresponds to fluctuations, which actually
play crucial parts and correlations between different fluctuations lead
to various mean-field effects.

Mathematically, once an averaging procedure has been defined, we have
the mean field $\meanBB$, indicated by an overbar.
Then, the difference between the actual and the mean field, $\BB$ and
$\meanBB$, gives the fluctuating field as $\bb\equiv\BB-\meanBB$.
The same procedure also applies to all other quantities.
This formal distinction between mean and fluctuating fields, which are
sometimes also called large-scale and small-scale fields, is important
in discussions with observers.
Coronal mass ejections, for example, are superficially reported as being
part of a large-scale field, but this may not be true anymore when we
think of averaging over the full solar circumference.
Thus, paradoxically, even if something is large by some standards,
it may not qualify as large-scale under this formal definition of an
azimuthal average.

Azimuthal averaging is not always a good recipe.
Some stars have nonaxisymmetric magnetic fields, and even the Sun
is believed to have what is known as active longitudes -- a weak
nonaxisymmetric magnetic field on top of a predominantly axisymmetric one.
Those nonaxisymmetric fields might best be described through low-order
Fourier mode filtering.
This is probably completely fine, but slightly problematic at the formal
level, because then the average of the product of mean and fluctuating
fields is no longer vanishing, as it would be in the case of an azimuthal
average.
This mathematical property is one of several rules that are called the
Reynolds rules.
However, as alluded to above, the violation of this particular Reynolds
rule is probably just a technicality that makes mean-field
predictions less accurate.
We refer here to the work of \cite{ZBC18} for a detailed investigation.
There are a number of other limitations in mean-field theories that will
be discussed below.

The purpose of defining mean fields is twofold.
On the one hand, they allow us to quantify large-scale magnetic, velocity,
and other fields that are observed or that are present in a simulation.
On the other hand, they allow us to develop predictive theories for
these averages.
In these theories, mean fields can sometimes emerge because of
instabilities and/or because of suitable boundary conditions.
This is possible because of certain mean-field effects, by which one
usually means the relations between correlations of fluctuations and
various mean fields.
Discussing those effects is an important purpose of this review.
The ultimate goal of mean-field dynamo theory is to understand and model
the Sun and other stars.
We therefore also discuss in this review the status of such attempts.
For a {\em basic} introduction to mean-field theory, which is not the
subject of this review, we refer to standard textbooks \citep{Mof78,
KR80, ZRS83} and other reviews \citep{BS05, KZ08, MT09, Charbonneau14,
Charbonneau10, Bran18, Tobias21, Bran+Ntor23, Karak23}.

\section{The golden years of dynamo theory}

The first mean-field model was constructed by \citet{Parker1955}.
In his model, the toroidal magnetic field is generated from the
dipole field by nonuniform rotation.
To overcome the restrictions of Cowling's theorem \citep{Cowling1933},
Parker suggested that the dipole magnetic field can be regenerated by
cyclonic convective motions which transform emerging toroidal magnetic
loops into poloidal magnetic field.
The coalescing loops can amplify the dipole magnetic field.
Studying the combined action of differential rotation and cyclonic
motions, he found a solution in the form of a dynamo wave and formulated
conditions for the equatorward propagation of dynamo waves.
\citet{SKR1966} and \citet{StKr1969} constructed the theoretical
basis of mean-field theory, introduced the notion of the mean
electromotive force (MEMF) of the turbulence
and showed that the Parker effect of the cyclonic convective motions
is equivalent to the effective MEMF along the large-scale field.
The 1970s can be considered the golden years of mean-field dynamo theory.
Back then, \cite{Schuessler1983I} stated: ``dynamo theory reached the
textbook state'', mentioning the famous monographs by \cite{Mof78},
\cite{Par79}, \cite{KR80}, \cite{VZR1980}, and \cite{ZRS83}.

Indeed, intensive theoretical and observational studies led to the
establishment of the basic solar dynamo scenario, identification of
key dynamo parameters, and the formulation of a general paradigm of the
nature of solar and stellar magnetism.
\cite{Schuessler1983I} summarized that mean-field dynamo models can
reproduce the ``physics of solar activity to a great extent'' including:
\begin{itemize}
\item Hale's polarity rule of sunspots groups,
\item the time-latitude evolution of sunspot activity (``butterfly diagram''),
\item reversals of the polar magnetic field,
\item the phase relationship between the evolution of poloidal and
toroidal magnetic fields and their consistence with the observed butterfly
diagrams \citep{Stix1976},
\item rigid rotation of the magnetic sector structure and coronal holes
\citep{Stix1974,Stix1977},
\item chaotic variations of dynamo activity either due to a random
$\alpha$ effect or the dynamo nonlinearity from the Lorentz force
\citep{Leighton1969,Yoshimura1978,Ruzmaikin1981}, and
\item a quantitative understanding of the solar torsional oscillations
\citep{Schuessler1981, Yoshimura1981}.
\end{itemize}
We note that the first and second items are based on the assumption
that sunspot groups are formed from the large-scale toroidal magnetic
field.
Already at the time it was recognized that the mean-field models need
to take into account the fibril state of the magnetic field which we
observed at the solar surface.
We return to this point later.

Classical mean-field dynamo models utilize the $\alpha\Omega$ scenario
using the differential rotation ($\Omega$ effect) as the source of
the toroidal magnetic flux production and the $\alpha$ effect for the
poloidal magnetic field generation. 
Since the seminal work of \cite{pouquet+76}, it started to become
clear that the magnetic helicity results in an important nonlinear
contribution to the $\alpha$ effect and turbulent magnetic field
generation \citep{Kleeorin1982}.

\section{Mean-field theory and avoiding some of its limitations}

We can never expect a mean-field theory to produce an accurate
representation of reality.
One reason is the fact that the underlying turbulence has stochastic
aspects, so each realization with slightly different initial conditions
would result in a somewhat different outcome.
However, there could be other reasons for discrepancies that we discuss
next.
Some of those discrepancies can nowadays be avoided.

\subsection{Mean-field electrodynamics}

In mean-field theory, one derives evolution equations for the averaged
fields, namely the mean magnetic field $\meanBB$, the mean velocity
$\meanUU$, and the mean thermodynamic variables such as mean specific
entropy $\meanS$ and the mean density $\meanrho$.
Often, one neglects the evolution of $\meanUU$, $\meanS$, and $\meanrho$,
which is then already an important limitation.

If one focuses on the evolution of the mean magnetic field only,
one often talks about the mean-fields electrodynamics or quasi-kinematic
mean-field theory, which can still be nonlinear if the various mean-field
transport coefficients depend on the mean fields.
If they are unaffected, one talks about kinematic mean-field theory, which
is linear.
Of course, once there is a dynamo, we have an exponentially growing
solution, so the magnetic field would grow without limit, i.e., it would
not saturate within kinematic mean-field theory.
Obviously, a correct mean-field theory must be nonlinear, but even within
the realm of linear theory, there are important lessons to be learnt.
Below, we discuss the aspects of nonlocality, which were often omitted
out of ignorance, but nowadays we know that this is often not possible
and this restriction can easily be relaxed.

\subsection{Nonlocality}
\label{nonloc}

The mean magnetic field is governed by the mean induction equation,
which is sometimes also referred to as the mean-field dynamo equation.
The most important term here is the mean electromotive force,
\begin{equation}
\meanEMF=\overline{\uu\times\bb},
\end{equation}
i.e., the averaged cross product of velocity and magnetic fluctuations.
In mean-field electrodynamics, it is often expressed as
\begin{equation}
\meanemf_i=\meanemf_{0i}+\alpha_{ij}\meanB_j+\eta_{ijk}\partial\meanB_j/\partial x_k+...,
\label{mult}
\end{equation}
where the ellipsis denotes higher derivative terms, of which there should
be infinitely many, and there should also be time derivatives.
The term $\meanemf_{0i}$ is a contribution that can exist already in
the absence of a mean field; see \cite{BR13} for details and numerical
experiments.
Including only a finite number of derivatives in \Eq{mult} and ignoring
time derivatives is another important approximation.
In fact, it is usually easier to express $\meanEMF$ as a convolution
between an integral kernel and the mean field.
Furthermore, it is instructive to split the integral kernel into two
pieces and write
\begin{equation}
\meanemf_i=\meanemf_{0i}+\hat{\alpha}_{ij}\ast\meanB_j
+\hat{\eta}_{ijk}\ast\partial\meanB_j/\partial x_k,
\label{conv}
\end{equation}
where the asterisks mean a convolution in space and time, and the hats
denote integration kernels.
In principle, the spatial derivative can be absorbed as being part of
the integral kernel, but separating the kernel into $\hat{\alpha}_{ij}$
and $\hat{\eta}_{ijk}$ has conceptual advantages, because they preserve
the similarity to \Eq{mult}.
Note also that, unlike \Eq{mult}, where we allowed for arbitrarily many
derivatives, here, we have no other terms, because all even derivatives
are already absorbed in $\hat{\alpha}_{ij}$ and all odd derivatives are
absorbed in $\hat{\eta}_{ijk}$.
Time derivatives can also be absorbed in both of them if the convolution
with the kernels is also over time.

For the benefit of better interpretation, both $\alpha_{ij}$ and
$\eta_{ijk}$ (and analogously also for $\hat{\alpha}_{ij}$ and
$\hat{\eta}_{ijk}$) can be broken down into further pieces.
The $\alpha_{ij}$ tensor can be split into a symmetric and an
antisymmetric tensor.
The latter is characterized by a vector,
$\gamma_i=-\half\epsilon_{ijk}\alpha_{jk}$,
which corresponds to a pumping velocity.
Having in mind that the magnetic gradient tensor can also be split
into symmetric and antisymmetric parts, where the latter is the
mean current density, $\meanJJ$, with $\meanJ_i=-\half\epsilon_{ijk}
\partial\meanB_j/\partial x_k$, we can separate the rank-3 tensor,
$\eta_{ijk}$, into a rank-2 tensor operating only on $\meanJJ$ and the
rest operating on the symmetric part of $\partial\meanB_j/\partial x_k$.

The convolution can only be replaced by a multiplication, as in
\Eq{mult}, if the mean
field is constant in time (which is normally never the case!) and if it
varies at most linearly in space (which is normally also not the case).
We return to this point further below.

\subsection{Beyond SOCA and scale separation limits}\label{bsoca}

An important question concerns the calculation of the $\alpha_{ij}$
and $\eta_{ijk}$ coefficients or kernels.
A problem arises from the fact that the differential equations for these
expressions are nonlinear and therefore hard to solve analytically.
A commonly used approximation is SOCA.
It neglects triple (and higher) correlations in the evolution equations
for the fluctuating velocity and magnetic fields.
This closure can be applied when either 
the magnetic Reynolds number, $\Rm$,
or the Strouhal number, $\St$, are much smaller than unity.
These limits are rather restrictive for astrophysical conditions.
For example, the convection zones (CZs) of the Sun and stars are in a
turbulent state with huge values of the fluid Reynolds number ($\Rey
\gtrsim 10^{12}$), the magnetic Reynolds number ($\Rm \gtrsim 10^8$),
Rayleigh number ($\Ra \gtrsim 10^{20}$), and an extremely low Prandtl
number ($\Pr \sim 10^{-4}$--$10^{-7}$); see, e.g., \cite{ossendrijver03}.

Results of \cite{schrinner+05} showed that SOCA does not work well
when $\Rm$ exceeds unity and $\St$ is not small.
There are analytical approaches, e.g., different variants of the so-called
$\tau$ approximation \citep{Kleeorin1996,FB02,BS05}, which can be applied
in the high Reynolds number limit.
The restrictions inherent to SOCA or the $\tau$ approximation no
longer apply when calculating solutions of the underlying differential
equations numerically.
This is done in the TFM \citep{schrinner+05, schrinner+07}.

Using a set of mean magnetic fields, the TFM allows one to determine
the turbulent transport coefficients for arbitrary velocity fields,
provided they can be computed or otherwise represented on the computer.
The velocity field can be determined either as a solution of the nonlinear
Navier-Stokes equations for a forced turbulent flow or it can be obtained
as a results of global convective dynamo (GCD) simulations.
To compute $\meanEMF$, the solution for the induction equation for the
fluctuating magnetic field is needed as well.
The original TFM of \cite{schrinner+05, schrinner+07} adopts the
scale-separation assumption.
It was shown that the TFM describes the dynamo processes for GCD
simulations at moderate Reynolds numbers of around $50$ rather well
\citep{Schrinner2011,Warnecke2018a,Viviani2018}.
The calculations within TFM have some technical restrictions and
are currently unable to meet the very high astrophysical limits of
$\Rey,\Rm>10^6$.
Nevertheless, the current applications of the TFM concern cases with
$\Rey,\Rm\gg 1$, which is well beyond the SOCA limits.
In recent developments of the TFM, \cite{Maarit2022} took compressibility
effects into account.
They also studied the effects of the small-scale dynamo on the turbulent
electromotive force; see also \cite{Rempel2023a}.

An alternative way of extracting the coefficients of the
mean-electromotive force employs a multi-dimensional regression method
\citep{BS02, racine+11, augustson+15,simard+16}.
In this approach, instead of solving the equations for the fluctuations
in the presence of different mean magnetic fields, as it is done in
the TFM, the regression methods try to exploit the form of \Eq{mult}
for the dynamo-generated large-scale magnetic field.
Detailed comparisons of the above method with TFM were done by
\cite{Warnecke2018a}.
It was found that the TFM gives a more accurate representation of the
mean-field coefficients than the multidimensional regression method.
We encourage the reader to consult this paper for further details.
We return to the problem of extracting turbulent transport coefficients
from GCDs in \Sec{dnssec}.

The limitations discussed so far are in principle all avoidable:
(i) The evolution equations for $\meanUU$, $\meanS$, and $\meanrho$
can be (and have been) included \citep{Brandenburg1992}, in addition to that for $\meanBB$,
but in practice, even this is still an approximation in the sense that
the full set of coefficients for the equations is not (or only approximately) known.
(ii) The electromotive force can be (and has been) expressed as a
convolution, which can most effectively be solved by rewriting the
equations as a differential equation, as will be described below.
(iii) Numerical solutions can be employed to find specific values for
$\alpha_{ij}$ and $\eta_{ijk}$; see \cite{warnecke18, War+21} for doing this
for solar simulations using the TFM.
It often turns out that analytical closure techniques are very useful
as a first orientation and they are often also accurate enough for a
qualitatively useful model.
In special cases, when a more accurate solution is required, the answer may
well be obtained numerically using the TFM.
The problem is then only that numerical solutions themselves are limited
in just the same way as those for a full numerical solution in the solar
and stellar dynamo problems.

\begin{figure}[t]
\centering\includegraphics[width=\columnwidth]{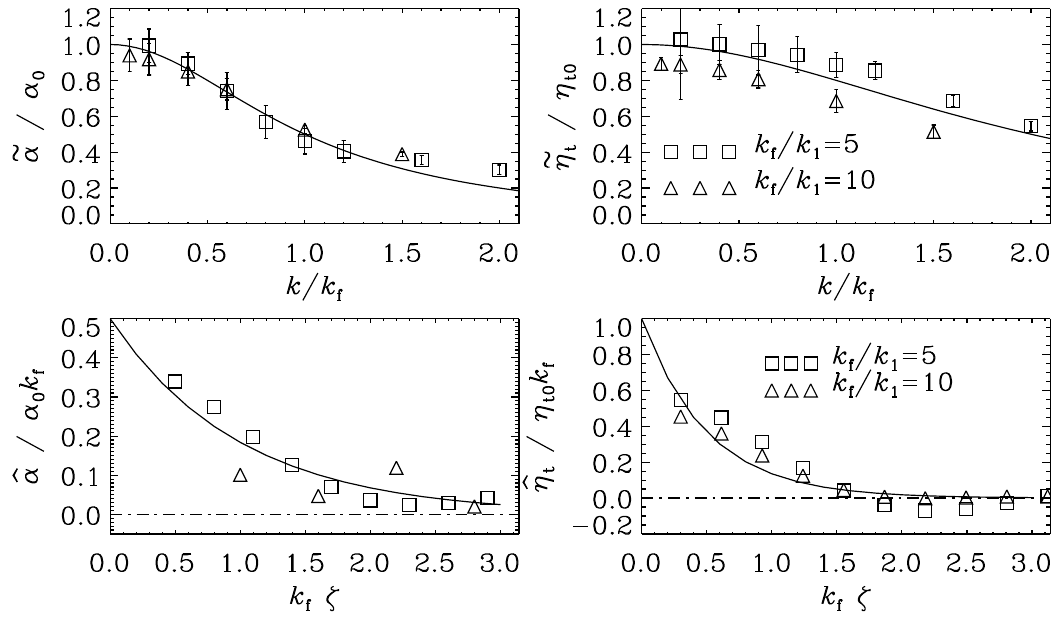}\caption{
Top: Dependences of the normalized $\tilde\alpha$ and $\tilde\eta_{\rm t}$
on the normalized wavenumber $k/k_{\rm f}$
for isotropic turbulence forced at wavenumbers $k_{\rm f}/k_1=5$ with $\Rm=10$ (squares)
and $k_{\rm f}/k_1=10$ with $\Rm=3.5$ (triangles), all with $\nu/\eta=1$,
using data from \cite{BRS08}.
The solid lines give the Lorentzian fits \eq{KernelsTurb}.
Bottom: Normalized integral kernels $\hat\alpha$ and $\hat\eta_{\rm t}$ versus
$k_{\rm f}\zeta$ for isotropic turbulence forced at wavenumbers $k_{\rm f}/k_1=5$ with $\Rm=10$ (squares)
and $k_{\rm f}/k_1=10$ with $\Rm=3.5$ (triangles), all with $\nu/\eta=1$.
The solid lines are defined by \eq{KernelsTurb2}.
Adapted from \cite{BRS08}.
}\label{ppk_both}\end{figure}

\FFig{ppk_both} shows results for $\tilde\alpha(k)$ and $\tilde\eta_{\rm t}(k)$
with $\nu / \eta = 1$.
Both $\tilde\alpha$ and $\tilde\eta_{\rm t}$ decrease monotonously with increasing $|k|$.
The functions $\tilde\alpha (k)$ and $\tilde\eta_{\rm t} (k)$
are well represented by Lorentzian fits of the form
\EQ
\tilde\alpha(k)\approx\frac{\alpha_0}{1+(k/k_{\rm f})^2} \, ,\quad
\tilde\eta_{\rm t}(k)\approx\frac{\eta_{\rm t0}}{1+(k/2k_{\rm f})^2} \, .
\label{KernelsTurb}
\EN

Also shown in the lower part of \Fig{ppk_both} are the kernels
$\hat\alpha (\zeta)$ and $\hat\eta_{\rm t} (\zeta)$, obtained
through Fourier transforms of the Lorentzian fits,
\EQ
\hat\alpha(\zeta)\approx\half\alpha_0 k_{\rm f} \exp(-k_{\rm f}|\zeta|) \, ,\quad
\hat\eta_{\rm t}(\zeta)\approx\eta_{\rm t0} k_{\rm f} \exp(-2k_{\rm f}|\zeta|) \, .
\label{KernelsTurb2}
\EN
We see that the profile of $\hat\eta_{\rm t}$ is half as wide as that
of $\hat\alpha$, but it is not known whether this is a general property.
It is important to realize that the suggested mean-field modifications
employing the Lorentzian forms of the integral convolution kernels are
based on empirical results.
Nevertheless, they are much more accurate than the approximation of
replacing the kernels by $\delta$ functions, which is done in
conventional approaches.

Under suitable conditions, the accuracy of the TFM can be so high that
discrepancies become apparent that are solely the result of having made
unjustified approximations in the comparison.
An example is the memory effect.
Comparing the growth rate for a supercritical dynamo with that obtained
theoretically from the coefficient obtained from the TFM can give
noticeable discrepancies if the memory effect is neglected; see Fig.~1
of \cite{HB09}.
The combined Fourier transformed integral kernel is of the form
\begin{equation}
\tilde\alpha(k,\omega)\approx\frac{\alpha_0}{1+(k/k_{\rm f})^2-\ii\omega\tau} \, ,\quad
\tilde\eta_{\rm t}(k,\omega)\approx\frac{\eta_{\rm t0}}{1+(k/2k_{\rm f})^2-\ii\omega\tau} \, ,
\label{KernelsTurb_om}
\end{equation}
where $\tau$ is well approximated by the turbulent turnover time.
Even for stationary flows, the memory effect can be dramatically important
\citep{Radl+11}.

In practice, it is cumbersome to solve the integral equation in time.
However, as alluded to above, it is possible to approximate this integral
equation by a differential equation for $\meanEMF$ with respect to space
and time $t$ of the form
\begin{equation}
\left(1+\tau\frac{\partial}{\partial t}-\ell^2\nabla^2\right)
\meanemf_i=\alpha_{ij}\meanB_j+\eta_{ijk}\partial\meanB_j/\partial x_k.
\label{evol}
\end{equation}
This has been done in several papers \citep{RB12,Rhei+14,BC18,Pipin23}. 
We return to this in \Sec{DynamosFromMemory}.

\subsection{The use of mean-field theory}\label{umft}

If mean-field theory cannot reliably be applied to a regime outside
that of the direct numerical simulations (DNS), one must ask what is
then the use of mean-field theory.
The answer lies in the fact that mean-field theory provides us with an
excellent diagnostic ``tool'' for approaching the problem.
Particular features of a solution can usually be attributed to particular
terms in the mean-field equations.
This would then allow us a more informed answer by saying that the main
dynamo mechanism is, for example, of $\alpha\Omega$ type, or of a specific
type of a shear flow dynamo, for example.
{\em Thus, mean-field theory may be regarded as a convenient tool for
understanding what is going on rather than predicting what might be
going on.}

\section{The catastrophic quenching problem}

Since the 1990s, a problem emerged in that 
numerical dynamo solutions were found to depend on the value of
the microphysical magnetic diffusivity.
Typically, the strength of the mean-fields then decreases with
increasing magnetic Reynolds number.
This is unusual and does not have any correspondence with ordinary
hydrodynamics where the large-scale dynamics is usually already
captured at moderate fluid Reynolds numbers.
In its original form, the catastrophic quenching problem refers to the
finding that the volume-averaged electromotive force scales with the
microphysical magnetic diffusivity, and thus goes to zero when $\eta\to0$.
To some extent, this is a problem related to the use of periodic boundary
conditions.
However, even for astrophysically more realistic boundary conditions,
numerical simulations reveal that there is still a problem.
Plasma relaxation experiments have identified the role of magnetic
helicity as the main culprit in causing $\eta$-dependent large-scale
dynamics and catastrophic quenching.
We therefore begin by briefly reviewing the essential findings.

\subsection{Lessons from plasma relaxation experiments}

The magnetic field is divergence-free and can therefore be written as
$\BB=\nab\times\AAA$, where $\AAA$ is the magnetic vector potential.
The magnetic helicity density is defined as $\AAA\cdot\BB$.
Its evolution equation follows directly from the uncurled induction
equation, $\partial\AAA/\partial t=-\EE-\nab\Phi$, or, using Ohm's law,
$-\EE=\UU\times\BB-\eta\mu_0\JJ$, so
\begin{equation}
\frac{\partial\AAA}{\partial t}=\UU\times\BB
-\eta\mu_0\JJ-\nab\Phi.
\end{equation}
It yields the evolution equation for the magnetic helicity density,
\begin{equation}
\frac{\partial}{\partial t}\left(\AAA\cdot\BB\right)
=-2\eta\mu_0\,\JJ\cdot\BB
-\nab\cdot\left(\EE\times\AAA+\Phi\BB\right).
\label{ABdens}
\end{equation}
It must here be emphasized that there is an important difference to the
equation for the magnetic energy density,
\begin{equation}
\frac{\partial}{\partial t}\left(\BB^2/2\mu_0\right)
=-\UU\cdot(\JJ\times\BB)
-2\eta\mu_0\,\JJ^2
-\nab\cdot\left(\EE\times\BB/\mu_0\right).
\label{B2dens}
\end{equation}
While both \Eqs{ABdens}{B2dens} have analogous terms such as dissipation
$\propto\JJ\cdot\BB$ versus $\propto\JJ^2$, respectively, and flux terms
$\EE\times\AAA$ versus Poynting vector $\EE\times\BB/\mu_0$, respectively,
there is the work against the Lorentz force, $-\UU\cdot(\JJ\times\BB)$
in \Eq{B2dens}, which would be $\UU\cdot(\BB\times\BB)$ in \Eq{ABdens},
but it obviously vanishes.
In statistical equilibrium, $\bra{2\eta\mu_0\,\JJ^2}$ must balance
$-\bra{\UU\cdot(\JJ\times\BB)}$, which implies that the current density
diverges like $|\JJ|\sim\eta^{-1/2}$.
By contrast, no magnetic helicity is being produced, and also its
dissipation converges to zero like $\propto\eta|\JJ\cdot\BB|\to\eta^{1/2}$
as $\eta\to0$.

Already since the 1970s, we know of the conjecture of J.\ B.\
\cite{Taylor74, Taylor86} that the magnetic field relaxes under the
constraint of magnetic helicity conservation to a nearly force-free
field to minimize dissipation.
The approximate conservation of magnetic helicity has been verified
experimentally in plasma relaxation experiments; see, e.g., \cite{Ji+95}.
There are obviously some differences between the solar convection zone
and laboratory plasmas, for example, the role of the electron pressure
in the generalized Ohm's law could play an important role in explaining
why magnetic helicity changes are observed to be faster in plasma
experiments than what is predicted by Ohm's law \citep{Ji+95}.
Furthermore, the $\alpha$ effect has been identified as the main agent
for converting magnetic helicity from the turbulent field to the mean
field \citep{Ji99}.

In the context of plasma relaxation experiments, it is useful to
distinguish between electromagnetic and electrostatic turbulence.
This distinction refers to the curl-free and divergence-free parts of
the electric field written as $\EE=-\nab\Phi-\partial\AAA/\partial t$.
In plasma relaxation experiments, turbulence is mostly electrostatic.
It can be affected by the electron pressure gradient $(e n_{\rm e})^{-1}
\nab p_{\rm e}$ in the generalized Ohm's law, where $e$ is the unit
charge, and $n_{\rm e}$ and $p_{\rm e}$ are the electron density and
electron pressure, respectively; see \cite{Ji99} for details.
This leads to a battery term $\propto\nab n_{\rm e}\times\nab p_{\rm e}$
in the equation for $\partial\BB/\partial t$ and to a magnetic helicity
flux, which transports magnetic helicity across physical space, as
opposed to wavenumber space \citep{Ji99}.

There is the possibility that the divergence of a magnetic helicity
flux, $\meanFFFFf$, itself can constitute an $\alpha$ effect.
This corresponds to $\alpha=-\half\BB^{-2}\nab\cdot\meanFFFFf$; see
\cite{Vishniac2001}, who have derived a specific form for such a flux.
The subscript `f' indicates that the flux originates from correlations
of the fluctuating magnetic field.
Mean-field models of the type described below have shown that a dynamo
can operate even without kinetic helicity, i.e., it is based only on
shear and current helicity fluxes, provided a nondimensional scaling
factor in front of the magnetic helicity flux exceeds a certain critical
value \citep{BS05c}.
However, there are so far no DNS that have supported this kind of
behavior, nor has the proposed flux been confirmed \citep{Hubbard2012}.
Nevertheless, the idea of an $\alpha$ effect being related to the magnetic
helicity flux divergence is certainly consistent with the laboratory
experiment presented in Fig.~1 of \cite{Ji99}.

The $\alpha$ effect reflects the physics of the inverse cascade of
magnetic helicity \citep{pouquet+76}.
In the absence of energy input, this is known to lead to a slower
turbulent decay of magnetic energy $\propto t^{-2/3}$ \citep{Hat84,
BM99}.
In hydrodynamics, by comparison, the kinetic energy density decays like
$t^{-10/7}$ or $t^{-6/5}$, depending on the initial subinertial range
energy spectrum \citep{Davidson00}; see also \cite{BL23} for a comparison
with the magnetic case.

In the presence of magnetic driving by applying a voltage drop along
the magnetic field, small-scale instabilities such as the tearing
instability develop.
This leads to a sawtooth-like time dependence in the mean toroidal
magnetic flux; see \cite{Ji+Prager02} for a review.
This can be associated with the resulting development of a mean
electromotive force, $\meanEMF=\overline{\uu\times\bb}$, along with an
$\alpha$ effect that accomplishes the helicity transport \citep{Ji99}.

Unlike astrophysical dynamos, which are generally understood as
self-excited ones, the plasma experiments operate in a regime where a
magnetic field is always present, but it is then redistributed by the
$\alpha$ effect.
The conceptional similarities and differences have been discussed in
detail by \cite{BJ06}.
In the following, we discuss in more detail the consequences imposed by
magnetic helicity evolution in astrophysical dynamos.
It is important to emphasize, however, that the same physics that is
used to explain the catastrophic quenching phenomenology also applies
to plasma experiments such as the reversed field pinch,
as was shown in corresponding mean-field simulations by \cite{Kemel+11}.

\subsection{Mean fields in periodic domains}

Under astrophysical conditions of interest, $\eta$ is so small that
the volume-averaged electromotive force would be negligibly small.
If this result was actually astrophysically relevant, it would be a
``catastrophe,'' i.e., it would not be possible to understand astrophysical
magnetic fields as mean-field dynamos.
The solution to this particular problem turned out to be that relating
the volume-averaged electromotive force to the volume-averaged mean
magnetic field is only of limited relevance to the problem of $\alpha$
effect dynamos.
Any dynamo would produce a non-uniform field.
Especially in a periodic domain, the mean magnetic flux through any of
the faces of the periodic domain is constant in time, so if it was zero
to begin with, it would always remain zero.
A dynamo problem can therefore not be formulated in that case.

A proper dynamo problem should always allow for the possibility of
the magnetic field to decay to zero if there is sufficient magnetic
diffusivity.
Simple examples of nontrivial mean fields in a periodic domain are
Beltrami fields of the form
\begin{equation}
\meanBB(x)\propto\begin{pmatrix}0\\ \sin kx\\ \cos kx\end{pmatrix},\quad
\meanBB(y)\propto\begin{pmatrix}\cos ky\\ 0\\ \sin ky\end{pmatrix},\quad\mbox{or}\quad
\meanBB(z)\propto\begin{pmatrix}\sin kz\\ \cos kz\\ 0\end{pmatrix},
\label{Belt}
\end{equation}
which can be solutions of the simple $\alpha^2$ dynamo problem,
$\partial\meanBB/\partial t=\alpha\nab\times\meanBB+\etaT\nabla^2\meanBB$.
Nevertheless, there is still a problem of catastrophic nature because
it turned out that the time required to reach the final solution scales
inversely with $\eta$.
This is demonstrated in \Fig{psat}, where we show the evolution of one of
the three planar averages.
In the beginning, all three mean fields grow in a similar fashion,
but at some point, only one of the three reaches a significant amplitude.
Note, however, that the ultimate saturation takes a resistive time,
$\tau_{\rm res}=1/(2\eta k_1^2)$.

\subsection{Quenching phenomenology}

To understand the reason for the catastrophically slow saturation, it
suffices to consider \Eq{ABdens} for the magnetic helicity density.
For periodic domains, we just have
\begin{equation}
\frac{\dd}{\dd t}\bra{\AAA\cdot\BB}=-2\eta\mu_0\bra{\JJ\cdot\BB}.
\label{dABdt}
\end{equation}
This equation is gauge-independent, because the gauge
transformation $\AAA\to\AAA'+\nab\Lambda$ yields
$\bra{\AAA\cdot\BB}=\bra{\AAA'\cdot\BB}$, with
$\bra{\BB\cdot\nab\Lambda}=\bra{\nab\cdot(\Lambda\BB)}
-\bra{\Lambda\nab\cdot\BB}=0$, using $\nab\cdot\BB=0$ and the fact that
the domain is periodic, so the average of a divergence vanishes.
In numerical calculations, it is often convenient to adopt the Weyl
gauge, which implies that $\partial\AAA/\partial t=-\EE$, i.e., the
scalar potential term drops out.

\begin{figure}[t]\begin{center}
\includegraphics[width=\columnwidth]{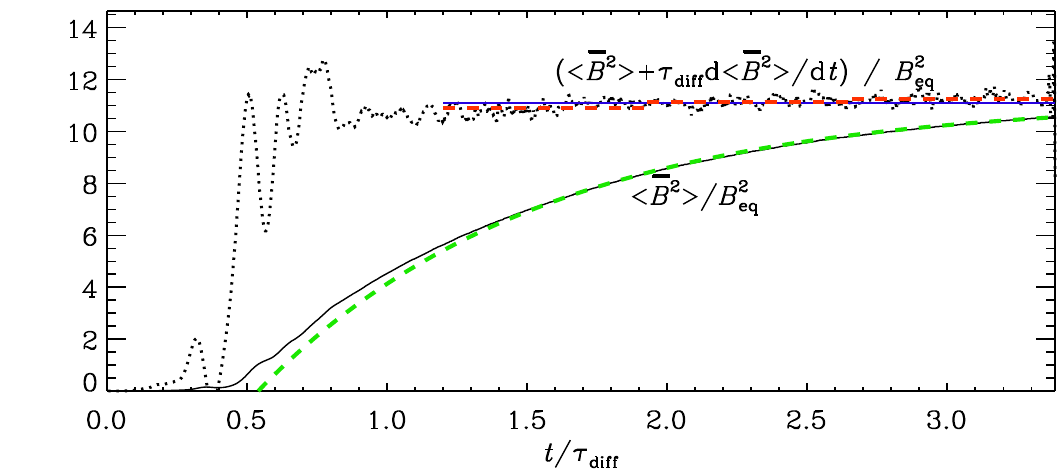}
\end{center}\caption[]{
Evolution of the normalized $\bra{\meanBB^2}$
and that of $\bra{\meanBB^2}+\tau_{\rm diff}\dd\bra{\meanBB^2}/\dd t$ (dotted),
compared with its average in the interval $1.2\leq t/\tau_{\rm diff}\leq3.5$
(horizontal blue solid line), as well as averages over three subintervals
(horizontal red dashed lines).
The green dashed line corresponds to \Eq{B2vol} with
$t_{\rm sat}/\tau_{\rm diff}=0.54$.
Adapted from \cite{CB13}.
}\label{psat}\end{figure}

For fully helical large-scale and small-scale magnetic fields
of opposite magnetic helicity, \Eq{dABdt} becomes \citep{Bra01}
\EQ
{\dd\over\dd t}\bra{\meanBB^2}=2\eta k_1\kf\Beq^2-2\eta k_1^2\bra{\meanBB^2},
\label{dB2voldt}
\EN
with the solution
\EQ
\bra{\meanBB^2}=\Beq^2{\kf\over k_1}
\left[1-e^{-2\eta k_1^2(t-t_{\rm sat})}\right].
\label{B2vol}
\EN
This agrees with the slow saturation behavior seen first in the simulations
of \cite{Bra01}; see \Fig{psat}.
Here, $t_{\rm sat}$ is the time when the slow saturation phase commences;
see the crossing of the green dashed line with the abscissa.
Interestingly, instead of waiting until full saturation is accomplished,
one can obtain the saturation value already much earlier simply by
differentiating the simulation data to compute \citep{CB13}
\EQ
B_{\rm sat}^2\approx
\bra{\meanBB^2}+\tau_{\rm res}{\dd\over\dd t}\bra{\meanBB^2}.
\EN
Since $\tau_{\rm res}$ involves the microphysical magnetic diffusivity,
the quenching is still in that sense catastrophic.

\subsection{The $\alpha$ quenching formula}

A more complete description is in terms of kinetic and magnetic $\alpha$
effects, i.e.,
\begin{equation}
\alpha=\alpK+\alpM\approx-\frac{\tau}{3}\left(\overline{\oo\cdot\uu}
-\overline{\jj\cdot\bb}/\meanrho\right),
\label{alphaFOSA}
\end{equation}
and observing the fact that the magnetic helicity evolution of averages
and fluctuations is given by
\begin{equation}
\frac{\dd}{\dd t}\bra{\meanAA\cdot\meanBB}=
+2\bra{\meanEMF\cdot\meanBB}-2\eta\mu_0\,\bra{\meanJJ\cdot\meanBB},
\label{dAmBmdt}
\end{equation}
\begin{equation}
\frac{\dd}{\dd t}\bra{\aaaa\cdot\bb}=
-2\bra{\meanEMF\cdot\meanBB}-2\eta\mu_0\,\bra{\jj\cdot\bb}\label{smhev}.
\end{equation}
\EEq{dAmBmdt} allows for the possibility that magnetic helicity
can be produced by the mean electromotive force, because, in general,
$\meanEMF\cdot\meanBB\equiv\overline{\uu\times\BB}\cdot\meanBB\neq0$.
(By contrast, of course, $(\uu\times\meanBB)\cdot\meanBB=0$.)
In particular, if $\meanEMF=\alpha\meanBB-\etat\mu_0\meanJJ$, then,
$\meanEMF\cdot\meanBB=\alpha\meanBB^2-\etat\mu_0\meanJJ\cdot\meanBB$,
which produces positive (negative) magnetic helicity of the mean field
when $\alpha>0$ ($\alpha<0$)

\EEq{smhev} is constructed such that the sum of \Eqs{dAmBmdt}{smhev}
yields \Eq{dABdt}.
Given that $\bra{\aaaa\cdot\bb}$ is related to $\bra{\jj\cdot\bb}$,
which, in turn, is related to a magnetic contribution to the $\alpha$
effect \citep{pouquet+76}, \Eq{smhev} can be rewritten as an evolution equation
for the total $\alpha$ \citep{Bra08AN},
\begin{equation}
\frac{\dd\alpM}{\dd t}=-2\etatz\kf^2 \left(
\frac{\alpha\meanBB^2-\etat\mu_0\meanJJ\cdot\meanBB}{\Beq^2}
+\frac{\alpM}{\Rm}\right),
\end{equation}
which can also be expressed in the form
\begin{equation}
\alpha(\meanBB)=\frac{\alpha_0+\Rm\times\mbox{``extra terms''}}
{1+\Rm\meanBB^2/\Beq^2}
\label{quenching}
\end{equation}
where
\begin{equation}
\mbox{``extra terms''}=
\etat\frac{\mu_0\meanJJ\cdot\meanBB}{\Beq^2}
-\frac{\nab\cdot\meanFFFFf}{2\kf^2\Beq^2}
-\frac{\partial\alpha/\partial t}{2\kf^2\Beq^2}.
\end{equation}
Note that the last term is here a time derivative.
\EEq{quenching} resembles the catastrophic quenching formula of \cite{VC92},
but it also shows that it needs to be extended in several important
ways: when the mean field is no longer defined as a volume average,
extra terms emerge that are of  the same order as those in the denominator.
They can therefore potentially offset the catastrophic quenching.
In practice, this is only partially true, because there are also other
terms, for example the aforementioned time derivative term.
It is responsible for the fact that a strong field state is only reached
after a resistively long time.

\subsection{Magnetic helicity fluxes and helicity reversals}

Magnetic helicity fluxes could in principle remove the catastrophic
quenching problem, but only if preferentially small-scale magnetic
helicity is being removed \citep{Kleeorin2000}.
To see this, let us first consider the problem of an $\alpha^2$
dynamo in insulating boundaries using the Weyl gauge, i.e.,
\begin{equation}
\frac{\partial}{\partial t}\meanAA=\alpha\meanBB-\etaT\mu_0\meanJJ,\quad
\mbox{with \; $\partial_z\meanA_x=\partial_z\meanA_y=\meanA_z=0$}.
\end{equation}
The boundary condition implies that $\meanB_x=\meanB_y=0$, and is
therefore also referred to as the vertical field condition.
In this 1-D problem, however, this boundary condition is equivalent
to a proper vacuum boundary condition.

\begin{figure}[t!]\begin{center}
\includegraphics[width=\columnwidth]{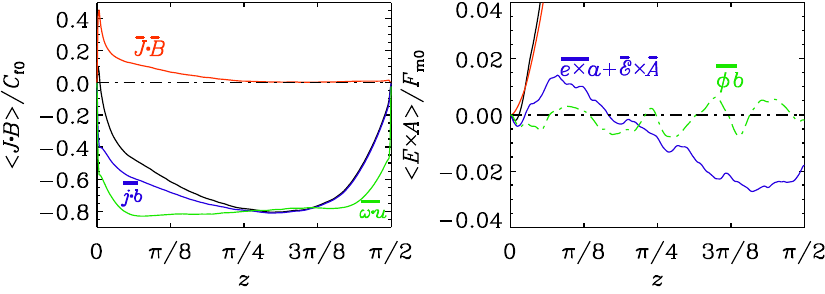}
\end{center}\caption[]{
Magnetic helicity, current helicity, and magnetic helicity fluxes
for Run~A of \cite{Brandenburg2018AN} with $\Rm=180$.
The kinetic helicity is shown in green and is found to be of similar
magnitude as the current helicity of the small-scale field.
The second panel shows $\overline{\EE\times\AAA}$ near zero.
The green line denotes $\overline{\phi\bb}$, which is seen
to fluctuate around zero.
}\label{pphelflux_2panels_M288h_cont2}\end{figure}

The $\alpha^2$ dynamo with this boundary condition was first considered
by \cite{GD94}, who found that the saturation field strength of such
a dynamo decreases with $\Rm$.
This was later confirmed by \cite{BD01}.
In \Fig{pphelflux_2panels_M288h_cont2}, we show the profiles
of magnetic helicity, current helicity, and the magnetic helicity fluxes
for Runs~A of \cite{Brandenburg2018AN} with $\Rm=180$.
The computational domain is $0\leq z\leq\pi/2$ with a perfect conductor
boundary condition on $z=0$ and a vertical field condition on $z=\pi/2$;
see \cite{Bra17} for the relevant mean-field models.
For normalization purposes, he defined the reference values
\EQA
C_{\rm f0}=\kf\Beq^2\quad\mbox{and}\quad
F_{\rm m0}=\etatz k_1^2\int_0^{\pi/2}\meanBB^2\,\dd z.
\ENA
He emphasized that the largest contribution to the magnetic helicity
density comes from the large-scale field.
Near the surface ($z=\pi/2$), the (negative) magnetic helicity flux from
small-scale fields is only about $0.02\,F_{\rm m0}$, which explains why
they are not efficient enough to alleviate the catastrophic dependence
of the resulting mean magnetic field \citep{DSordo2013MN, Rincon21}.

\begin{figure}[t!]\begin{center}
\includegraphics[width=\columnwidth]{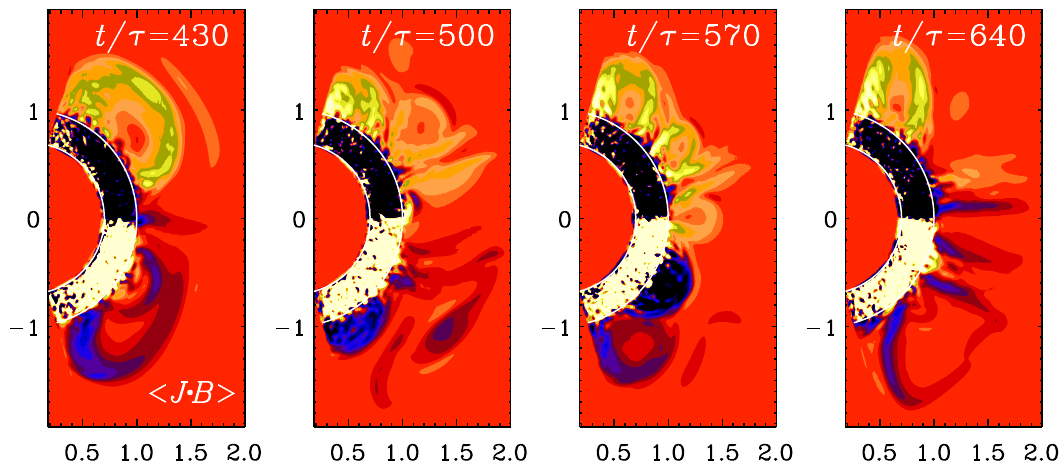}
\end{center}\caption[]{
Current helicity $\bra{\JJ\cdot\BB}$ in the plane of the observer
at four different times.
Yellow and white shades denote positive values and
blue and black shades denote negative values;
adapted from \cite{BAJ17}.
}\label{ppfft3_pi_times}\end{figure}

Subsequent simulations with an outside corona indicated that the magnetic
helicity changes sign at or near the outer surface \citep{BCC09}.
This was just a speculation and needs to be reconsidered with the
help of global models of the type considered by \cite{War+11,War+12}
and \cite{BAJ17}.
This is shown in \Fig{ppfft3_pi_times}, where we present the line-of-sight
averaged current helicity density, $\bra{\JJ\cdot\BB}$ in the plane of
the sky using a simulation of \cite{BAJ17}.
The quantity $\bra{\JJ\cdot\BB}$ is a proxy of magnetic helicity at
small scales and shows clearly the reversal of sign between the dynamo
interior and the exterior.

\subsection{Radial magnetic helicity reversal in the solar wind}

If the idea of alleviating catastrophic quenching by magnetic helicity
fluxes is to make sense, we would expect to see signs of the expelled
magnetic helicity in the solar wind.
The magnetic helicity spectrum can be measured in the solar wind by
determining the parity-odd contribution to the magnetic correlation
tensor, which, in Fourier space, takes the form
\begin{equation}
\bra{\tilde{B}_i(\kk)\tilde{B}^*_j(\kk)}
=\left(\delta_{ij}-\hat{k}_i\hat{k}_j\right)\,2E(k)
-\ii\hat{k}_k\epsilon_{ijk}H(k).
\end{equation}
This would allow one to compute $H(k_z)=\Imag(\tilde{B}_x\tilde{B}_y^*)$
and $E(k_z)=\half(|\tilde{B}_x|^2+|\tilde{B}_y|^2)$, which also obeys
the realizability condition $k_z|H(k_z)|\leq 2E(k_z)$.

\begin{figure}[t]
\begin{center}
\includegraphics[width=\textwidth]{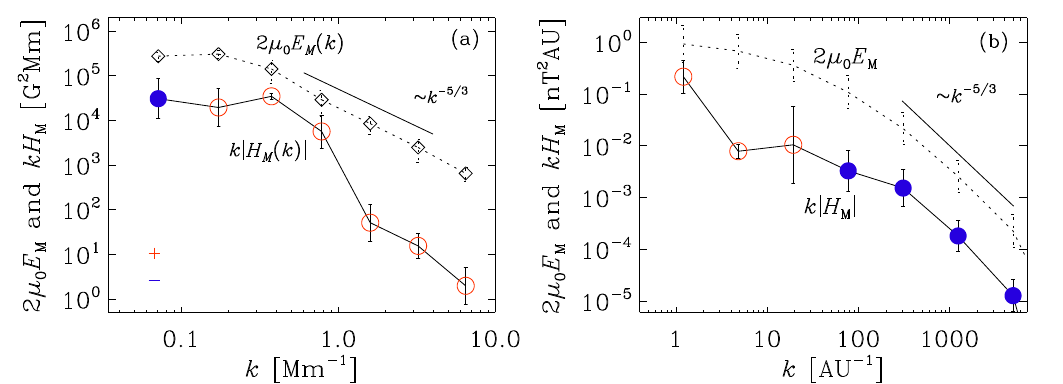}
\end{center}
\caption{
Magnetic energy and magnetic helicity spectra for southern latitudes
(a) at the solar surface in active region AR~11158, and
(b) in the solar wind at $\sim1\AU$ distance ($1\AU\approx149,600\Mm$).
Positive (negative) signs are shown as red open (blue filled) symbols.
Positive signs are the solar surface at intermediate and large $k$
correspond to positive values in the solar wind at small $k$.
Note that $1\G=10^{-4}\T=10^5\nT$.
}\label{phelicity_plat4b_Lidingo}
\end{figure}

The Ulysses spacecraft was the only one to cover high heliographic
latitudes, where a non-vanishing sign of magnetic helicity can be
expected.
It turned out that $H(k)$ has, as expected from dynamo theory, different
signs in the northern and southern hemispheres.
It also has different signs at small and large wavenumbers.
This, in itself, is also expected from an $\alpha^2$ dynamo, because
the $\alpha$ effect produces no net magnetic helicity, but it separates
magnetic helicity in wavenumber space.
However, the signs are opposite to what is seen at the solar surface,
where the helicity in the north is negative at small length scales.
In the solar wind, however, it is positive in the north and at small
scales.
Of course, the meaning of small is here relative and has to be with
respect to larger scales, where a sign change in $k$ has been seen.
If one just assumed a linear expansion of all scales from the solar
surface (radius $r=700\Mm$, to the location of the Earth at $1\AU$,
we expect a corresponding expansion ratio so that a wavenumber of
$1\Mm^{-1}$ corresponds to $1/700\AU^{-1}$.
In particular, $20\Mm^{-1}$ corresponds to $2/70\AU^{-1}$,
which is close to the wavenumber where we see a sign-change in
\Fig{phelicity_plat4b_Lidingo}.
It is unexpected, however, that at the solar surface
(\Fig{phelicity_plat4b_Lidingo}b), the sign in the northern hemisphere
changes from minus to plus as $k$ increases, while in the solar wind,
it changes from plus to minus.
This apparent mismatch may not just be a measurement error, but it may
actually be a real result and would tell us that the simpleminded picture
of expelling magnetic helicity of one sign all the way to infinity may
not be accurate.

When the domain is inhomogeneous, and especially when there are boundaries,
magnetic helicity fluxes are possible and \Eq{smhev} takes the form
\begin{equation}
\frac{\partial}{\partial t}\overline{\aaaa\cdot\bb}=
-2\meanEMF\cdot\meanBB-2\eta\mu_0\,\overline{\jj\cdot\bb}
-\nab\cdot\meanFFFFf.
\label{smhev_flux}
\end{equation}
In the steady state, we have
\begin{equation}
\nab\cdot\meanFFFFf=\underbrace{{-2\alpha\meanBB^2
+2\etat\mu_0\meanJJ\cdot\meanBB}}_{-2\meanEMF\cdot\meanBB}
-2\eta\mu_0\,\overline{\jj\cdot\bb}.
\end{equation}
In the dynamo interior at the northern hemisphere, $\alpha>0$,
and, assuming $\alpha\BB^2$ to dominate the EMF, we expect
$-2\meanEMF\cdot\meanBB$ to be negative.
However, a negative flux divergence of a negative quantity would
eventually make this quantity positive, which is what has been observed.

Whether or not this is really the right interpretation remains still
an open question.
It would clearly be useful to have an independent assessment of this
interpretation.

\subsection{Nonlocal effects of $\meanEMF$ and catastrophic quenching}

Catastrophic quenching in large-scale dynamos is a rather general property.
It is a consequence of the build-up of magnetic helicity of the mean magnetic field.
It has been conjectured that catastrophic quenching would be prevented
if the sources of toroidal field generation are spatially separated from
the sources of the poloidal field; see, e.g., \cite{TobiasWeiss2007}.
This would be the case in what is known as interface dynamos \citep{Par93}.
It could also be through a nonlocal $\alpha$ effect.
Such a nonlocal $\alpha$ effect is an essential ingredient of the
Babcock--Leighton and flux-transport dynamo models; see, \cite{Hazra2023}.
The studies of \cite{BranKap2007} and \cite{Chatterjee2010G} showed that
a spatial separation between shear and $\alpha$ effects does in general
not help to avoid catastrophic quenching for such types of dynamo models.
It is interesting, however, that \cite{KitOle2011AN} and later also
\cite{BranHubKap2015} found that the inclusion of diamagnetic downward
pumping of the toroidal magnetic field can alleviate the catastrophic
quenching in the Babcock--Leighton dynamo model with a strongly
nonlocal $\alpha$.

The catastrophic quenching models are reasonably well reproduced by
DNS when the geometries of the setups are sufficiently simple.
It would therefore be worthwhile to apply DNS to conditions where
turbulent pumping and a strongly nonlocal mean electromotive force can
be expected.
At present, however, even just the physical reality of a nonlocal
$\alpha$ effect of Babcock--Leighton type through the decay of active
regions rests mainly on the interpretation of solar observations.
Turbulence simulations have so far not been able to make contact
with such concepts.

\section{Alternative large-scale dynamo effects}

Given the difficulties encountered with $\alpha$ effect dynamos, there
have been various attempts to construct large-scale dynamos that are
not based on the $\alpha$ effect.
A common misconception here is the idea that catastrophic quenching
would not apply if just because there is no $\alpha$ effect. This is
not true, because an $\alpM$ term can always emerge regardless of whether
or not there existed an original $\alpha$ effect.
An example is the shear--current effect.
It is due to the presence of shear and boundaries that a helicity can
be introduced.
Shear of the form $\meanUU=(0,Sx,0)$ implies a finite vorticity,
$\nab\times\meanUU=(0,0,S)$ and boundaries would lead to a gradient
vector of turbulent intensity near the boundaries.
Thus, while there can be hope that catastrophic quenching may not be as
strong, this may turn out not to be the case.
An example of this was presented in \cite{BS05c}.

\subsection{R\"adler and shear--current effects}

The R\"adler effect is another large-scale dynamo effect \citep{Raedler1969}.
In the simplest representation it leads to an EMF proportional to $\OO\times\meanJJ$.
It is similar to the shear--current effect. In this case it cannot change the
magnetic energy of the mean field.
Indeed, the energy equation for the mean field is given by
\begin{equation}
\frac{\dd}{\dd t}\bra{\meanBB^2/2}=
\underbrace{\meanJJ\cdot(\OO\times\meanJJ)}_{=0}
+\underbrace{\bra{\nab\cdot[(\OO\times\meanJJ)\times\meanBB]}}_{\mbox{$=0$ under periodicity}}.\label{engdelt}
\end{equation}
In the general case, the generation effects due to global
rotation and mean currents can be written as follows
(see \citealp{KR80, Kitchatinov1994,Raedler2003a,Pipin2008a}):
\begin{equation}
\meanEMF^{(\delta)}=\delta_{1}\boldsymbol{\Omega}\times\meanJJ+\delta_{2}\boldsymbol{\nabla}\left(\boldsymbol{\Omega}\cdot\meanBB\right)
+\delta_{3}\frac{\boldsymbol{\Omega}\left[\boldsymbol{\Omega}\cdot \boldsymbol{\nabla}\left(\boldsymbol{\Omega}\cdot\meanBB\right)\right]
}{\boldsymbol{\Omega}^{2}},\label{deltas}
\end{equation}
where the coefficients $\delta_{1,2,3}$ depend on the spatial profiles of
the turbulent parameters such as the typical convective turnover time, 
the convective velocity $\urms$, etc.
The last two terms in this equation may lead to an $\delta^2$ dynamo
\citep{Pipin2009}.
For the solar case, the $\delta$ effect can provide an additional
non-helical source of poloidal magnetic field generation.
Interestingly, \cite{Pipin2009} found that for the solar-type dynamos,
i.e., those with equatorward propagation of the dynamo waves, the $\delta$
dynamo effect does not dominate the contributions of the $\alpha$-effect.
We will discuss the available scenario in the next section.

\subsection{Dynamos from negative turbulent magnetic diffusivity}

There are two other effects that are noteworthy, although it is not
clear that either of them can play a role in stellar convection zones.
One is the negative turbulent magnetic diffusivity and the other is
the memory effect in conjunction with a pumping effect.

When modeling a negative turbulent magnetic diffusivity dynamo,
high wavenumbers must not be destabilized at the same time.
\cite{BC20} studied classes of dynamos with a very low critical $\Rm$.
The Willis dynamo \citep{Willis12} has a critical $\Rm$ of 2.01, which
is small compared to 6.3 for the Roberts flow and 17.9 for the ABC flow.
In this dynamo, one of the two horizontally averaged field components
grows exponentially, because the total magnetic diffusivity in that
direction is negative \citep{BC20}.
The other component decays and is not coupled to the former one.

\begin{figure}[t!]\begin{center}
\includegraphics[width=0.9\columnwidth]{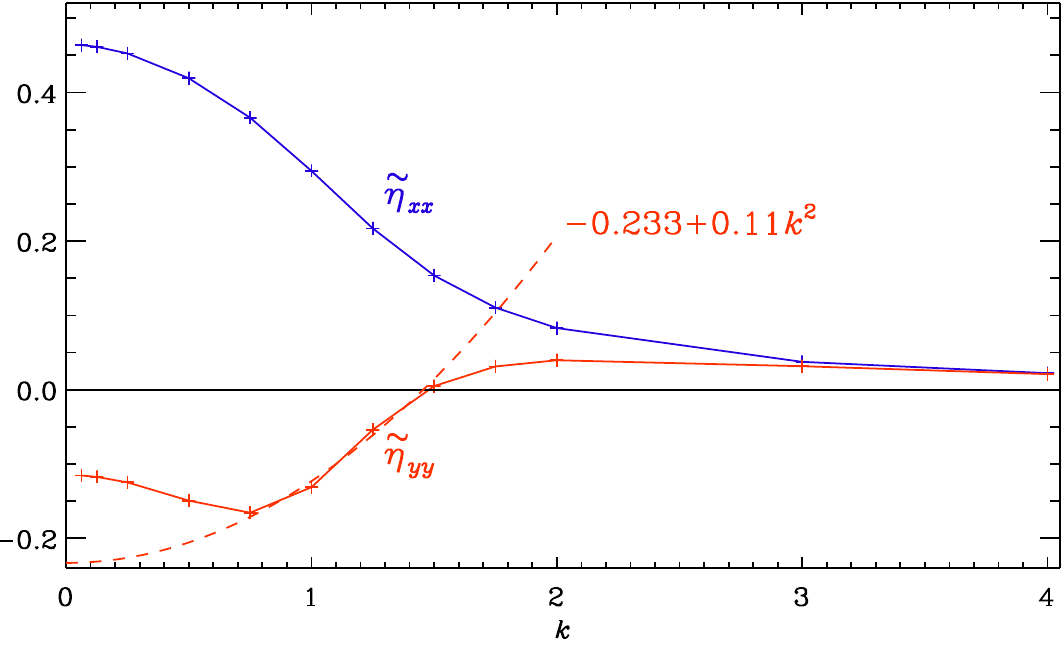}
\end{center}\caption[]{
Dependence of $\tilde{\eta}_{xx}$ (blue) and $\tilde{\eta}_{yy}$ (red) on $k$
for the Willis flow in the marginally exited case with $\eta=0.403$.
The red dashed line denotes the fit $-0.233+0.11\,k^2$.
Adapted from \cite{BC20}.
}\label{pom0k}\end{figure}

As we see from \Fig{pom0k}, $\etat$ is negative only for $k\la1.5$.
The $k$ dependence of the turbulent magnetic diffusivity can be expanded
up to second order as
\begin{equation}
\tilde{\eta}_{yy}(k)=\tilde{\eta}_{yy}^{(0)}+\tilde{\eta}_{yy}^{(2)}k^2+
\ldots,
\end{equation}
where the tildes indicate Fourier transformed quantities.
In the proximity of $k=1$, which corresponds to the
largest scale in the computational domain of $2\pi$,
we have $\tilde{\eta}_{yy}^{(0)}\approx-0.233$ and
$\tilde{\eta}_{yy}^{(2)}\approx0.11$.
In addition, there is still the microphysical magnetic diffusivity,
which is positive ($\eta=0.403$).
To a first approximation, one can just consider the equation
for $\meanA_{yy}$, which can then be written as
\begin{equation}
{\partial\meanA_{yy}\over\partial t}=
\left[\eta+\tilde{\eta}_{yy}^{(0)}\right]{\partial^2\meanA_{yy}\over\partial z^2}
-\tilde{\eta}_{yy}^{(2)}{\partial^4\meanA_{yy}\over\partial z^4}.
\label{dAmyydt}
\end{equation}
We recall that the minus sign in front of the fourth derivative corresponds
to positive diffusion if $\tilde{\eta}_{yy}^{(2)}$ is positive, and so
does the plus sign in front of the second derivative, unless the term in
squared brackets is negative, which is the case we are considering here.

\subsection{Dynamos from pumping and memory effects}
\label{DynamosFromMemory}

Pumping effects alone cannot usually lead to interesting dynamo effects,
unless there is also a memory effect.
This effect means that the mean electromotive force depends not just
on the instantaneous mean magnetic field at that time, but also on the
mean magnetic field at earlier times.
It is therefore described as a convolution between a pumping kernel and
the mean magnetic field.
This can lead to dynamo action, as has been demonstrated by \cite{Rhei+14}
for the case of two of the four flow fields studied by \cite{Rob72}.
These are flows~II and III with $\UU_{\rm II}(x,y)$ and
$\UU_{\rm III}(x,y)$, respectively, and are given by
\begin{equation}
\UU_{\rm II}=\begin{pmatrix}
 u_0 \sin Kx \cos Ky \\
-u_0 \cos Kx \sin Ky \\
 u_0 \cos Kx \cos Ky
\end{pmatrix},\quad
\UU_{\rm III}=\begin{pmatrix}
 u_0 \sin Kx \cos Ky \\
-u_0 \cos Kx \sin Ky \\
 \half u_0 (\cos2Kx+\cos2Ky)
\end{pmatrix},
\label{RobF}
\end{equation}
where $K$ is the wavenumber of the flow.
Both flows have zero kinetic helicity and no $\alpha$ effect,
but flow~II is also pointwise nonhelical.
A supercritical three-dimensional magnetic field with growth
rate $\gamma$ and wavenumber $k$ in the $z$ direction of the
form $\BB=\bb_0(x,y)\,e^{\gamma t+\ii kz}$ is possible when
$\Rm\equiv u_0/\eta K>4.58$ and 2.9 for flows~II and III,
respectively; see \cite{Rhei+14}.
Here, $\bb_0(x,y)$ is the eigenfunction.

For both flows, there are $xy$-averaged mean fields $\meanB_x(z,t)$
and $\meanB_y(z,t)$, with waves traveling in opposite directions for
flow~II and in the same direction for flow~III; see Figures~6 and 8 of
\cite{Rhei+14}, respectively.
These dynamos appear to be atypical, because there is so far no other
known example of a flow where pumping produces a memory effect that is
strong enough to lead to dynamo action.
This may well be due to the absence (until recently) of
computational tools for determining the memory effect.
Indeed, it was only with the development of the TFM \citep{schrinner+05,
schrinner+07} that the importance of the memory effect was noticed
\citep{HB09} and applied to pumping.

The dispersion relation for a problem with turbulent pumping
$\gamma$ and turbulent magnetic diffusion $\etat$ is given by
$\lambda=-\ii k\gamma-\etat k^2$.
Since $\Rey\lambda<0$, the solution can only decay, but it is
oscillating with the frequency $\omega=\Imag\lambda=\gamma$.
In the presence of a memory effect, $\gamma$ is replaced by
$\gamma/(1-\ii\omega\tau)$, where $\tau$ is the memory time.
Then, $\lambda\approx-\ii k\gamma\,(1-\ii\omega\tau)-\etat k^2$,
and $\Rey\lambda$ can be positive.
This is the case for the Roberts flows II and III.

We return to nonlocality and memory effects further below in this article
when we discuss concrete solar models; see \cite{Pipin23}.
One of the most obvious consequences of the memory effect is a lowering
of the critical excitation conditions for the dynamo, which was already
reported by \cite{RB12}.
Interestingly, for the nonlocal mean electromotive force, the lowering
of the critical threshold can be accompanied by multiple instabilities
of different dynamo modes that have different frequencies and spatial
localization; see \cite{Pipin23}.

\subsection{Dynamos from cross-helicity}

An alignment of velocity and magnetic field, i.e., cross helicity, plays
a key role in numerous processes and phenomena of astrophysical plasmas.
\cite{KR80} showed that the saturation stage of the turbulent generation
is characterized by an alignment of the turbulent convective velocity
and the magnetic field.
This consideration does not account for the effects of cross-helicity
that take place in the strongly stratified subsurface layers of the
stellar convective envelope.
For example, the direct numerical simulations of \cite{Matthaeus2008}
showed a directional alignment of velocity and magnetic field fluctuations
in the presence of gradients of either pressure or kinetic energy.

The mean electromotive force in this case is along to the mean vorticity,
\begin{equation}
   \meanEMF^{\Upsilon}= \Upsilon\nabla\times\overline{\mathbf{U}}+\dots, \label{crhl}
\end{equation}
where, $\Upsilon=\tau_c\left\langle
\mathbf{u}^{(0)}\cdot\mathbf{b}^{(0)}\right\rangle$ is the cross helicity
pseudoscalar, and $\tau_c$ is the turbulent turnover time.
The superscripts ${(0)}$ indicate quantities of the background turbulence,
which exists in the absence of a mean magnetic field and a mean flow;
see our comment after \Eq{mult} about the $\meanEMF_0$ term.
In the standard mean-field framework, it is assumed that $\Upsilon=0$;
see \cite{KR80}.
\cite{Yoshizawa1993} generalized the framework assuming $\Upsilon\ne0$,
see the comprehensive review of \cite{Yokoi2013}.

Dynamo scenarios based on cross helicity have been suggested in a
number of papers \citep{Yoshizawa1993,Yoshizawa2000,Yokoi2013}.
\cite{Yokoi2018} showed that the large-scale dynamo instability does
not require the existence of a global axisymmetric mean.
The mix of axisymmetric and nonaxisymmetric magnetic fields can be
produced even in the case $\overline{\Upsilon=0}$, where the overbar
means azimuthal averaging.
The surface magnetic field of the Sun and other similar stars tends
to be organized in sunspots, plagues, ephemeral regions, super-granular
magnetic network, etc.
These structures tend to demonstrate the alignment of local
velocity and magnetic fields \citep{Ruediger2011s}.
Therefore, the cross helicity dynamo instability can contribute to
dynamo generation effects that operate near the stellar surface.
Stellar observations, for example those of \cite{Katsova2021G},
require such dynamo effects to be working in situ at the stellar surface.
The solar analogs show an increase of the spottiness with an increase
of the rotation rate \citep{Berdyugina2005LRSP}.
In that case, cross helicity dynamo effects can be considered as a
relevant addition to the standard turbulent generation by means of
convective helical motions.
Rapidly rotating M-dwarfs show the highest level of magnetic activity
\citep{Kochukhov2021AAr}.
There is a population of rapidly rotating M-dwarfs that show a rather
strong dipole type magnetic field.
These stars show a rather small level of differential rotation.
For solid body rotation, an $\alpha^2$ dynamo generates a nonaxisymmetric
magnetic field \citep{Chabrier2006AA,Elstner2007AN}.
At high rotation rates, the $\alpha$ effect is highly anisotropic
\citep{Ruediger1993AA}.
It cannot employ the component of the large-scale magnetic field along
the rotation axis for the generation of an axial electromotive force.
Results of \cite{Yokoi2018} show that the $\alpha^2\Upsilon^2$ scenario
can produce a strong constant dipole magnetic field.
The model predicts the existence of large-scale cross helicity patterns
occupying the stellar surface.
We hope that this can be tested either in observations or in GCDs.

The nonlinear theory for the cross helicity effect is not yet developed.
\cite{Sur2009MN} showed that the turbulent generation due to $\Upsilon$
is quenched by large-scale vorticity in a way that is similar to
catastrophic quenching given by \Eq{quenching}, i.e.,
\begin{equation}
    \Upsilon \sim \frac{1}{1+\Rm \tau_c^2 (\nab\times\overline{\mathbf{U}})^2}.
\end{equation}
One should remember that for the initialization of the cross-helicity
dynamo instability we have to seed both the cross helicity and the
magnetic field.
The solar type model scenarios based on cross helicity require an
$\alpha$ effect, which produces poloidal magnetic field and cross helicity
at the top of the dynamo domain \citep{Yokoi2016}.

\begin{figure}[t!]\begin{center}
\includegraphics[width=.98\textwidth]{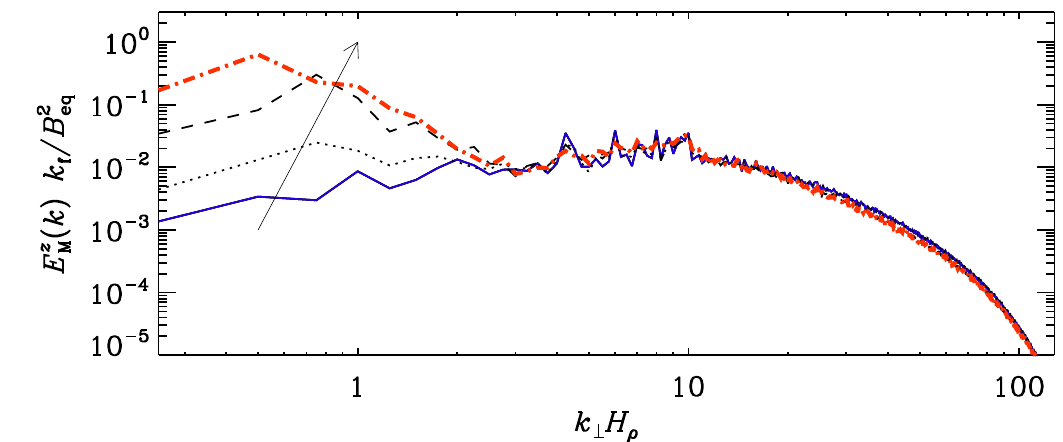}
\end{center}\caption[]{
Normalized spectra of $B_z$ from a simulation of MHD turbulence
with strong gravity at turbulent diffusive times
$t\etat/H_\rho^2\approx0.2$, 0.5, 1, and 2.7
with $\kf H_\rho=10$ and $k_1 H_\rho=0.25$.
Adapted from \cite{BGJKR14}.
}\label{pBz_spec2__pBzm_top_comp_gkf_bern23}\end{figure}

Given that cross helicity is an ideal invariant of the MHD equations,
it is natural to ask whether systems with strong cross helicity exhibit
inverse cascading.
The answer seems to be yes; see \cite{BGJKR14}.
In \Fig{pBz_spec2__pBzm_top_comp_gkf_bern23} we demonstrate the gradual
build-up of magnetic fields in the vertical direction when the system
has significant cross helicity owing to the presence of a magnetic field
along the direction of gravity \citep{Ruediger2011s}.

\section{Mean-field dynamo models}
\label{MeanFieldModels}

In general, the $\alpha$ effect, as well as any other turbulent
generation effect, including the $\delta$ effect \citep{Raedler1969}, the
shear-current effect \citep{Kleeorin2000}, and the cross-helicity effect
\citep{Yokoi2013} can generate both toroidal and poloidal magnetic
fields.
Therefore there can be a number of possibilities for solar-types
dynamo models \cite{KR80, Yokoi2016, Pipin2018c}.
Some of them skip the $\alpha$ effect altogether.
For example, \cite{Seehafer2009} studied $\delta^{\Omega}\Omega$ and
$\delta^{W}\Omega$ scenarios, where turbulent generation of the poloidal
magnetic field is due to $\OO\times\meanJJ$ and shear-current effect,
respectively.
These scenarios show oscillating solutions and the correct time-latitude
diagram of the toroidal magnetic field if the meridional circulation is
included.
A similar possibility was mentioned earlier by \cite{KR80} for the
$\delta\Omega$ scenario.
However, the given scenarios result in an incorrect phase relation
between activity of the toroidal and poloidal magnetic fields.
The aim to search for $\alpha$ effect alternatives pursues double
benefits.
First, the nonhelical source of dynamo generations avoids the
above mentioned catastrophic quenching problem.
This issue is less important currently.
Secondly, and it was already mentioned earlier by \cite{Koehler1973}
as well as \cite{StKr1969}, the mixing length estimate of the $\alpha$ effect
for the solar convection zone parameters results in a very strong $\alpha$
effect with a magnitude as strong as the convective velocity rms.
Solar observations of the ratio between the typical strength of the
toroidal and poloidal fields and the solar cycle period, favor an order
of magnitude smaller $\alpha$ effect.
In addition, the turbulent generation sources in the $\alpha\Omega$
scenario help reduce the given constraints.
We must stress that the GCD simulations of \cite{Schrinner2011},
\cite{Schrinner2011a}, and \cite{War+21} showed that the mean-field
models need a full spectrum of turbulent effects to describe DNS.

In the case of a solar-like star, i.e., with solar-like
stratification, differential rotation, and meridional circulation
profiles, the turbulent sources of the poloidal magnetic field generation
due to $\delta$, shear-current, and cross-helicity effects are likely
complimentary to the $\alpha$ effect.

We thus arrive at the conclusion that the $\alpha^2\Omega$ dynamo is,
probably, the simplest scenario for the solar dynamo.
Also, this scenario seems to fit well with observations of stellar activity of young solar-type stars.

\subsection{Basic model\label{MFMsubsec}}

We discuss some results of the state-of-the-art mean-field dynamo model of a
solar dynamo developed recently by \citet{Pipin2019c} (hereafter PK19).
The magnetic field evolution is governed by the mean-field induction
equation: 
\begin{equation}
\frac{\partial \meanBB}{\partial t} =\mathbf{\nabla}\times
\left(\meanEMF+\meanUU\times\meanBB-\eta\mu_0\meanJJ\right).
\label{eq:mfe}
\end{equation}
The expression for the components of $\meanEMF$ reads
\begin{equation}
\overline{\mathcal{E}}_{i}=\left(\alpha_{ij}+\gamma_{ij}\right)\meanB{}_{j}-\eta_{ijk}\nabla_{j}\meanB{}_{k}.\label{eq:Ea}
\end{equation}
Here, $\alpha_{ij}$ describes the turbulent generation by the $\alpha$ effect,
$\gamma_{ij}$ represents turbulent pumping, and $\eta_{ijk}$ is the eddy
magnetic diffusivity tensor.
The $\alpha$ effect tensor includes
the small-scale magnetic helicity density contribution, i.e., the
pseudoscalar $\left\langle \mathbf{a}\cdot\mathbf{b}\right\rangle $,
\begin{eqnarray}
\alpha_{ij} & = & C_{\alpha}\psi_{\alpha}(\beta)\alpha_{ij}^{\rm K}
+\alpha_{ij}^{\rm M}\psi_{\alpha}(\beta)\frac{\left\langle \mathbf{a}\cdot\mathbf{b}\right\rangle \tau_{c}}{4\pi\overline{\rho}\ell_{c}^2},\label{alp2d}
\end{eqnarray}
where $C_{\alpha}$ is the dynamo parameter characterizing the magnitude
of the kinetic $\alpha$ effect, and $\alpha_{ij}^{\rm K}$ and
$\alpha_{ij}^{\rm M}$ are the anisotropic versions of the kinetic and
magnetic $\alpha$ effects, as described in PK19.
Note that, unlike \Eq{alphaFOSA}, where the two $\alpha$ contributions are
pseudoscalars and have the same dimension, they are here tensorial where
only $\alpha_{ij}^{\rm K}$ is a pseudotensor, but $\alpha_{ij}^{\rm M}$
is not, and they have here different dimensions.

The radial profiles of $\alpha_{ij}^{(H)}$ and
$\alpha_{ij}^{(M)}$ depend on the mean density stratification, the profile
of the convective velocity $\urms$, and on the Coriolis number,
\begin{equation}
  {\rm Co} = 2\Omega_0 \tau_c, \label{eq_M8}
\end{equation}
where $\Omega_{0}$ is the global angular velocity of the star and
$\tau_{c}$ is the convective turnover time.
The magnetic quenching function $\psi_{\alpha}(\beta)$ depends on
the parameter $\beta=|\meanBB|/(\sqrt{4\pi\overline{\rho}}\urms)$.
In this model the magnetic helicity is governed by the global conservation
law for the total magnetic helicity,
$\left\langle \mathbf{A}\cdot\mathbf{B}\right\rangle=\left\langle \mathbf{a}\cdot\mathbf{b}\right\rangle +\meanAA\cdot\meanBB$
\citep[see][]{Hubbard2012,Pipin2013c}:
\begin{equation}
\left(\frac{\partial}{\partial t}+\meanUU\cdot\boldsymbol{\nabla}\right)
\left\langle \mathbf{A}\cdot\mathbf{B}\right\rangle
=-\frac{\left\langle \mathbf{a}\cdot\mathbf{b}\right\rangle }{\Rm\tau_{c}}
-2\eta\meanBB\cdot\meanJJ-\mathbf{\nabla\cdot}\meanFFFFf,
\label{eq:helcon}
\end{equation}
where we have used $2\eta\mathbf{\left\langle j\cdot b\right\rangle}=\left\langle \mathbf{a}\cdot\mathbf{b}\right\rangle/{\Rm\tau_{c}}$
\citep{Kleeorin1999}.
Also, we have introduced a diffusive
flux of the small-scale magnetic helicity density,
$\meanFFFFf=-\eta_{\chi}\mathbf{\nabla}\left\langle
\mathbf{a}\cdot\mathbf{b}\right\rangle $, and $\Rm$ is the magnetic
Reynolds number, for which we employ $\Rm=10^{6}$.
Following results of \citet{Mitra2010}, we put $\eta_{\chi}=0.1\,\eta_{T}$.
Here, the turbulent fluxes of the magnetic helicity are approximated by the only term which is related to the diffusive flux. 
Besides the diffusive helicity flux, the other turbulent
fluxes of the magnetic helicity can be important for the
nonlinear dynamo regimes and the catastrophic quenching problem
\citep{Kleeorin2000,Vishniac2001,Pipin2008a,Chatterjee2011,BS05,Kleeorin2022,Gopalakr2023}.
The relative importance of different kinds of magnetic helicity fluxes for the
dynamo should be studied further.

The above ansatz for the helicity evolution differs from that given by
Eq.~(\ref{smhev}); see also papers by \citet{Kleeorin1982} and \cite{Kleeorin1999}.
\cite{Hubbard2012} had studied the magnetic helicity evolution
for shearing dynamos.
They found that employing \Eq{smhev} in the dynamo problem can result
in nonphysical fluxes of magnetic helicity over spatial scales.
For the ansatz given by \Eq{smhev}, the nonlinear dynamo models can
show sharp magnetic structures inside the dynamo domain. Such
structures are connected with concentrations of the magnetic helicity;
see, e.g., \cite{Chatterjee2011} and \cite{BC18}.
Even a strong diffusive helicity flux does not seem to correct for those
irrelevant features from the numerical solution.
The technical point is that the helicity fluxes, which are omitted in
\Eq{smhev}, should be consistent with the turbulent effects involved
in the mean electromotive force, e.g., the rotationally induced anisotropy
of the $\alpha$ effect, the magnetic eddy diffusivity, etc.
Such calculations are currently absent.
Also, we have to take into account the modulation of the magnetic helicity
density by the magnetic activity.
On the other hand, with the magnetic helicity evolution equation
\Eq{eq:helcon}, \citet{Pipin2013c} found that magnetic helicity density
follows the large-scale dynamo wave.
This alleviates the catastrophic quenching of the $\alpha$ effect.
They showed that if we write the Eq.~(\ref{eq:helcon}) in the form of
Eq.~(\ref{smhev}), we get an additional helicity flux due to the global dynamo, 
Rewriting Eq.~(\ref{eq:helcon}) in the form of Eq.~(\ref{smhev}) we get
\begin{equation}
\frac{\partial \left\langle \mathbf{a}\cdot\mathbf{b}\right\rangle  }{\partial t} =  -2\left(\meanEMF\cdot\overline{\bm{B}}\right)
-\frac{\left\langle \mathbf{a}\cdot\mathbf{b}\right\rangle }{\Rm\tau_{c}}
+\boldsymbol{\nabla}\cdot\left(\eta_{\chi}\boldsymbol{\nabla}
\left\langle \mathbf{a}\cdot\mathbf{b}\right\rangle \right)
 -\eta\overline{\mathbf{B}}\cdot\mathbf{\overline{J}}-\boldsymbol{\nabla}\cdot\left(\meanEMF\times\meanAA\right)+\dots,\label{eq:hel-1}
\end{equation}
where the ellipsis refers to additional helicity transport terms due to the large-scale flow.
The term $\meanEMF\times\overline{\mathbf{A}}$ consists of the
counterparts of the sources of magnetic helicity, which are represented by
$-2\meanEMF\cdot\overline{\mathbf{B}}$, and the fluxes which result from
pumping of the large-scale magnetic fields.
The sources of magnetic helicity in the term
$-2\meanEMF\cdot\overline{\bm{B}}$ are partly compensated
in \Eq{eq:hel-1} by the counterparts in $\meanEMF\times\meanAA$.
This results in the spatially homogeneous quenching of the large-scale
magnetic generation and alleviation of the catastrophic quenching problem.
The effect of $\meanEMF\times\meanAA$ was not unambiguously confirmed
in DNS because of limited numerical resolution; see \cite{DSordo2013MN}
and \cite{Brandenburg2018AN}.

The turbulent pumping is expressed by the antisymmetric tensor
$\gamma_{ij}$. The tuning of $\gamma_{ij}$ for the solar-type mean-field
dynamo model was discussed by \citet{Pipin2018b}. We define it as
follows, 
\begin{eqnarray}
\gamma_{ij} & = & \gamma_{ij}^{(\Lambda\rho)}+\frac{\alpha_{\mathrm{MLT}}\urms}{\gamma}\mathcal{H}\left(\beta\right)\mathrm{\hat{r}_{n}\varepsilon_{inj}},\label{eq:pump0}\\
\gamma_{ij}^{(\Lambda\rho)} & = & 3\nu_{T}f_{1}^{(a)}\left\{ \left(\mathbf{\boldsymbol{\Omega}}\cdot\boldsymbol{\Lambda}^{(\rho)}\right)\frac{\Omega_{n}}{\Omega^{2}}\varepsilon_{\mathrm{inj}}-\frac{\Omega_{j}}{\Omega^{2}}\mathrm{\varepsilon_{inm}\Omega_{n}\Lambda_{m}^{(\rho)}}\right\}, \label{eq:pump1}
\end{eqnarray}
where $\mathbf{\boldsymbol{\Lambda}}^{(\rho)}=\boldsymbol{\nabla}\log\overline{\rho}$
, $\mathrm{\alpha_{MLT}}=1.9$ is the mixing-length theory parameter,
$\gamma$ is the adiabatic law constant.
In \Eq{eq:pump0}, the first term takes into
account the mean drift of large-scale field due the gradient of the
mean density, and the second one does the same for the mean-field
magnetic buoyancy effect. The function $\mathcal{H}\left(\beta\right)$
takes into account the effect of the magnetic tensions.
It is $\mathcal{H}\left(\beta\right)\sim\beta^{2}$ for small $\beta$
and it saturates as $\beta^{-2}$ for $\beta\gg1$; see P22.

We employ an anisotropic diffusion tensor following the formulation
of \citet{Pipin2008a} (hereafter, P08): 
\begin{eqnarray}
\eta_{ijk} & = & 3\eta_{T}\left\{ \left(2f_{1}^{(a)}-f_{2}^{(d)}\right)\varepsilon_{ijk}+2f_{1}^{(a)}\frac{\Omega_{i}\Omega_{n}}{\Omega^{2}}\varepsilon_{jnk}\right\} ,\label{eq:diff}
\end{eqnarray}
where the functions $f_{1,2}^{(a,d)}\left(\Omega^{*}\right)$ are determined
in P08.
Analytical calculations of $\meanEMF$ in the above cited paper include the
effects of a small scale dynamo.
In the above expressions for $\meanEMF$, we assume
equipartition between kinetic energy of the turbulence and
magnetic fluctuations which stem from the small-scale dynamo.
It was found that for the case of slow rotation (${\rm Co}\ll 1$),
the part of $\meanEMF$ that depends on the gradients of $\meanBB$ consists
of an isotropic eddy diffusivity and R\"adler's $\OO\times\meanJJ$
effect due to the small-scale dynamo (see also \citealp{Raedler2003a}).
In the case of rapid rotation, the fluctuating magnetic fields from the
small-scale dynamo contribute both to isotropic and anisotropic parts
of the diffusivity.
The effect appears already in the terms of order $\Omega^2$ in the global
rotation rate \citep{Raedler2003a}.
In particular, the part of EMF which corresponds to Eq(\ref{eq:diff}
can be written as
\begin{eqnarray}
\meanEMF^{\eta}&=& -3\eta_{T}\left(2f_{1}^{(a)}-f_{2}^{(d)}\right)\meanJJ+
6\eta_{T}f_{1}^{(a)}\OO\frac{\OO\cdot\meanJJ}{\Omega^2}.\label{eq:difJ}
\end{eqnarray}
It is noteworthy that the full expression for $\meanEMF$ obtained in P08 is complicated and includes other
contributions due to the effects of global rotation $\OO$, mean shear, mean current density
$\meanJJ$, and the magnetic deformation tensor $(\nabla\meanBB)$. 
We skip them in the application to the solar dynamo model.
The analytical results about the relations of the specific effects of
$\meanEMF$ and the global rotation rate show qualitative agreement
with the DNS of \cite{Kaepylae2009AA} and \cite{Brandenburg2012AA}.
Yet, a more detailed comparison of the analytical results and the GCD
simulations is needed; for further discussions, see \Sec{dnssec}.

We assume that the large-scale flow is axisymmetric. It is decomposed
into the sum of meridional circulation and differential rotation,
$\mathbf{\overline{U}}=\mathbf{\overline{U}}^{m}+r\sin\theta\Omega\left(r,\theta\right)\hat{\mathbf{\boldsymbol{\phi}}}$,
where $r$ is the radial coordinate, $\theta$ is the polar angle, $\hat{\mathbf{\boldsymbol{\phi}}}$ is
is the unit vector in the azimuthal direction, and $\Omega\left(r,\theta\right)$
is the angular velocity profile. The angular momentum conservation
and the equation for the azimuthal component of large-scale vorticity,
$\mathrm{\overline{\omega}=(\boldsymbol{\nabla}\times\overline{\mathbf{U}}^{m})_{\phi}}$,
determine the distributions of differential rotation and meridional
circulation: 
\begin{equation}
\frac{\partial}{\partial t}\overline{\rho}r^{2}\sin^{2}\theta\Omega  =  -\boldsymbol{\nabla\cdot}\left[r\sin\theta\overline{\rho}\left(\hat{\mathbf{T}}_{\phi}+r\sin\theta\Omega\mathbf{\overline{U}^{m}}\right)\right]
  + \boldsymbol{\nabla\cdot}\left[r\sin\theta\frac{\meanBB\meanB_{\phi}}{4\pi}\right],\label{eq:angm} 
\end{equation}
\begin{eqnarray}
\frac{\partial\omega}{\partial t} & = & r\sin\theta\boldsymbol{\nabla}\cdot\left[\frac{\hat{\boldsymbol{\phi}}\times\boldsymbol{\nabla\cdot}\overline{\rho}\hat{\mathbf{T}}}{r\overline{\rho}\sin\theta}-\frac{\mathbf{\overline{U}}^{m}\overline{\omega}}{r\sin\theta}\right]\label{eq:vort}
  +  r\sin\theta\frac{\partial\Omega^{2}}{\partial z}-\frac{g}{c_{p}r}\frac{\partial\overline{s}}{\partial\theta}\nonumber \\
 & + & \frac{1}{4\pi\overline{\rho}}\left(\meanBB\boldsymbol{\cdot\nabla}\right)\left(\boldsymbol{\nabla}\times\meanBB\right)_{\phi}-\frac{1}{4\pi\overline{\rho}}\left[\left(\boldsymbol{\nabla}\times\meanBB\right)\boldsymbol{\cdot\nabla}\right]\meanB{}_{\phi},\nonumber 
\end{eqnarray}
where $\hat{\mathbf{T}}$ is the turbulent stress tensor: 
\begin{equation}
\hat{T}_{ij}=\left\langle u_{i}u_{j}\right\rangle -\frac{1}{4\pi\overline{\rho}}\left(\left\langle b_{i}b_{j}\right\rangle -\frac{1}{2}\delta_{ij}\left\langle \mathbf{b}^{2}\right\rangle \right);
\label{eq:rei}
\end{equation}
see the detailed description in \cite{Pipin2018c} and PK19.
Also, $\overline{\rho}$ is the mean density, $\overline{s}$ is the mean entropy;
$\partial/\partial z=\cos\theta\partial/\partial r-\sin\theta/r\cdot\partial/\partial\theta$
is the gradient along the axis of rotation. The mean heat transport
equation determines the mean entropy variations from the reference
state due to the generation and dissipation of the large-scale magnetic field  and large-scale flows: 
\begin{equation}
\overline{\rho}\overline{T}\left[\frac{\partial\overline{\mathrm{s}}}{\partial t}+\left(\overline{\mathbf{U}}\cdot\boldsymbol{\nabla}\right)\overline{\mathrm{s}}\right]=-\boldsymbol{\nabla}\cdot\left(\mathbf{F}^{c}+\mathbf{F}^{r}\right)-\hat{T}_{ij}\frac{\partial\overline{U}_{i}}{\partial r_{j}}-\meanEMF\cdot\meanJJ,\label{eq:heat}
\end{equation}
where $\overline{T}$ is the mean temperature, $\mathbf{F}^{r}$ is
the radiative heat flux, $\mathbf{F}^{c}$ is the anisotropic convective
flux; see PK19.
The last two terms in \Eq{eq:heat} take into account the convective
energy gain and loss caused by the generation and dissipation of
large-scale magnetic fields and large-scale flows.
The reference profiles of mean thermodynamic parameters, such as entropy,
density, and temperature are determined from the stellar interior model
MESA \citep{Paxton2015}.
The radial profile of the typical convective turnover time, $\tau_{c}$,
is determined from the MESA code, as well. We assume that $\tau_{c}$
does not depend on the magnetic field and global flows.
The convective rms velocity is determined from the mixing-length
approximation,
\begin{equation}
u_{\rm c}=\frac{\ell_{c}}{2}\sqrt{-\frac{g}{2c_{p}}\frac{\partial\overline{s}}{\partial r}},\label{eq:uc-1}
\end{equation}
where $\ell_{c}=\alpha_{\mathrm{MLT}}H_{p}$ is the mixing length,
$\alpha_{\mathrm{MLT}}=1.9$ is the mixing length parameter, and $H_{p}$
is the pressure height scale. Equation~(\ref{eq:uc-1}) determines the
reference profiles for the eddy heat conductivity $\chi_{T}$, eddy
viscosity $\nu_{T}$, and eddy magnetic diffusivity $\eta_{T}$,
\begin{eqnarray}
\chi_{T} & = & \frac{\ell^{2}}{6}\sqrt{-\frac{g}{2c_{p}}\frac{\partial\overline{s}}{\partial r}},\label{eq:ch-1}\\
\nu_{T} & = &  \mathrm{Pr_{\rm T}}\,\chi_{\rm T},\label{eq:nu-1}\\
\eta_{T} & = & \mathrm{Pm_{\rm T}}\,\nu_{\rm T}.\label{eq:et-1}
\end{eqnarray}

It should be noted that stellar convection might well have convection zones
with slightly subadiabatic stratification in some layers.
In those cases, the enthalpy flux can no longer be transported entirely
by the mean entropy gradient, but there can be an extra term that is nowadays
called the Deardorff term; see \cite{Deardorff72}.
Such convection can be driven through the rapid cooling in the surface layers
and is therefore sometimes referred to as entropy rain \cite{B16}.
It is useful to stress that the Deardorff term is distinct from the
usual overshoot, because there the enthalpy flux points downward, while
entropy rain still produces an outward enthalpy flux.
It is instead more similar to semiconvection.

\textbf{Boundary conditions.} At the bottom of the tachocline, $r_{\rm i}=0.68\,R$,
we assume solid body rotation and perfect conductor boundary conditions.
Following to the MESA solar interior model, we put the bottom of the
convection zone to $r_{\rm b}=0.728\,R$.
At this boundary we fix
the total heat flux, $F_{\rm r}^{\rm conv}+F_{\rm r}^{\rm rad}=L_{\star}(r_{b})/4\pi r_{b}^{2}$.
We introduce the decrease by $\exp\left(-100\,z/R\right)$ for
all turbulent coefficients (except the eddy viscosity and eddy diffusivity),
where $z$ is the distance from the bottom of the convection zone.
The decrease of the eddy viscosity and eddy diffusivity is
at most one order of magnitude for numerical stability.
The meridional circulation is restricted to the convection zone.
Therefore, we put the azimuthal component of the large-scale vorticity
to zero, i.e., we set $\mathrm{\overline{\omega}}=0$ at $r_{\rm b}$.
At the top,
$r_{\rm t}=0.99\,R$, we employ a stress free and black body radiating
boundary. Following ideas of \citet{Moss1992}, we formulate the top
boundary condition in a form that allows penetration of the toroidal
magnetic field to the surface: 
\begin{eqnarray}
\delta\frac{\eta_{T}}{r_{\mathrm{top}}}B\left[1+\left(\frac{\left|B\right|}{B_{\mathrm{esq}}}\right)\right]+\left(1-\delta\right)\mathcal{E}_{\theta} & = & 0,\label{eq:tor-vac}
\end{eqnarray}

\textbf{Free parameters.} The model employs a number of free parameters,
including $C_{\alpha}$, the turbulent Prandtl numbers $\PrT$ and $\PmT$,
$\delta$, $B_{\mathrm{esq}}$, and the global rotation rate $\Omega_{0}$.
For the solar case we use a period of rotation of the solar tachocline determined
from helioseismology, $\Omega_{0}/2\pi=434\nHz$ \citep{Kosovichev1997}.
The best agreement of the angular velocity profile with helioseismology
results is found for $\mathrm{Pr}_{T}=3/4$. Also, the dynamo model
reproduces the solar magnetic cycle period, $\sim20$ years, if $\mathrm{Pm}_{T}=10$.
Results of \citet{Pipin2011a} showed that the parameters $\delta$
and $B_{\mathrm{esq}}$ affect the drift of the equatorial drift of
the toroidal magnetic field field in the subsurface shear layer and
magnitude of the surface toroidal magnetic field. Solar observations
show the magnitude of the surface toroidal field to be about 1-2 G \citep{Vidotto2018}.
To reproduce, it we use $\delta=0.99$ and $B_{\mathrm{esq}}=50$G.
In what follows, we present the results of the solar dynamo model for a
slightly supercritical parameter $C_{\alpha}$ (10\% above the threshold).
Further details of the dynamo model can be found in PK19.

\begin{figure}[t]
\centering \includegraphics[width=0.95\columnwidth]{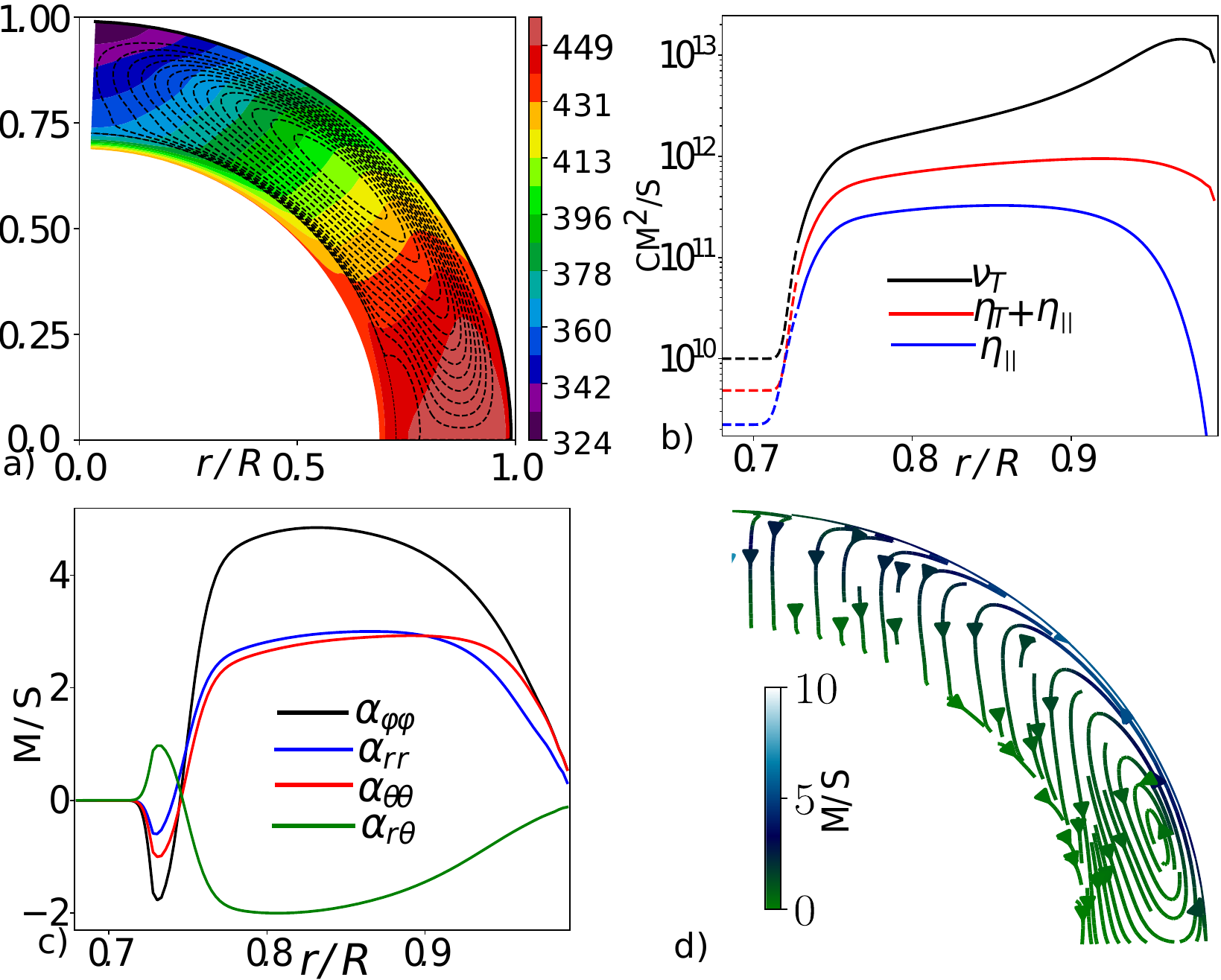}
\caption{\label{fig1}
(a) Streamlines of meridional circulation and the angular velocity
distribution; the magnitude of circulation velocity is of 13 m/s on the
surface at the latitude of $45^\circ$.
(b) Radial profiles of $\etaT+\eta_{||}$, the rotationally
induced part $\eta_{||}$, as well as $\nuT$.
(c) Radial profiles of the $\alpha$ tensor
at $45^\circ$ latitude.
(d) Streamlines of effective drift velocity from magnetically affected
pumping and meridional circulation.
Reproduced by permission from \cite{Pipin2022}.}
\end{figure}

\FFig{fig1} shows profiles of the basic turbulent effects
and large-scale flow distributions for the nonmagnetic case.
The amplitude of the meridional circulation on the surface is about $13\m\s^{-1}$.
In the lower part of the convection zone, the equatorward flow is about $1\m\s^{-1}$.
The angular velocity profile is in agreement with helioseismology data.

Interestingly, the stagnation point of the meridional circulation is near
the lower boundary of the subsurface shear layer, i.e., at $r=0.9\,R$.
This is in agreement with observations of \cite{Hathaway2012} and the
helioseismic inversions of \cite{Stej2021}.
The structure of meridional circulation and turbulent pumping promotes
an effective equatorward drift of the toroidal magnetic field below
the subsurface shear layer; see \Fig{fig1}(d).

\subsection{Parker--Yoshimura dynamo waves and extended cycle} 

\begin{figure}[t]
\includegraphics[width=0.86\textwidth]{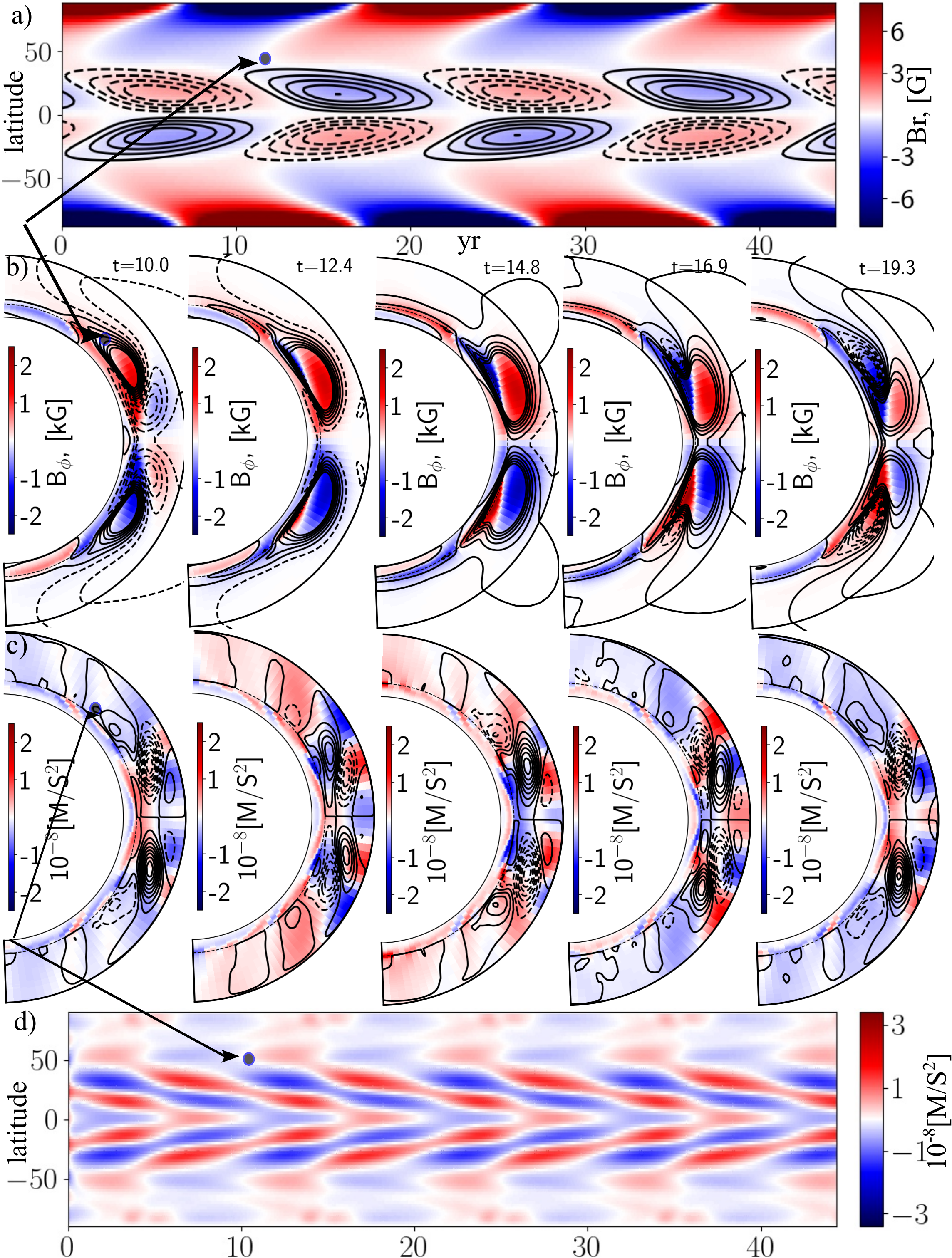}
\caption{\label{fig2}a) The surface radial magnetic field evolution (color
image) and the toroidal magnetic field at $r=0.9R$ (contours in range
of $\pm$1kG); b) snapshots of the magnetic field distributions inside
the convection zone for half dynamo cycle, color shows the toroidal
magnetic field and contours show streamlines of the poloidal field;
c) snapshots of the dynamo induced variations of zonal acceleration
(color image) and streamlines of the meridional circulation variations
(contours); d) variations of zonal velocity acceleration at the surface.
This Figure was prepared using the data of the dynamo model of PK19.
}\end{figure}

The dynamo shown in \Fig{fig2} demonstrates the numerical solution of the dynamo
system including Eqs.~(\ref{eq:mfe}) and (\ref{eq:angm})--(\ref{eq:heat}).
The time latitude diagrams of the surface radial
magnetic field and the toroidal magnetic field in the upper part of
the convection zone show agreement with observations of the evolution of
the large-scale magnetic field of the Sun (\citealp{Hathaway2015,Vidotto2018},
see also the review of Righmire in this volume). The dynamo waves propagate
to the surface- and equatorward.
The radial direction of propagation follows the Parker--Yoshimura rule
because of the positive sign of the $\alpha$ effect in the main part of the
convection zone and a positive latitudinal shear.
It is noteworthy that at high latitudes, the model shows another dynamo wave
family which propagates poleward along the convection zone boundary.
This family follows the Parker--Yoshimura rule as well. 
Further we will see that the latitudinal shear plays a dominant
role in this dynamo model and perhaps in the solar dynamo as well
(see also \citealp{CS15}). The latitudinal drift of the toroidal
magnetic field in this model results from turbulent pumping and
meridional circulation; see Fig\ref{fig1}(d).
The GCD simulations of \citealp{warnecke18, War+21} show the crucial role of
turbulent pumping in the solar type dynamo model, as well.
The extended mode of the dynamo cycle is another feature of their
model. The toroidal magnetic dynamo wave starts
at the bottom of the convection zone at around 50$^{\circ}$ latitude
(see the marks in \Fig{fig2}).
It disappears near the solar equator after a full dynamo cycle.
On the surface, the extended mode
of the solar cycle is seen in the radial magnetic field evolution,
in the torsional oscillations of zonal flow, and in the variations of the
meridional circulation as well \citep{Getling2021}. The origin of
the extended mode of the dynamo cycle is due to the distributed character
of the large-scale dynamo and the interaction of the global dynamo modes,
where the low order dynamo modes, e.g., dipole and octupole modes,
are mainly generated in the deep part of the convection zone. The high
order modes are predominantly generated in the near surface level.
The phase difference between the models results in a dynamo mode of
the extended length \citep{Sten1992,Obridko21}.

\subsection{Torsional oscillations}

Solar zonal variations of the angular
velocity (``torsional oscillations'') were discovered by \citet{Howard1980}.
Since that time it was found that torsional oscillations represent
a complicated wave-like pattern which consists of alternating zones
of accelerated and decelerated plasma flows \citep{Snodgrass1985,Altrock2008,Howe2011}.
\citet{Ulrich2001} found two oscillatory modes of these variations
with periods of 11 and 22 years. Torsional oscillations were linked
to ephemeral active regions that emerge at high latitudes during the
declining phase of solar cycles, but represent the magnetic field of the
subsequent cycle \citep{Wilson1988}. It is interesting that in their original
paper, \citet{Howard1980} conjectured that the solar torsional oscillation
can shear magnetic fields and induce the dynamo cycle. This idea was
further elaborated upon in a number of papers. However, the idea looks unreasonable
because it conflicts with Cowling's theorem. Also, the magnitude
of the torsional oscillations of $3$--$6\m\s^{-1}$ is too small in comparison
with the magnitude of the magnetic field generated by a dynamo.
The first papers by \citet{Schuessler1981} and \cite{Yoshimura1981}
suggested that the 11-year solar torsional oscillation can be explained
by the mechanical effect of the Lorentz force.
The double frequency of the zonal variation
results from the $B^{2}$ modulation of the large-scale flow due to
the dynamo activity. On the basis of a flux-tube dynamo model, \citet{Schuessler1981}
used the simple estimated of a large-scale Lorentz force and found
both the 11 and 22 year mode of the torsional oscillations. This result
was further elaborated upon by \citet{Kleeorin1991}. 
The further development of the mean-field theory of the solar differential
rotation showed that, in addition to the large-scale Lorentz force, the
dynamo induced $B^{2}$ modulation of the turbulent angular momentum
fluxes is also an essential source of the torsional oscillations
\citep{Ruediger1990, Kitchatinov1994a, Kleeorin1996, Kueker1996,
Ruediger2012}.
Global convective dynamo simulations \citep[e.g.,][]{Beaudoin2013,
Kapyla2016, Guerrero2016} confirmed this conclusion.
The strength of the solar torsional oscillations is more than two
orders of magnitude less than the differential rotation.
It looks like the theory of the torsional oscillations can
be constructed using perturbative approximations. Models of
this type (see, e.g., \citealt{Tobias1996,Covas2000,Bushby2007,Pipin2015,Hazra2017})
were inspired by results of \citet{Malkus1975}. Yet, the constructed
models are incomplete because they ignored the Taylor-Proudman balance,
which is a key ingredient of solar differential rotation theory
(see \citealt{Kitchatinov2013}, also contribution of Hazra
et al, this volume). Complete mean-field dynamo models, which take
into accounts the Taylor-Proudman balance (hereafter TPB), were constructed
by \citet{Brandenburg1992}, \citet{Rempel2007ApJ}, and PK19. 
Figure \ref{fig2} shows variations of the zonal
acceleration for our mean-field model in following the PK19 line of work.
Similar to the results of helioseismology \citep{Howe2011,Kosovichev2019}
and the results of \citet{Rempel2007ApJ}, snapshots of the model show
that in the main part of the convection zone, the acceleration patterns
are elongated along the rotation axis. This is caused by the Taylor-Proudman
balance. Near the convection zone boundaries, these patterns deviate
in the radial direction, which is in agreement with the above cited
helioseismology results, as well. The given observation on the role
of TPB shows the importance of the meridional circulation and the
dynamo-induced heat transport perturbation \citep{Spruit2003,Rempel2007ApJ}
in the theory of torsional oscillations. This fact does not deny the
importance of the large-scale Lorentz force and the magnetic modulation
of the turbulent angular momentum transport. Results of Figures \ref{fig2}(b)
and (c) show that the positive sign of the zonal acceleration propagates
from high latitude bottom of the convection zone toward equator
sticking to the equatorial edge of the dynamo wave. The torsional
oscillation wave is accompanied by corresponding variations of
the meridional circulation. These variations are induced by magnetic
perturbations of the heat transport (see details in PK19).
We emphasize that the given dynamo models also show overlapping magnetic
cycles; see \Fig{fig2}(b), similarly to what was originally proposed
by \citealp{Schuessler1981}].
In this case, the $B^{2}$ effect of the dynamo on the heat transport and the
TPB results in about 4 to 5 meridional circulation cells along latitude.
This tracks the zonal variations of angular
velocity, which are caused by the mechanical action of the large-scale
Lorentz force and magnetic quenching of the turbulent stresses, from
polar regions to the equator. PK19 found that the induced zonal acceleration
is $\sim(2$--$4)\times10^{-8}~$m$\,$s$^{-2}$, which is in agreement
with the observational results of \citet{Kosovichev2019}.
However, the individual forces in the angular momentum balance such that
the large-scale Lorentz force, the variations of the angular momentum transport
due to meridional circulation, the inertial forces, and others are
by more than an order of magnitude stronger than their combined action
and can reach a magnitude of $\sim10^{-6}~$m$\,$s$^{-2}$.
Therefore, the resulting pattern of the torsional oscillations forms in nonlinear
balance, which includes the forces driving the angular momentum transport,
the TPB, and heat perturbations due to magnetic activity in the convection
zone (see details in PK19).

\section{Mean-field models based on the EMF obtained from DNS\label{dnssec}}

Here we provide an example of how the mean-field theory is utilized as
{\it a tool for understanding what is going on} (see \Sec{umft}).
We discuss recent studies of mean-field dynamo models constructed
based on the electromotive force (EMF) obtained from direct numerical
simulation (DNS) of rotating stratified convection, especially focusing on
``semi-global'' models.
The properties of solar and stellar convection, and the various methods
for extracting the information of the EMF from DNS are also summarized.

\subsection{Properties of solar and stellar convection}

A quantitative physical description of solar and stellar dynamos, which
should be the result of the nonlinear interaction of turbulent flows and
magnetic fields, is a great challenge for us and constitutes a significant
milestone on the long way to a full understanding of turbulence.
Even with state-of-the-art supercomputers, it is impossible to numerically
simulate solar and stellar convection and its interaction with the
magnetic field and to observe/analyze numerical data in detail with
realistic parameters.
Therefore, to say with confidence that one has fully understood
the solar and stellar dynamo problem, it should be necessary to find a
universal law of magneto-hydrodynamic (MHD) turbulence, build a reliable
sub-grid scale (SGS) turbulence model, and then reproduce the magnetic
activities of the Sun and stars quantitatively in an integrated framework
by numerical models with incorporating the SGS model.
This is because fluid quantities that may be verified in future
observations should include the meridional distributions of fluid velocity,
vorticity, kinetic helicity, and thus the turbulence model constructed
on the basis of these profiles \citep[e.g.,][]{hanasoge+16}.
Only when the correctness of the turbulence model is observationally
validated should our understanding of the solar and stellar dynamos as
a consequence of the turbulent dynamo process be completed.
In the near future, a very exciting time may come when we will be able
to test and verify various turbulence models under extreme conditions
inside the solar and stellar interiors.

What physical characteristics should be taken into account when
constructing a turbulence model of thermal convection in the Sun
and stars?
Let us summarize some essential features:
\begin{enumerate}
\item {\bf Extremely low dissipation}: turbulent state with ${\rm Re} \gtrsim 10^{12}$, $\Rm \gtrsim10^{8}$,
and a large P\`{e}clet number, ${\rm Pe} \sim 10^6-10^9$ (where ${\rm Pe} = {\rm Re} \cdot {\rm Pr})$.
\item {\bf Huge separation of dissipation scales}: ${\rm Pr} \sim 10^{-4}$--$10^{-7}$, ${\rm Pr_M} \sim 10^{4}$ 
\item {\bf Compressibility}: high Mach number $\mathcal{O}(1)$ in the upper convection zone makes the convective motion compressible.
\item {\bf Anisotropy}: spin of stars (i.e., Coriolis force in a rotating system) makes fluid motions anisotropic. 
\item {\bf Inhomogeneity}: density contrast of $10^6$ between top and bottom CZs results in multi-scale properties of fluid motion. 
\item {\bf Non-locality}: Radiative energy loss at the CZ surface (open system), allowing the growth of cooling-driven downflow. 
\end{enumerate}
In view of these features, it can be seen that the characteristics of
thermal convection operating inside the Sun and stars are quite different
from those of isotropic turbulence.
Those can be considered to some extent in DNS even with the current
computing performance, as listed under 3--6, while the others, 
(items 1 and 2) are unreachable with current grid-based simulations.
It should be emphasized, however, that higher resolution simulations
using state-of-the-art supercomputers is a classical way forward in
turbulence research, and the knowledge obtained from such studies in
unexplored low-dissipation regimes will greatly expand the horizon of
our understanding of turbulence \citep[e.g.,][]{kaneda+03,hotta+21}.
Moreover, if sufficient scale separation between the turbulent and mean
fields is ensured and the inertial range of the turbulent cascade is
captured appropriately, there is the possibility that the evolution
of mean-field components, such as large-scale flow and large-scale
magnetic field, can be approximately reproduced even by simulations with
enhanced dissipation compared to the actual solar and stellar values
\citep[e.g.,][]{ossendrijver03}.
It should be remembered, however, that in spite of the rapid increase
in computing power, some rather basic questions about the solar dynamo
still remain, for example the equatorward migration of the sunspot belts
and the formation of sunspots themselves.

\subsection{Semi-global simulation of rotating stratified convection}

On our way toward a reliable SGS turbulence model for solar and stellar
interiors, numerical models of convection and its dynamo should be
studied, while keeping the characteristic features of solar and stellar
convection, as listed under items 3--6 above, in mind.
It should be noted that the underlying necessity for numerical modeling
is an important component of earlier studies that applied mixing-length
type concepts to the dynamo theory, which never successfully explained the
magnetic activities of the Sun and stars \citep[e.g.,][]{brandenburg+88}.

\begin{figure}[t]
\includegraphics[width=0.96\textwidth]{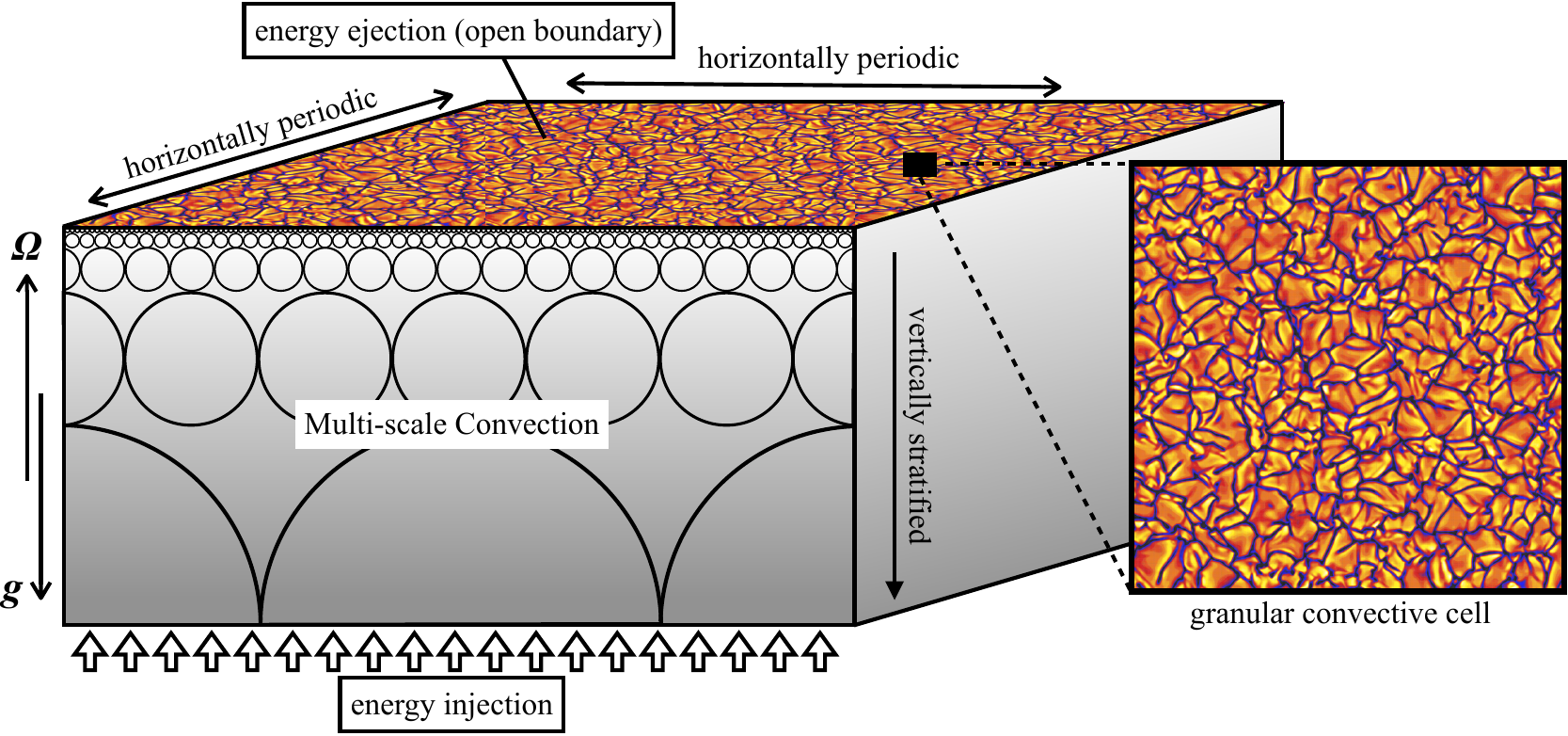}
\caption{Numerical setup typical for semi-global simulation of rotating
stratified convection.
Since the CZs of the Sun and stars are strongly stratified, there is
a large separation of time scales from minutes (upper CZs) to months
(bottom CZs).}
\label{fig_YM1}
\end{figure}

In recent years, significant progress has been made in GCD
simulations \citep[e.g.,][]{browning+06, ghizaru+10, kapyla+12,
masada+13, fan+14, augustson+15, hotta+16, warnecke18}, there is also
a growing effort to extract the information of turbulent transport
processes from so-called ``semi-global'' (or local model) MHD convection
simulations with the aim of quantifying the dynamo effect of rotational
stratified convection \citep[e.g.,][]{brandenburg+90, brandenburg+96,
nordlund+92, brummell+98, brummell+02, ossendrijver+01, kapyla+06a,
kapyla+09, masada14a, masada14b, masada+16, bushby+18, masada+22}.
A typical numerical setup of the semi-global model is shown in
\Fig{fig_YM1} schematically.
In this setting, the gas is gravitationally stratified in the vertical
direction, while periodicity is assumed in the horizontal directions.
The governing equations (mostly compressible MHD equations) are solved
in a rotating Cartesian frame, and the rotation axis is usually set to
be parallel or anti-parallel to the gravity vector.
Several studies have simulated the model with the tilt of the rotation
axis with respect to the gravity vector, and the latitudinal dependence
of the convection has been investigated \citep[e.g.,][]{ossendrijver+01,
kapyla+04, kapyla+06a}.

\subsection{Extraction of information of dynamo effects}
In the semi-global studies, four-types of approaches have been used typically to extract the information of dynamo effects veiled in the convective motion.
The starting point of all the four methods is common, the decomposition of the flow field (${\bm U}$) and magnetic field (${\bm B}$) into a spatially large-scale,
slowly-varying mean-component, and a small-scale, rapidly varying fluctuating component, as introduced in \S~1, i.e.,
${\bm U} = \overline{{\bm U}} + {\bm u}$ and ${\bm B} = \overline{{\bm B}} + {\bm b}$,
where the lower-case represent the fluctuating component and the overbars denote the mean component. 
In the case of a semi-global model, a temporal and horizontal average
is often used for deriving the mean component.
Then, the equation of mean-field electrodynamics can be derived
\begin{equation}
  \frac{\partial \overline{\bm B}}{\partial t} = \nabla \times (\overline{{\bm U}} \times \overline{{\bm B}} + \meanEMF - \eta \nabla \times \overline{{\bm B}} ) \;, \label{eq_M2}
\end{equation}
where $\meanEMF = \overline{ {\bm u}^\prime \times {\bm b}^\prime} $ is the mean electromotive force (EMF) due to the fluctuation of the flow and
the magnetic field. The mean EMF can be described as a power series about the large-scale magnetic component and its derivatives as
\begin{equation}
  \meanEMF = \overline{ {\bm u} \times {\bm b}} = {\bm \alpha} \cdot \overline{{\bm B}} + {\bm \gamma}\times \overline{{\bm B}} - {\bm \beta}\cdot (\nabla \times \overline{{\bm B}}) + \cdots \;, \label{eq_M3}
\end{equation}
where ${\bm \alpha}$ represents (tonsorial form of) the $\alpha$-effect, ${\bm \gamma}$ is the turbulent pumping, and ${\bm \beta}$ denotes the turbulent diffusion. 

To obtain the information of the dynamo coefficients, such as
${\bm \alpha}$, ${\bm \gamma}$, and ${\bm \eta_T}$, from the MHD
convection simulation, there are the following four methods 
(see general discussion in \Sec{bsoca}):
\begin{enumerate}
\renewcommand{\labelenumi}{(\roman{enumi})}
\item Methods based on results of analytical theories, e.g., SOCA (or first-order smoothing approximation, FOSA) expressions
\item Imposed-field method
\item Test-field method
\item The multi-dimensional regression method based on the dynamo generated magnetic field
\end{enumerate}

Method (i) involves the the estimation of dynamo coefficients based on
results of analytical theory. It also exploit the mixing-length
approximation in final results.
There, the distributions of, for example, the fluctuating components of
the convection velocity (${\bm u}$), vorticity (${\bm \omega} = \nabla
\times {\bm u}$), and the resulting kinetic helicity ($\mathcal{H} =
{\bm \omega}\cdot{\bm u}$), are directly extracted from the simulation
results and used to reconstruct the turbulent $\alpha$ and $\beta$
via their analytic forms, derived under SOCA, such as
\Eq{alphaFOSA} and $\beta =(\tau/3)\overline{ {\bm u}^2}$,
where $\tau$ is the correlation time of the turbulence and
is often replaced by the convective turnover time, $\tau_c$.
Note that anisotropy effects are often neglected in the expressions above,
but see \cite{BS07}, who included them.

Method (ii) is mainly used in the analysis of the numerical results
without self-sustained magnetic field.
In this method, a uniform external magnetic field is imposed as the
mean component to the computational domain, artificially.
The $\alpha$ effect and turbulent pumping are then
inferred from $\meanEMF = \overline{ {\bm u} \times {\bm b}}$ by
directly calculating from simulation data
\begin{equation}
\alpha = \meanEMF \cdot \overline{{\bm B}} /\overline{{\bm B}}^2  \;, \quad
\ggamma = -\meanEMF \times \overline{{\bm B}} /\overline{{\bm B}}^2  \;. \label{eq_M5}
\end{equation} 
Furthermore, one might be tempted to compute
$\etat=\meanEMF\cdot\meanJJ/(\mu_0 \meanJJ^2$), but this would assume
that $\meanJJ\cdot\meanBB$ is vanishing, which is generally not the case
for $\alpha$-effect dynamos; see \cite{Hubbard+09} for details.

Method (iii) utilizes a so-called test field, as introduced by
\citet{schrinner+05,schrinner+07} for the spherical case and
\cite{BRS08} for the Cartesian case, allowing for scale dependence.
In this method, the evolution equation of ${\bm b}^\prime_{\rm T}$, the
fluctuating component of the test field ${\bm B}_{\rm T}$, which are passive
to the velocity field taken from the simulation, is solved additionally
to the basic (MHD) equations.
From the linear evolution of the test-field, the mean EMF is evaluated
and then the full set of turbulent transport coefficients can be obtained.
For example, in the case without the large-scale flow,
the test-field equation is, for ${\bm b}_{\rm T}$,
\begin{equation}
\frac{\partial {\bm b}_{\rm T}}{\partial t}
= \nabla \times ({\bm u} \times {\bm B}_{\rm T} 
+ {\bm u} \times {\bm b}_{\rm T} 
- \overline{{\bm u}\times {\bm b}_{\rm T}} 
- \eta \nabla \times {\bm b}_{\rm T}) \;, \label{eq_M6}
\end{equation} 
with a chosen test field ${\bm B}_{\rm T}$ while taking ${\bm u}$ from the MHD simulation. 
The original method does not work in cases when magnetic fluctuations
are driven through an artificially added electromotive force.
In such cases, one needs to use a more general nonlinear method explored
by \cite{RB10} and \cite{Maarit2022}.
Theses magnetically driven systems are also examples where the
imposed-field method gives reliable results in two dimensions with volume
averaged mean fields, where no turbulent diffusion can act.

Method (iv) can be used only in the analysis of the numerical results
with self-sustained magnetic fields.
Since the fluctuating and mean components are all known quantities
in such simulations, the mean {\it emf}, $\meanEMF = \overline{ {\bm u} \times {\bm b}}$, and the mean magnetic component, $\overline{{\bm B}}$, can be directly
calculated from the simulation data. Then, the mean profiles of dynamo coefficients are inferred based on a fitting procedure via the relationship,  
\begin{equation}
  \mathcal{E}_i = \alpha_{ij}\overline{B}_j + \epsilon_{ijk}\gamma_j\overline{B}_k 
+\mbox{higher derivative terms}\;. \label{eq_M7}
\end{equation}
Given $\mathcal{E}_i$ and $\overline{B}_i$ which are calculated from
simulation data, and then find $\alpha_{ij}$ and $\gamma_i$ such that
the residual of Eq.~(\ref{eq_M7}) is minimized.
In the equation above, the contributions from the derivatives of the
mean magnetic component to the mean {\it emf} are neglected
\citep[see, e.g.,][for the fitting based analysis of
the dynamo coefficient with including the contribution
from the first-order derivative of the mean magnetic
component]{racine+11,simard+13,simard+16,shimada+22}.
As we have mentioned in \Sec{bsoca}, the results of \cite{Warnecke2018a}
show that the TFM is more accurate compared to the regression method in
obeying \Eq{eq_M7}; see the detailed comparison in above cited paper.

In all cases, however, the first (and often higher) derivative terms are
of the same order as the first term and can therefore not be neglected.
This was already done in the work of \cite{BS02}, who typically found small
diffusion coefficients in the cross-stream direction.
This, however, turned out to be a shortcoming of the method and has not
been borne out by more advanced measurements \citep{Karak+14}.

\subsection{Transport coefficients from semi-global turbulence simulations}

Here, we briefly review the results of previous semi-global simulations,
with a particular focus on the studies that have been dedicated for
extracting information about dynamo coefficients.

\citet{brandenburg+90}, hereafter B90, performed turbulent 3-D
magneto-convection simulations under the influence of the rotation for
the semi-global model whose depth is equivalent to about one pressure
scale height.
They found that, due to the effect of the rotation, a systematic
separation of positive and negative values of the kinetic helicity was
developed in the vertical direction of the CZ, i.e., in the upper CZ,
negative (positive) helicity in the northern (southern) hemisphere,
while positive (negative) helicity in the northern (southern) hemisphere.
Using the imposed field method, they evaluated the magnitude of the
turbulent $\alpha$-effect with anisotropic properties as $\alpha_V/(\tau
\mathcal{H}) \sim \mathcal{O}(0.1)$ and $\alpha_H/(\tau \mathcal{H})
\sim \mathcal{O}(0.01)$, where $\mathcal{H} = {\bm \omega}\cdot{\bm u}$
and $\meanEMF = \alpha_{\rm H} \overline{\bm B}_{\rm H} + \alpha_{\rm V}
\overline{\bm B}_{\rm V}$.
It is interesting to note that these values are about one to two orders
of magnitude smaller than $\alpha \sim \Omega d$, which is the estimation
based on the mixing-length theory.
Additionally, it was also suggested that the magnetic helicity showed
a similar depth variation, but the sign was opposite to that of the
kinetic helicity.

While $\alpha_{\rm H}$ had the expected sign (opposite to that
of the kinetic helicity), $\alpha_{\rm V}$ was found to have the
`wrong' sign (same as that of the kinetic helicity).
Such a result was subsequently also obtained by \cite{Ferriere+93}.
The theoretical possibilities for such effects should be studied further.
For example, \cite{Ruediger2000AA} found that large-scale shear can affect
both the sign of the $\alpha$ effect and kinetic helicity in magnetically
driven compressible turbulence in such a way that they have the same sign,
e.g., for Keplerian accretion disks.
These ideas were also applied to understanding the finding of a
negative $\alpha$ effect in stratified accretion disk simulations
\citep{Brandenbur1998proc}.

\citet{ossendrijver+01} also performed the semi-global simulation with a similar model as B90. They showed that, even in the regime where the condition justifying the
FOSA (or SOCA) is not satisfied, i.e., in the situation where ${\rm St} = u_{\rm rms}\tau /d \gtrsim 1$ and $Re > 1$, the kinetic helicity was clearly separated into positive
and negative values at the lower and upper CZs when taking temporal average of the convective motion over sufficiently long time. Using the imposed field method, they also
measured the magnitude of the turbulent $\alpha$-effect and obtained similar values to B90 in terms of $\alpha_H$ and $\alpha_V$. The rotational dependence of the $\alpha$-effect
was also investigated in this work for the first time. They showed that, in the larger ${\rm Co}$ regime, the $\alpha_V$ underwent a rotational quenching, while the $\alpha_H$
was saturated, where {\rm Co} is the Coriolis number [see \Eq{eq_M8}]. 
The turnover time was defined, in this work, as $\tau = d/u_{\rm rms}$.
While the depth-dependence or rotational dependence
of the $\alpha$, which was obtained from the simulation, agreed, to some extent, with a theoretical model based on the mixing-length theory (R\"udiger \& Kitchatinov 1993), their
amplitudes were one to two orders of magnitude smaller than those predicted from the theoretical model. 
Noteworthy, the critical threshold of the $\alpha$ effect parameter in
mean-field dynamo models (see \Sec{MFMsubsec}) is about same
magnitude less than the mixing-length models of the solar convection
zone predicts; see \Sec{MeanFieldModels}.

In \citet{kapyla+04,kapyla+06a}, additionally to the rotational dependence, the latitudinal dependence of the turbulent $\alpha$-effect was studied in the semi-global convection
simulations with varying the inclination of the rotation axis with respect to the gravity vector. With the imposed field method, they found that, for slow and moderate rotation
with ${\rm Co} < 4$, the latitudinal dependence of the $\alpha $ followed $\cos\theta $ profile with a peak at the pole \citep[see also,][]{egorov+04}, while, in the rapid rotation
regime with ${\rm Co} \approx 10$, it rather peaked much closer to the equator at $\theta \simeq 30^\circ$. Additionally, the vertical profile of the $\alpha$ directly evaluated from
simulation was found to be qualitatively consistent with analytic expression derived under the FOSA even when changing the latitude. A practical application of these results was 
the development of a kinematic mean-field solar dynamo model in \citet{kapyla+06b}. In it, the rotation profile deduced from the helioseismic observation and the meridional
profiles of the $\alpha$-effect and turbulent pumping obtained with the semi-global simulation of \citet{kapyla+06a} are integrated into the framework of the $\alpha$--$\Omega$
dynamo, and then the solar dynamo mean-field model was constructed. It is interesting that their kinematic dynamo model correctly reproduced many of the general features of the
solar magnetic activity, for example realistic migration patterns and correct phase relation.

The existence of large-scale dynamo, i.e., self-excitation of the mean
magnetic component, in rigidly-rotating convection was demonstrated for
the first time in the semi-global simulation by \citet{kapyla+09}.
By changing the angular velocity, they showed that the large-scale
dynamo could be excited only when the rotation is rapid enough, i.e.,
${\rm Co} \gtrsim 60$, with Eq.~(\ref{eq_M8}) as the definition of
${\rm Co}$ which is same as that used in \citet{ossendrijver+01} and
\citet{kapyla+06a}; see, e.g., \citet{tobias+08} and \citet{cattaneo+06},
and \citet{favier+13}, for unsuccessful large-scale dynamo in
rigidly-rotating convection probably due to slow rotation, and/or short
integration time.
From the measurements of the turbulent $\alpha$-effect and the turbulent
diffusivity by the TFM, they also suggested that while the magnitude
of the $\alpha$-effect stayed approximately constant as a function
of rotation, the turbulent diffusivity decreased monotonically with
increasing the angular velocity, resulting in the excitation of the
large-scale dynamo in the higher ${\rm Co}$.
The reliability of the dynamo coefficients extracted with the test-field
method from the simulation was validated with the one-dimensional
mean-field dynamo model in which the test-field results for $\alpha$
and $\beta$ were used as input parameters by studying the excitation of
the large-scale magnetic field at the linear stage.
Note that the oscillatory properties of the large-scale dynamo
in rigidly-rotating convection and its possible relationship with
$\alpha^2$ dynamo mode with inhomogeneous $\alpha$ profile were also
found in \citet{kapyla+13}; see, e.g., \citet{baryshnikova+87} and \citet{mitra+10} for the oscillatory $\alpha^2$ dynamo.

\subsection{Mean-field dynamo models linked with DNSs}
\subsubsection{Weakly-stratified Model}

Below we review recent mean-field dynamo models linked with semi-global
MHD convection simulations, where the large-scale dynamo is successfully
operated; see \citet{masada14b, masada14a, masada+16, masada+22}
for a series of numerical studies.

While \citet{kapyla+09,kapyla+13} were the first to demonstrate that
rigidly-rotating convection can excite the large-scale dynamo as reviewed
above, their simulation model was a so-called ``three-layer polytrope''
consisting of top and bottom stably-stratified layers and the CZ in
between them.
Therefore, it was suspected for a while that the essential factor for
the successful large-scale dynamo observed there might be the presence
of the stably-stratified layer assumed in their model rather than the
rapid rotation \citep[e.g.,][]{favier+13}.
To pin down the key requirement for the large-scale dynamo, the impact of the stably-stratified layers on the
large-scale dynamo was studied in \cite{masada14a}, hereafter MS14a, in which two-types of semi-global
models with and without stably-stratified layers are compared with the same control parameters and the same grid
spacing. It was found in this study that a large-scale dynamo was successfully operated even in the model without
the stably-stratified layer, and confirmed that the key requirement for it should be a rapid rotation if we
evolved the simulation for a sufficiently long time than the ohmic diffusion time. Note that a relatively weak
density stratification (the density contrast between the top and bottom CZs is about $10$) was assumed in the
simulation model employed in this study as well as \citet{kapyla+09,kapyla+13}.

\begin{figure}[t]
\includegraphics[width=0.96\textwidth]{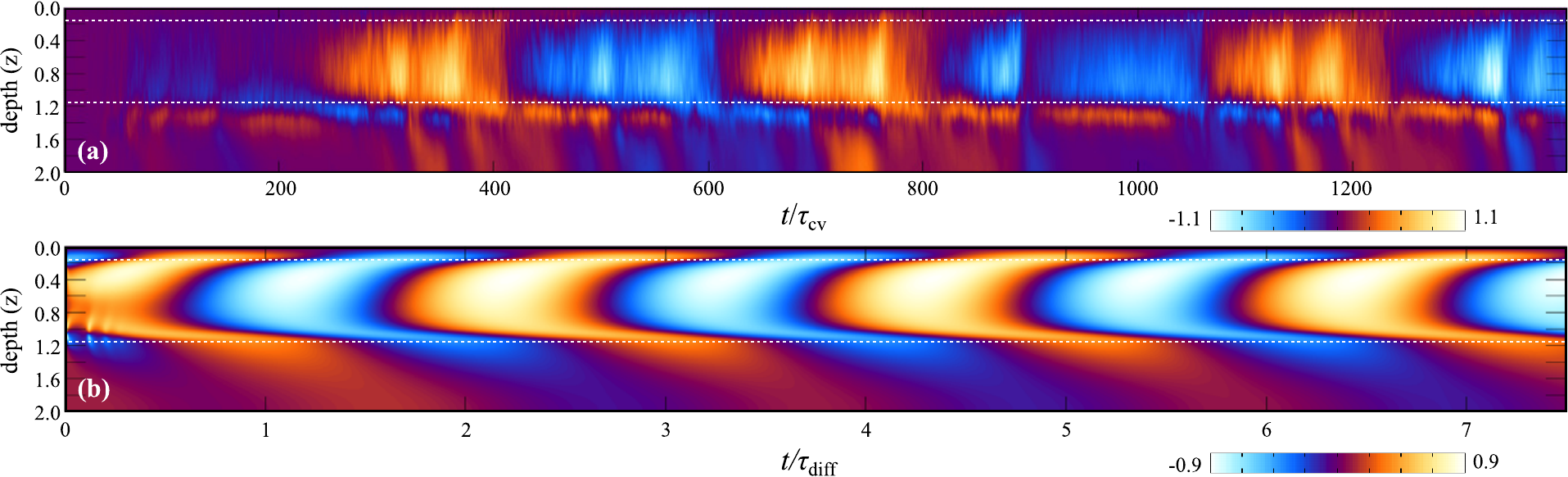}
\caption{Time-depth diagram $\overline{B}_x(z,t)$ for the MF model (panel~(b)) and its DNS counterpart (panel~(a)). 
For DNS result, the horizontal average of the magnetic field is shown. The orange and blue tones represent 
positive and negative $\overline{B}_x$ in units of $B_{\rm cv} \equiv (\overline{\rho {\bm u}^2})^{1/2}$. 
Time is normalized by $\tau_c$. Note that $\overline{B}_y$ shows a similar cyclic behavior with $\overline{B}_x$ yet 
with a phase delay of $\pi/2$; see MS14a,b for details.}
\label{fig_YM2}
\end{figure}

With these results, \cite{masada14b}, hereafter MS14b, explored the mechanism of the large-scale dynamo
operated in the rigidly-rotating stratified convection by linking the mean-field (MF) dynamo model with the DNS. In this study, the FOSA based approach was adopted in the MF modeling. The mean
vertical profiles of the kinetic helicity and root-mean square velocity were directly extracted from the
simulation data and then the vertical profiles of the turbulent $\alpha$, turbulent pumping ($\gamma $) and
turbulent diffusivity ($\beta$) were reconstructed according to the analytic expressions of 
\begin{equation}
\alpha (z) = - \tau_c (\overline{u_z\partial_x u_y} - \overline{u_z\partial_y u_x}) \;, \ \ \ \gamma (z) = -\tau_c \partial_z \overline{(u_z)^2} \;, \ \ \ \beta(z) = \tau_c \overline{(u_z)^2} \;, \label{eq_M9}
\end{equation} 
in anisotropic forms of dynamo coefficients under the FOSA \citep[e.g.,][]{kapyla+06a}. 
Although recent numerical studies indicate that the small-scale current helicity, i.e., 
${\bm j} \cdot {\bm b}$, is important for the $\alpha$-effect when the magnetic field is dynamically important \citep[][]{pouquet+76,brandenburg+05b}, its contribution was ignored in this study. 
As the correlation time
$\tau_c$, the convective turnover time defined by $\tau = H_\rho(z)/u_{\rm rms }$ was chosen there ($H_\rho$ is the
density scale-height as a function of the depth).
By solving one-dimensional MF $\alpha^2$ dynamo equation in which these
profiles were used as input parameters, i.e.,
\begin{equation}
  \frac{\partial \overline{\bm B}_h}{\partial t} = \nabla \times ( \meanEMF - \eta \nabla \times \overline{{\bm B}}_h ) \;, \label{eq_M10}
\end{equation}
with 
\begin{equation}
  \meanEMF = \alpha (z)\overline{\bm B}_h + \gamma (z) {\bm e}_z \times \overline{\bm B}_h - \beta (z) \nabla \times \overline{\bm B}_h \;, \label{eq_M11}
\end{equation}
the time-depth diagram for the mean (horizontal) magnetic component
($\overline{\bm B}_h$) was obtained.
In Fig.~\ref{fig_YM2}, we show $\overline{B_x}(z,t)$ for the MF model
(panels~(b)) and its DNS counterpart (panels~(a)).
Note that, for ensuring the saturation of the magnetic field growth,
the quenching effect was also taken into account.
Since the DNS results were quantitatively reproduced by the MF $\alpha^2$
dynamo, MS14b concluded that the large-scale magnetic field organized
in the rigidly-rotating turbulent convection was a consequence of the
oscillatory $\alpha^2$ dynamo.

Reproducing the DNS results with mean-field models using coefficients
from the original DNS is an important verification of the whole approach.
This has been done on many occasions in the past; see, for example, the
work by \cite{Gressel10} and \cite{War+21}.

\subsubsection{A strongly stratified model}

\begin{figure}[t]
\includegraphics[width=0.96\textwidth]{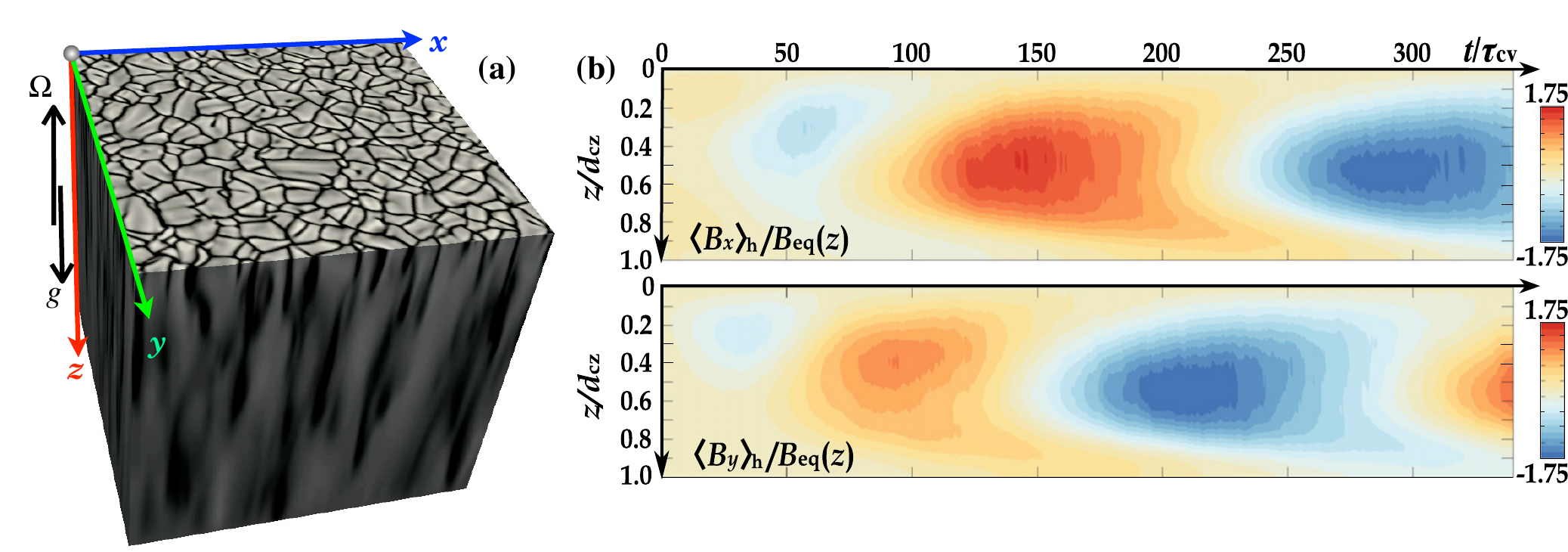}
\caption{(a) 3-D view of the strongly stratified convection (for the
progenitor run without rotation). The black (gray) tone denotes 
downflows (upflows). (b) Time-depth diagrams for $\overline{B}_x$ and $\overline{B}_y$. The normalization is the equipartition field strength,
$B_{\rm eq} \equiv (\overline{\rho {\bm u}^2})^{1/2}$. 
In MS16, a one-layer polytrope with a super-adiabaticity of
$\delta \equiv \nabla -\nabla_{\rm ad} = 1.6\times 10^{-3}$ was used; see
MS16 and MS22 for details.}
\label{fig_YM3}
\end{figure}

\begin{figure}[t]
\includegraphics[width=0.96\textwidth]{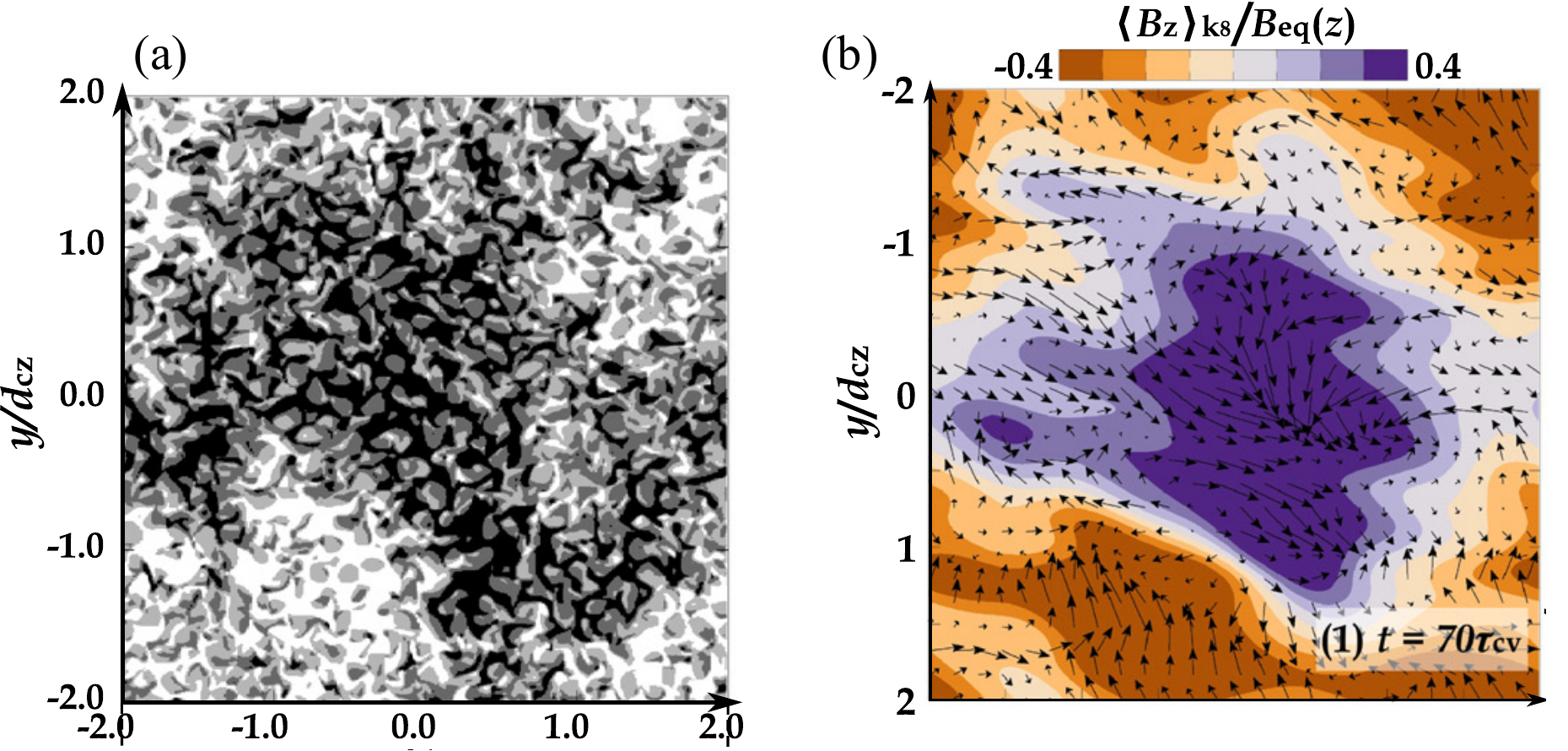}
\caption{Snapshots for (a) the horizontal distribution of $B_z$ at the CZ 
surface, and (b) its Fourier filtered image. In panel~(b), the small-scale 
structures with $k/k_c \gtrsim 8$ are eliminated for casting light on the 
large-scale pattern ($k$ is the wavenumber and $k_c = 2\pi/L_h$ with the horizontal box size $L_h$). }
\label{fig_YM4}
\end{figure}

In MS14a,b, a weakly-stratified model, in which the density contrast 
between top and bottom CZs is about $10$, was adopted. However, the 
actual Sun has very strong stratification with a density contrast of $\sim 10^6$ 
between top and bottom CZs, resulting in a large segregation of time scales 
from minutes to months. Aiming at the application to solar and stellar 
interiors, \citet{masada+16}, hereafter MS16, performed a convective dynamo
simulation in a strongly stratified atmosphere with a density contrast
of $700$ in a semi-global setup. Due to the strong solar- and stellar-like 
density stratification, multi-scale convection with a strong up-down asymmetry, 
i.e., slower and broader upflow regions surrounded by a network of faster and
narrower downflow lanes, was developed in this simulation, as shown in 
Fig.~\ref{fig_YM3}(a). Even in such a situation, a large-scale dynamo was 
found to operate. As shown in \Fig{fig_YM3}(b), the mean magnetic field components observed there showed a time--depth evolution similar to that seen 
in the weakly-stratified model (MS14a,b), suggesting that an oscillatory 
$\alpha^2$ dynamo is responsible for it. It was intriguing that, in addition
to the mean horizontal component, the large-scale structures of the vertical magnetic field were spontaneously organized at the CZ surface in the case of 
the strongly stratified atmosphere, as shown in \Fig{fig_YM4}.

A possible physical origin of such surface magnetic structure formation is
the negative magnetic pressure instability (NEMPI; see \S~8 for details).
NEMPI is a mean-field process in the momentum equation, where the Reynolds
and Maxwell stresses attain a component proportional to the square of
the mean magnetic field, which acts effectively like a negative pressure
by suppressing the turbulent pressure.
Since its growth rate becomes larger for stronger density stratification
\citep[e.g.,][]{jabbari+14}, one can imagine that it may play an important
role in organizing sunspot-like large-scale magnetic field structures
in the upper part of the solar CZ.
Although its presence has been confirmed numerically in forced MHD
turbulence \citep[e.g.,][]{brandenburg+11,warnecke+13}, it does not
play a significant role in organizing the surface magnetic structure
seen in MS16 because of their relatively rapid rotation; $\Ro = 0.02$
was assumed there, while, according to \citet{losada+12}, $\Ro \gtrsim 5$ is
required to excite the NEMPI.

The large-scale structure of the vertical magnetic field observed in MS16
is similar to that observed in the large-scale dynamo by forced turbulence
in a strongly stratified atmosphere \citep[][]{mitra+14,jabbari+16}.
This implies that there would be an as-yet-unknown mechanism for the
self-organization of large-scale magnetic structures, which would be
inherent in a strongly stratified atmosphere. 

In \citet{masada+22}, hereafter MS22, with varying angular velocity
as a control parameter, the rotational dependence of the large-scale dynamo 
was explored in a series of DNSs of rigidly-rotating convection. 
They linked its cause through MF dynamo models with DNSs where a strongly
stratified polytrope was adopted as a model of the convective atmosphere,
as in MS16. 

\begin{figure}[t]
\includegraphics[width=0.96\textwidth]{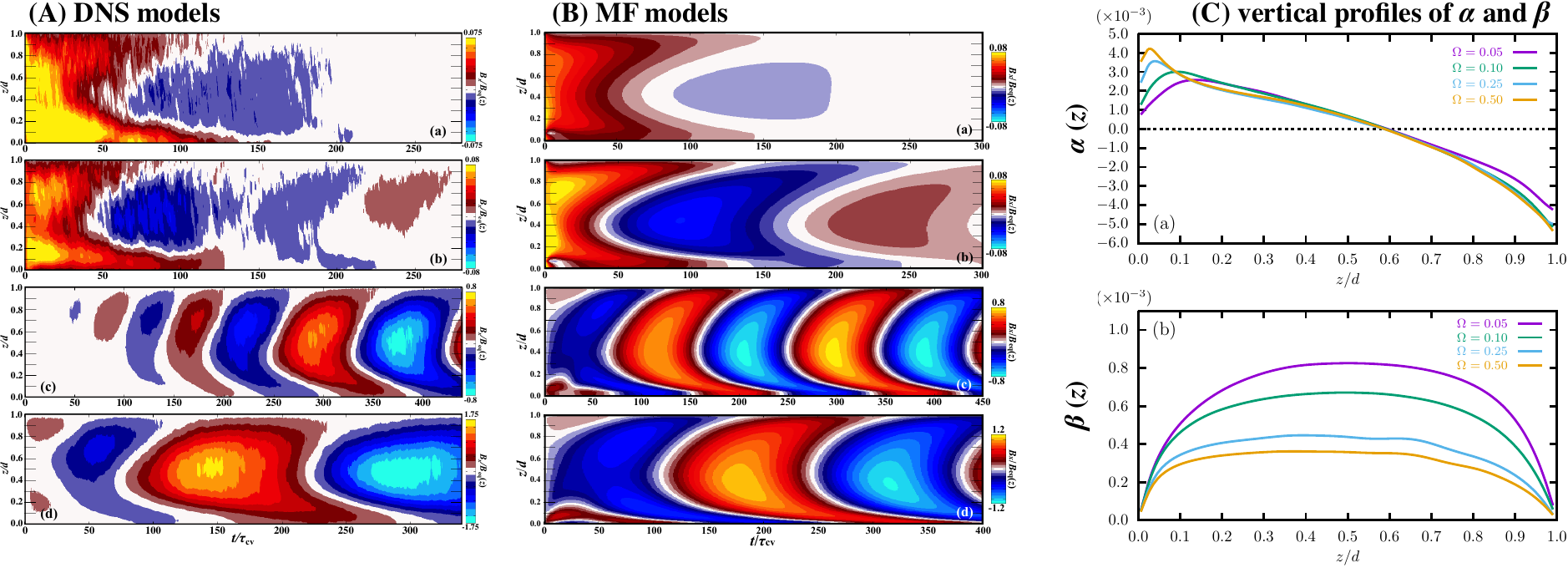}
\caption{Time-depth diagrams of $\overline{B}_x$ for (A) DNS models and 
(B) MF models. (C) Vertical profiles of the turbulent $\alpha$-effect (top)
and turbulent diffusivity $\beta$ (bottom) which are reconstructed with the analytic expressions of \Eq{eq_M9} from the information, such
as kinetic helicity and rms velocity, directly extracted from DNSs.} 
\label{fig_YM5}
\end{figure}

In \Fig{fig_YM5}(a), DNS results are shown where a time-depth diagram
of $\overline{B}_x$ is depicted for models with different values of $\Co$.
While in the slowly rotating model with low $\Co$, 
the large-scale magnetic component fails to grow,  
it is found to be spontaneously organized in the rapidly-rotating models 
with high $\Co$. 
It was found from DNS that the large-scale dynamo was excited when
$\Co \gtrsim \Co_{\rm crit}$, where $\Co_{\rm crit}$ is the critical
Coriolis number in the range $25 \lesssim \Co_{\rm crit} \lesssim 80$,
with \Eq{eq_M8} as the definition of the Coriolis number.
It is remarkable that $\Co_{\rm crit}$, which determines the success or
failure of the large-scale dynamo, was almost the same -- independent of the
density contrast \citep[see][]{kapyla+09} or the geometry of the simulation 
model \citep[see, e.g.,][for $\Co_{\rm crit}$ in the global simulations]
{kapyla+12,warnecke18}; see MS22 for the quantitative comparison between
models.

To explore the underlying mechanism of the rotational dependence of the
large-scale dynamo, the influence of the rotation on the turbulent transport 
coefficients was also studied in MS22 with the FOSA-based approach similar to 
that adopted in MS14b. In \Fig{fig_YM5}(c), the vertical profiles of the 
turbulent $\alpha$ effect and turbulent diffusivity $\beta$ reconstructed 
with the analytic expressions of \Eq{eq_M9} were shown. 
It can be found that, as the spin rate increases, the turbulent diffusivity 
decreases significantly, but the profile of the $\alpha$ effect remains almost 
unchanged. This result suggested that the rotational dependence of the large-
scale dynamo observed in MS22 may be primarily due to a change in the magnitude 
of the turbulent diffusion. 
In fact, this insight was confirmed by the evidence that the MF dynamo model 
with incorporating the dynamo coefficients shown in \Fig{fig_YM5} reproduced 
quantitatively the result of the DNS; see \Fig{fig_YM5}(b) for the time-depth 
diagram of $\overline{B}_x$ obtained in the MF models with using different 
dynamo coefficients extracted from the corresponding DNSs. 

MS22 concluded that the independence (dependence) of the turbulent
$\alpha$ effect (turbulent diffusivity) on the rotation was the essence
of the rotational dependence of the dynamo.
This is not only the same as the conclusion obtained by \citet{kapyla+09}
from weakly-stratified convective dynamo simulations using the TFM,
but also the same as that obtained by \citet{shimada+22} from global
solar dynamo simulation with using the ``self-sustained field method''.
Although we don't know whether the independence (dependence) of the 
$\alpha$-effect (turbulent diffusion) on the rotation, seen in these studies, 
is universal or not, it may give an important suggestion not only on the 
turbulence modeling but on the solar dynamo modeling.

\section{Solar-stellar connections and questions beyond
standard mean-field theory}

The title of our review is ``Turbulent processes and mean-field dynamo''
Obviously, there are mean-field effects in turbulence that are
not just of dynamo type, and the mean-field dynamo is not just related
to the solar dynamo problem, but its relevance goes much beyond.
Here, we highlight, just some effects, but we refer to the reviews of
\cite{Kapyla+23} for additional examples.

\subsection{Origin of sunspots and active regions}\label{sec:sp}

An important goal in solar dynamo theory is to compute synthetic
butterfly diagrams.
The question then emerges from which depth to take the mean toroidal
field, for example.
The usual argument here is to invoke Parker's theory of sunspot
formation and to postulate that the field at some depth translates
directly to one at the surface.
This is critical because the final result depends on the assumed depth.

It is possible that sunspots are not deeply rooted, but are actually
a surface phenomenon.
No successful and self-consistent model of shallow formation of active
regions or sunspots exists as yet.
Noteworthy in this context is NEMPI, which is a mean-field theory of
the Reynolds and Maxwell stresses.
This theory is extremely successful in that its results agree remarkably
well with direct numerical simulations (DNS).\footnote{DNS means that
viscous and diffusive operators are assumed to be the physical ones, but
with coefficients that are enhanced relative to the physical ones, but
as small as possible.
The ordering of these coefficients is often preserved so as to access
the relevant regimes with small magnetic and thermal Prandtl numbers.
Large eddy simulations (LES) or implicit LES, by contrast, use just
numerical schemes to keep the code stable.
Such schemes are often too complicated to state them as an explicit
term in the equations, as if they are negligible, but they never are.}
The problem is only that the effect is not strong enough to make real
sunspots or active regions.
Because of this remarkable agreement between theory and simulations,
and because it is an important mean-field process, we shall discuss
here a bit more detail.

The essence of the effect is the contribution of the turbulent
hydromagnetic pressure to the horizontal force balance.
The turbulent pressure is a small-scale effect, but it reacts to
the large-scale magnetic field.
As the magnetic field increases, it suppresses the turbulence locally,
disturbing therefore the horizontal force balance.
Although this large-scale magnetic field itself contributes with its
own magnetic pressure to the horizontal force balance, the effect from
the suppression of the turbulence is often stronger, so the net effect
is a negative one.
This is why the mean-field effect from a large-scale magnetic field is
a negative effective magnetic pressure.
This idea goes back to early work of \cite{KRR89,Kleeorin1996}, who
developed the mean-field theory for this effect.

In the beginning, it was not clear what kind of numerical experiments
one could try to test the a negative effective magnetic pressure effect.
The first mean-field simulations were done with a uniform horizontal
magnetic field \citep{BKR10}.
This led to the development of magnetic flux concentrations near the
surface, but those began to sink downward as time went on.
A similar effect was soon also seen in DNS \citep{brandenburg+11}.
The sinking of such structures was explained by the {\em negative}
effective magnetic pressure: a positive magnetic pressure would lead to
the rise of structures \citep{Parker67} while a negative one leads to
a sinking.
The sinking of magnetic structures had the side effect that the
structures disappeared from the surface and became less prominent.

\begin{figure}[t]
\includegraphics[width=\textwidth]{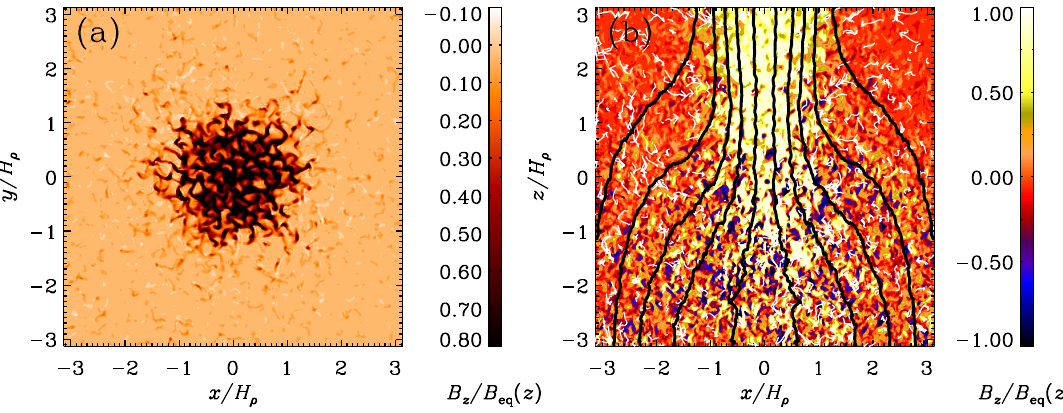}
\caption[]{Cuts of $B_z/\Beq(z)$ in the $xy$ plane at the top boundary
($z/H_\rho=\pi$) and the $xz$ plane through the middle of the spot at $y=0$.
In the $xz$ cut, we also show magnetic field lines and flow vectors
obtained by numerically averaging in azimuth around the spot axis.
Adapted from \cite{BKR13}.}
\label{pslices_V256k30VF_Bz002}\end{figure}

Subsequent experiments with a vertical field had a more dramatic effect
on the general appearance of structures.
Because the ambient field was vertical, the downflow had little effect
on the magnetic flux concentrations themselves \citep{BKR13}.
\Fig{pslices_V256k30VF_Bz002} shows the spontaneous development of a
magnetic spot.
\cite{BKR13} found that NEMPI saturates slightly below equipartition value.

Most of the numerical experiments where done with forced turbulence,
where one had explicit control over the degree of scale separation.
For the results presented here $\Rm\sim 19$ and $\Pm=0.5$.
The value of $\Rm=\urms/\eta\kf$ is very small because it is based on
the forcing wavenumber $\kf$, which is chosen to be large.
Furthermore, $\Pm$ is chosen to be less than unity, because NEMPI is
not expected to work for $\Pm\ge14$.
This is rather different in actual stellar convection, where
the development of magnetic structures takes different shapes
\citep{SN12,masada+16,KBKKR16}.

\subsection{Dynamo flux budget and impact of the surface activity on the deep dynamo}
\label{DynamoFluxBudget}

\begin{figure}[t]
\includegraphics[width=0.99\textwidth]{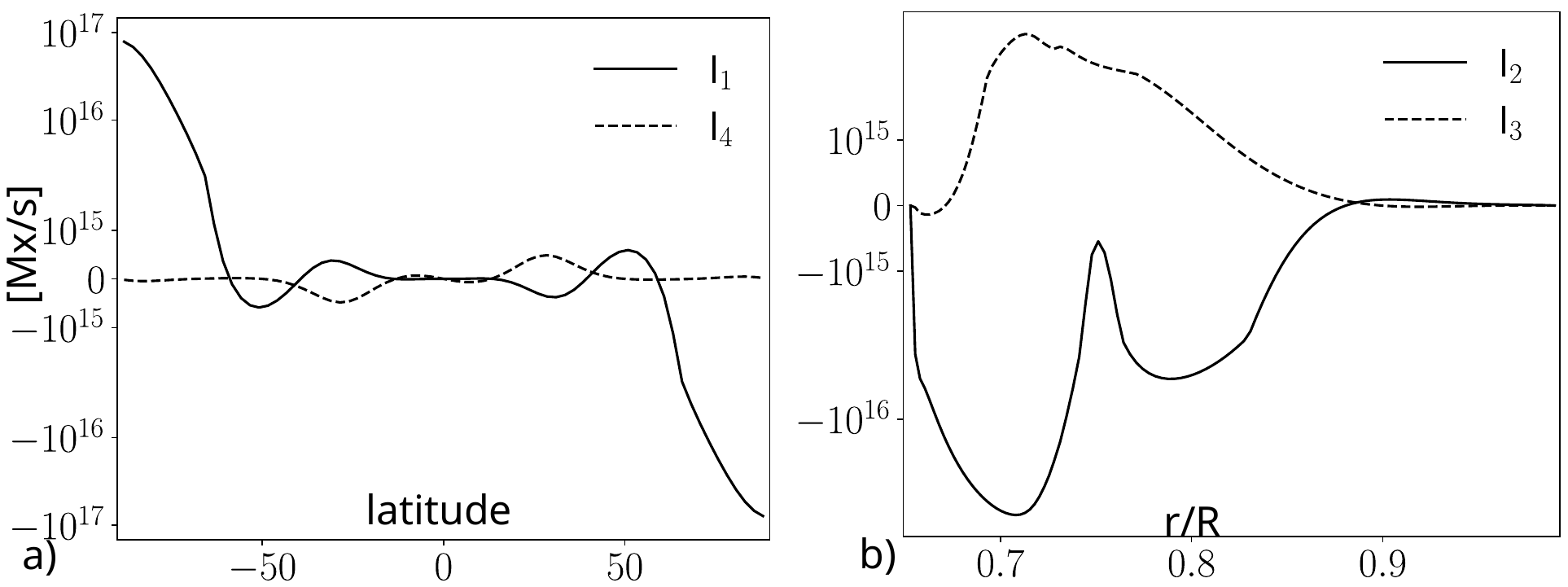}\caption{
\label{fig:bdgs}Estimation of contributions of the budget equation;
see \Eq{eq:integ}, for the time of cycle minimum.}
\end{figure}

Following \cite{CS15} (hereafter, CS15, also see the chapter by Cameron \&
Sch\"ussler) we now estimate the budget of the toroidal magnetic flux
in the dynamo region.
Using the Stokes theorem and the induction equation \Eq{eq:mfe},
we define the derivative of the toroidal magnetic field flux in the
northern hemisphere of the Sun as
\begin{equation}
\frac{\partial\Phi_{\mathrm{tor}}^{\mathrm{N}}}{\partial t}=\oint_{\delta\Sigma}\left(\overline{\mathbf{U}}\times\mathbf{\overline{B}}+\boldsymbol{\mathcal{E}}\right)\cdot\mathrm{d\mathbf{l}},\label{eq:St}
\end{equation}
where $\Phi_{\mathrm{tor}}^{\mathrm{N}}=\int_{\Sigma}\overline{B}_{\phi}\mathrm{dS}$,
$\Sigma$ is the meridional cut of the northern hemisphere of the
solar convection zone, $\delta\Sigma$ stands for the contour confining
the cut and the differential $\mathrm{dl}$ is the line element of
$\delta\Sigma$. The same can be written for the southern hemisphere
flux $\Phi_{\mathrm{tor}}^{\mathrm{S}}$.
Similarly to CS15, we use
the boundary conditions, and we estimate the RHS of the Eq.~(\ref{eq:St})
in the coordinate system which is co-rotating with angular velocity
of the solar equator, $\overline{U}_{0\phi}=R\sin\theta\Omega_{0}$,
and $\Omega_{0}$ the surface angular velocity at the equator,
\begin{eqnarray}
\frac{\partial\Phi_{\mathrm{tor}}^{\mathrm{N}}}{\partial t} & = & \int_{0}^{\pi/2}\overset{\mathrm{I_{1}}}{\overbrace{\left(\overline{U}_{\phi}-\overline{U}_{0\phi}\right)\overline{B}_{r}\mathrm{r_{t}}}}\,\mathrm{d}\theta+\int_{\mathrm{r_{i}}}^{\mathrm{r}_{\mathrm{t}}}\overset{\mathrm{I_{2}}}{\overbrace{\left(\overline{U}_{\phi}^{(\mathrm{\frac{\pi}{2}})}-\overline{U}_{0\phi}\right)\overline{B}_{\theta}^{(\mathrm{\frac{\pi}{2}})}}}\mathrm{dr}\label{eq:integ}\\
 & + & \int_{\mathrm{r_{i}}}^{\mathrm{r}_{\mathrm{t}}}\overset{\mathrm{I_{3}}}{\overbrace{\left(\mathcal{E}_{r}^{(\mathrm{0})}-\mathcal{E}_{r}^{(\mathrm{\frac{\pi}{2}})}\right)}}\mathrm{dr}+\int_{0}^{\pi/2}\overset{\mathrm{I_{4}}}{\overbrace{\left(\mathcal{E}_{\theta}^{(\mathrm{t)}}\mathrm{r_{t}}-\mathcal{E}_{\theta}^{(\mathrm{i})}\mathrm{r_{i}}\right)}}\mathrm{d}\theta\nonumber 
\end{eqnarray}
here, $r_{t}=0.99\,R$, $r_{i}=0.67\,R$, are the radial boundaries of the dynamo region.
In compare to CS15 we have
additional contributions in the budget equation. Figure \ref{fig:bdgs} shows profiles of the
kernels $I_{1-4}$ for the period of the magnetic cycle minimum. The
estimations are based on results and parameters of the mean-field
model presented above. Noteworthy, the south hemisphere should show
the profiles of the opposite sign (see CS15). The results for $I_{1,4}$
qualitatively similar to CS15. This is because the mean-field model
show the qualitative agreement with solar observations for the time
latitude evolution of the surface radial magnetic field. The diffusive
decay of the toroidal magnetic flux is captured as well because of
the phase shift between evolution of the poloidal and toroidal magnetic
field in dynamo model and presumably in the solar dynamo as well.
The model show the sharp poleward increase of $I_{1}$. This effect
produces the winding of the toroidal magnetic field from poloidal
component by the latitudinal shear. The effect of the radial shear,
$I_{2}$, has maximum near the bottom of the convection zone, where
its magnitude is less than the $I_{1}$.

\begin{figure}[t]
\includegraphics[width=0.96\textwidth]{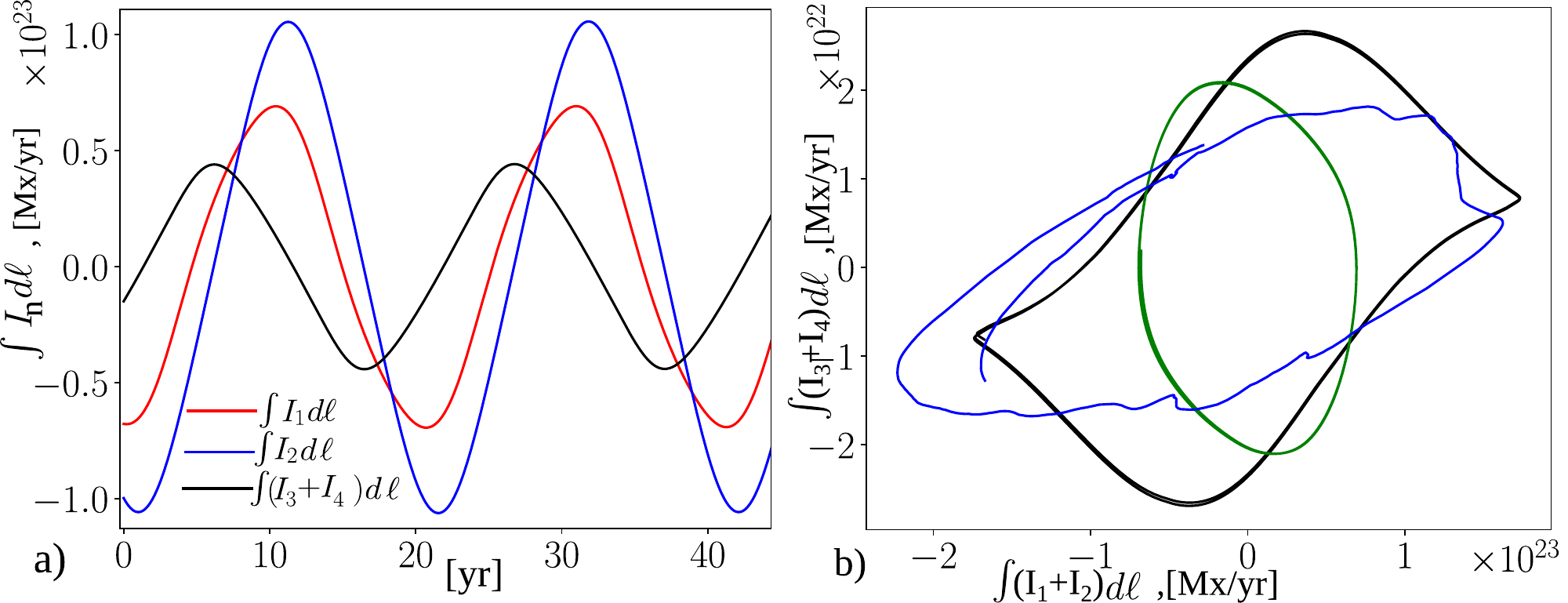}\caption{\label{fig:bdg}
(a) Time evolution of the RHS contributions of Eq.~(\ref{eq:integ});
(b) the dynamo models budget, black line show the standard mean-field
model, green line -- the budget which includes only the surface contributions
($I_{1,4}$) and the blue line shows the model
with accounts of surface spot-like activity effects.}
\end{figure}

Figure~\ref{fig:bdg}(a) shows the time evolution of the RHS contributions
of Eq.~(\ref{eq:integ}). In our model the We see that $I_{2}$ has the same magnitude as  $I_{1}$. Therefore, for the distributed mean-field dynamo model the latitudinal and radial shear produce equal effect on the net toroidal field in the bulk of the convection zone.
This is different to applications  of simple 1-D Babcock--Leighton dynamos to
the solar observations as argued by CS15.
Together with the fact of the poleward increase of
$I_{1}$ it explain the relative success of correlation of the polar
magnetic field strength and the magnitude of the subsequent magnetic
cycle for the solar cycle prediction \citep{Choudhuri2007}.

Figure~\ref{fig:bdg}(b) shows the budget of the toroidal flux generation
rate ($I_1+I_2$) and loss rate($I_3+I_4$) for our dynamo model. 
The parameters of the budget
are larger than those deduced by CS15 from solar observations. The
difference is because of additional generation and loss terms. The
budget which includes only the surface activity contributions (green
line in Fig.\ref{fig:bdg}b) is less than the full case. Also, the
magnitude of the generation rate by the latitudinal shear can be larger
than in the solar observations because of difference in the latitudinal
profiles of the surface radial magnetic field. We guess that in the
dynamo model the radial magnetic field increase poleward steeper than
in observations. This issue have to be studied further. Figure \ref{fig:bdg}b
shows the budget for another dynamo model which include the effects of the surface magnetic activity in form of bipolar magnetic regions. The model shows an increase of the toroidal field generation rate.  

\begin{figure}[t]
\includegraphics[width=0.95\textwidth]{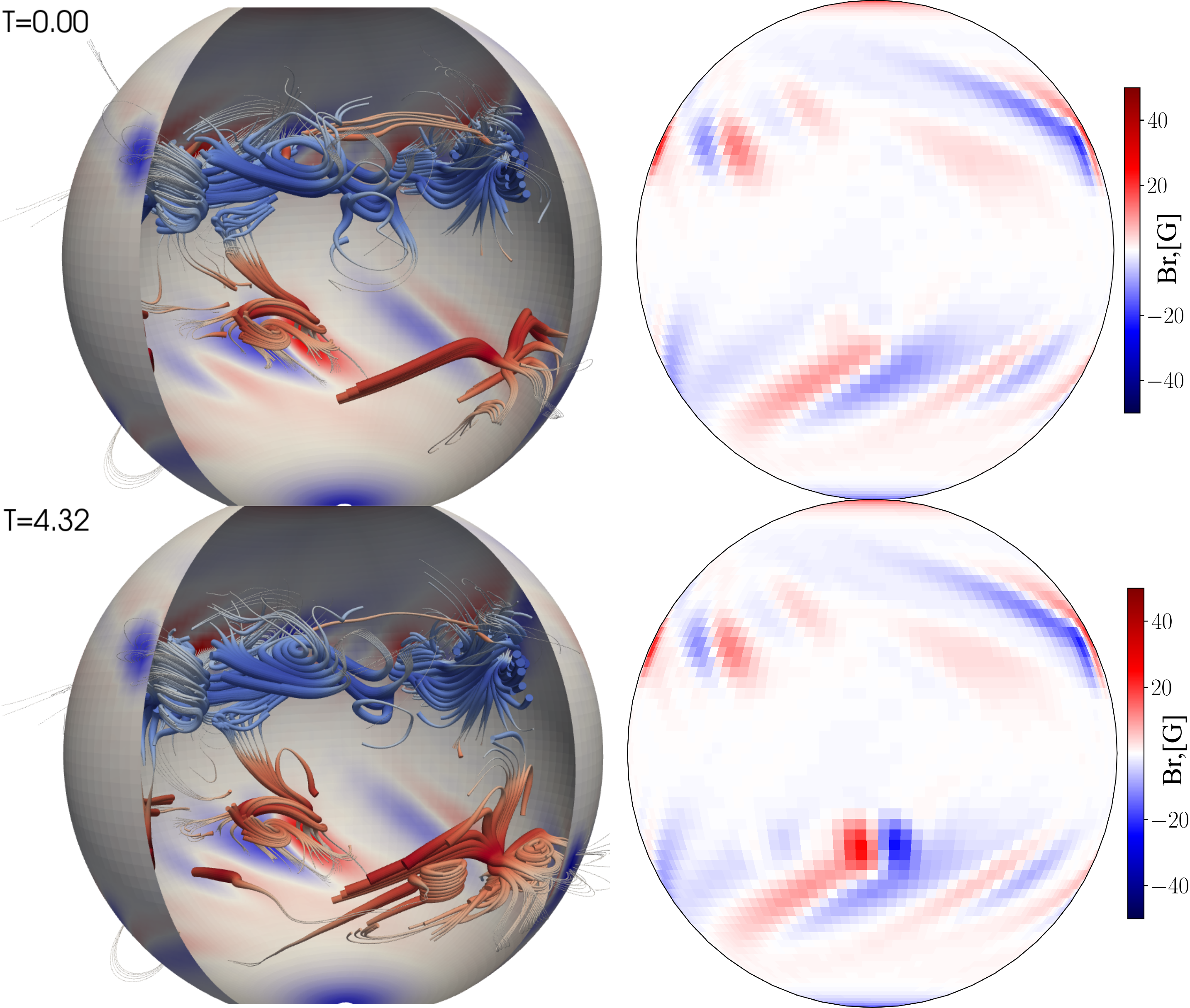}
\caption{\label{fig:bip}Snapshots of magnetic regions in the south hemisphere in ascending phase of the magnetic cycle.
The left column shows  the nonaxisymmetric magnetic
field lines, time is shown in days.
The right column shows the radial magnetic field on the top boundary. 
Reproduced by permission from \cite{Pipin2023a}.}
\end{figure}

The above analysis shows the importance of surface activity for the
dynamo model and perhaps for the solar dynamo as well.
Sunspot activity in the form of magnetic bipolar regions is one of
the most important aspects of magnetic surface activity.
A consistent approach to include it in dynamo models is at present absent. 
Also, the origin of sunspots and their bipolar magnetic field is
not well known; see \Sec{sec:sp}.
The Babcock--Leighton type and flux-transport dynamo models use a
phenomenological approach.
It is also applicable to mean-field models.
\citet{Pipin2022} studied the effect of surface activity on convection
zone dynamos. 
Here, we briefly discuss some results of the paper.
The emergence of bipolar magnetic regions (BMRs) is modeled using
the mean electromotive force which is represented by the $\alpha$ and
magnetic buoyancy effects acting on the unstable part of the axisymmetric
magnetic field as follows:
\begin{equation}
\mathcal{E}_{i}^{(\mathrm{BMR})}=\alpha_{\beta}\delta_{i\phi}\left\langle B\right\rangle _{\phi}+V_{\beta}\left(\hat{\boldsymbol{r}}\times\left\langle \mathbf{B}\right\rangle \right)_{i},\label{eq:ep}
\end{equation}
where the first term takes into account the BMR's tilt and the second
term models the surface magnetic region in the bipolar form.
To produce the bipolar like regions we have to restrict spatially
$V_{\beta}$ in Eq.~(\ref{eq:ep}) to the small scales that are typical
for the solar BMR; see details in the above cited paper.
Position and emergence time are chosen to be random and
modulated by the large-scale magnetic activity.
The BMR's $\alpha$-effect parameters are random as well; see details in
\citep{Pipin2022,Pipin2023a}. The given approach could be refined further
using the 3D hydrodynamics, effects of the Coriolis force and the theory
of the Joy's law developed recently by \cite{Kleeorin2020m}.
Figure~\ref{fig:bip} illustrates the formation of BMR simulated in the
dynamo model.
It was found that the BMR effects on the dynamo are restricted to the
shallow layer below the surface.
At the surface, the effect of the BMR on the magnetic field generation
is dominant.
Compare to the standard axisymmetric mean-field model discussed in
the subsections above, the nonaxisymmetric dynamo, which includes the
emergence of tilted BMR, can result in additional dynamo generation of
the large-scale poloidal magnetic field and to an increase of the polar
magnetic field.
The red line in \Fig{fig:bdg}(b) shows the budget for this nonaxisymmetric
dynamo model.
We see an increase of the toroidal flux generation rate in the
nonaxisymmetric model because of the surface BMR activity.
Similar to \citet{CS15}, we can conclude that sunspot surface activity
seems to play an important part in the solar large-scale dynamo.

\subsection{Effect of corona on the dynamo}

Usually, dynamo models are limited to the star embedded in a vacuum,
which is described by boundary conditions on the stellar surface. 
However, the boundary conditions have a determining influence on the
global solutions, such as the symmetry about the equator.
With the assumption of an external vacuum, all induction effects in the
corona are neglected.
Since the solar surface rotates differentially, the highly conductive
plasma in the corona also causes induction effects through shear.
Observations of coronal rotation are very scarce. 
There is evidence from extended coronal holes of rigid rotation in latitude \citep{Timothy75, Bagashvili17}.  
Kinematic dynamo models involving the corona with various assumptions 
on its rotation and conductivity give a wide range of solutions \citep{Elstner20}. 
A notable influence of the corona on the dynamo in the convection zone 
was also observed in DNS by \citet{Warnecke16}. 
A too weak density contrast and too strong viscous coupling of the corona to the star 
in their model probably underestimates the effect of the Lorentz force in the corona. 
Considering a dynamical situation with dominant Lorentz force in the corona, 
the solution in the Sun corresponds to that with vacuum boundary condition independent of rotation and conductivity in the corona. 
The magnetic field in the corona varies in time to a nearly force-free solution. 
Further investigations of the star-corona coupling are needed to clarify the exchange of magnetic energy and helicity. 

\subsection{Stellar cycle periods}
\label{StellarCyclePeriods}

\cite{NWV84} developed an early understanding of the observed stellar
cycle periods, $P_{\rm cyc}$.
In those years, there where just six stars with measured rotation
and cycle periods.
Remarkably, for the stars of \cite{NWV84}, those values have
not changed much with the more accurate data of \cite{Bal+95};
see \Tab{Tstar1} for a comparison of their cycle periods and the more
recent data sets \citep{Olspert2018,BoroSaikia2018, BC22}.
However, for the many stars of \cite{Bal+95}, the modern analysis of
\cite{Olspert2018} and \cite{BoroSaikia2018} showed considerable changes
in the results for stellar cycle periodicities.
In particular, many of the double periodicities of \cite{Bal+95} vanished
when extended data became available, and different methods were used.

\begin{table}[t]
\caption{Comparison of cycle periods $P_{\rm cyc}$ (in years) from
\cite{NWV84} (NWV84), \cite{Bal+95} (Bal+95), and \cite{BC22} (BC22).
The last two columns compare the seismic age given by BC22 and the
gyrochronological age as listed by \cite{Brandenburg2017A} (BMM17).
The latter differ significantly, but the determined cycle periods
were remarkably stable over the decades.
}\label{Tstar1}
\centerline{\begin{tabular}{rccccc}
\hline\noalign{\smallskip}
       & \multicolumn{3}{c}{------ $P_{\rm cyc}$ [yr] ------\;} & \multicolumn{2}{c}{\;--- age [Gyr] ---} \\
   HD  & NWV84 & Bal+95 & BC22  & BC22 & BMM17 \\
\noalign{\smallskip}\svhline\noalign{\smallskip}
  3651 &  10   & 13.8   & 14.70 & ---  & 7.2 \\
  4628 &   8.5 &  8.37  &  8.47 & 3.33 & 5.3 \\
 16160 &  11.5 & 13.2   & 12.68 & ---  & 6.9 \\
160346 &   7   &  7.00  &  7.19 & ---  & 4.4 \\
201091 &   7   &  7.3   &  7.11 & 6.10 & 3.3 \\
201092 &  11   & 11.7   &  ---  & ---  & 3.2 \\
\noalign{\smallskip}\hline\noalign{\smallskip}
\end{tabular}}
\end{table}

\begin{table}[t]
\caption{Comparison of stellar cycle properties from the samples of
\cite{BC22} and \cite{Brandenburg2017A} (indicated as ``old'').
The blue italics and red roman letters refer to the stars
discussed in \cite{Brandenburg2017A} and are also indicated in
\Fig{pKG_BonannoCorsaro}.
}\label{Tstar2}
\centerline{\begin{tabular}{rcccccccc}
\hline\noalign{\smallskip}
HD~~ & Sym & $\log\bra{R_{\rm HK}'}$ & $\log\bra{R_{\rm HK}^{'\rm old}}$ & 
$P_{\rm rot}$ [d] & $P_{\rm rot}^{\rm old}$ [d] &
$P_{\rm cyc}$ [yr] & $P_{\rm cyc}^{\rm I}$ [yr] & $P_{\rm cyc}^{\rm A}$ [yr] \\
\hline\noalign{\smallskip}
100180 & \blue{\it h} & $-4.83$ & $-4.92$ & 14.06 & 14.00 &  3.60 &  3.60 & 12.90\\
103095 & \blue{\it i} & $-4.90$ & $-4.90$ & 32.51 & 31.00 &  7.07 &  7.30 &  --- \\
 10476 &      \red{c} & $-4.97$ & $-4.91$ & 35.40 & 35.20 & 10.45 &  9.60 &  --- \\
146233 & \blue{\it l} & $-4.95$ & $-4.93$ & 22.66 & 22.70 & 11.59 &  7.10 &  --- \\
160346 &      \red{m} & $-4.86$ & $-4.79$ & 34.20 & 36.40 &  7.19 &  7.00 &  --- \\
 16160 &      \red{d} & $-4.94$ & $-4.96$ & 48.29 & 48.00 & 12.68 & 13.20 &  --- \\
165341 &      \red{n} & $-4.61$ & $-4.55$ & 19.51 & 19.00 &  5.09 &  5.10 & 15.50\\
166620 &      \red{o} & $-5.00$ & $-4.96$ & 42.25 & 42.40 & 16.81 & 15.80 &  --- \\
219834 &      \red{s} & $-4.93$ & $-4.94$ & 38.89 & 43.00 &  9.48 & 10.00 &  --- \\
 26965 &      \red{f} & $-4.96$ & $-4.87$ & 40.83 & 43.00 & 10.24 & 10.10 &  --- \\
  3651 &      \red{a} & $-5.06$ & $-4.99$ & 40.50 & 44.00 & 14.70 & 13.80 &  --- \\
  4628 &      \red{b} & $-4.95$ & $-4.85$ & 37.82 & 38.50 &  8.47 &  8.60 &  --- \\
 81809 &      \red{g} & $-4.89$ & $-4.92$ & 40.93 & 40.20 &  8.05 &  8.20 &  --- \\
219834 &      \red{r} & $-5.10$ & $-5.07$ & 43.40 & 42.00 & 16.29 & 21.00 &  --- \\
201091 &      \red{p} & $-4.56$ & $-4.76$ & 35.62 & 35.40 &  7.11 &  7.30 &  --- \\
   Sun & \blue{\it a} & $-4.94$ & $-4.90$ & 25.55 & 25.40 & 10.70 & 11.00 & 80.00\\
149661 &      \red{K} & $-4.61$ & $-4.58$ & 20.92 & 21.10 & 12.38 &  4.00 & 17.40\\
152391 & \blue{\it M} & $-4.46$ & $-4.45$ & 11.01 & 11.40 & 11.94 &  ---  & 10.90\\
156026 &      \red{L} & $-4.56$ & $-4.66$ & 18.85 & 21.00 & 19.31 &  ---  & 21.00\\
190406 & \blue{\it N} & $-4.76$ & $-4.80$ & 14.01 & 13.90 & 18.61 &  2.60 & 16.90\\
 76151 & \blue{\it F} & $-4.68$ & $-4.66$ & 14.70 & 15.00 & 16.34 &  2.50 &  --- \\
 78366 & \blue{\it G} & $-4.57$ & $-4.61$ &  9.60 &  9.70 & 14.26 &  5.90 & 12.20\\
114710 & \blue{\it J} & $-4.74$ & $-4.75$ & 11.99 & 12.30 & 14.12 &  9.60 & 16.60\\
 22049 &      \red{E} & $-4.46$ & $-4.46$ & 11.09 & 11.10 & 11.00 &  2.90 & 12.70\\
\noalign{\smallskip}\hline\noalign{\smallskip}
\end{tabular}
}\end{table}

In recent work of \cite{BC22}, new cycle data were collected for
altogether 67 stars.
Their new sample includes stars with less accurate data points, so the
existence of different branches was no longer a pronounced feature.
In addition, many of the new data points are different from the earlier
ones of \cite{Brandenburg2017A}; see \Tab{Tstar2}.
As in their paper, we denote G and F dwarfs by the same blue italic symbols
and K dwarfs by the same red roman symbols.

\begin{figure}[t]
\begin{center}
\includegraphics[width=.9\textwidth]{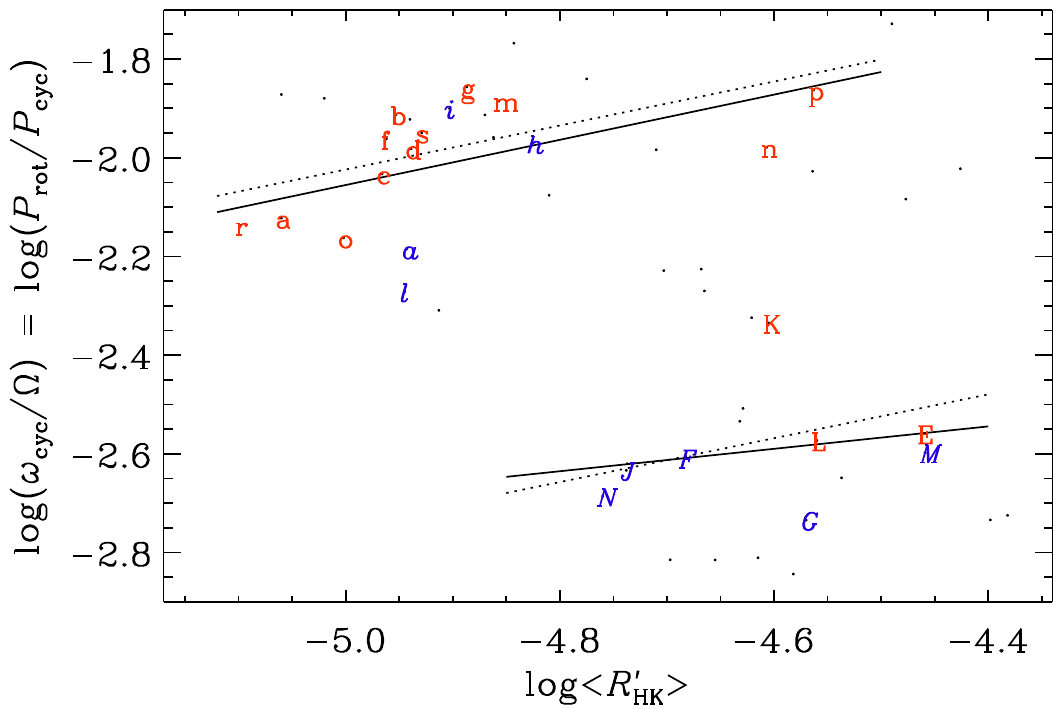}
\end{center}
\caption{$P_{\rm rot}/P_{\rm cyc}$ versus $\log\bra{R_{\rm HK}'}$ for all
stars of \cite{BC22} (small black symbols).
Lowercase (uppercase) letters denote data points of \cite{BC22}
that were also included in the sample of \cite{Brandenburg2017A}. There are refer to 
the data of  stars given in the Table \ref{Tstar2}.
The dotted lines denote the fits determined by \cite{Brandenburg2017A}
while the upper (lower) solid lines denote fits to the stars of
\cite{BC22} with lowercase  (uppercase) letters.}
\label{pKG_BonannoCorsaro}\end{figure}

To see how strong this revision of the data is, we plot in 
\Fig{pKG_BonannoCorsaro} the ratios $P_{\rm rot}/P_{\rm cyc}$
versus $\log\bra{R_{\rm HK}'}$ for all stars of \cite{BC22}
and highlight with lowercase and uppercase letters the stars
that were also included in the sample of \cite{Brandenburg2017A}.
We see that the new data are remarkably consistent with the
old ones.
Out of the eight stars on the branch of active stars,
five where listed by \cite{Brandenburg2017A} as having two periods.
Of the 16 inactive stars, three were listed with two periods, but
the case of the Sun was classified by \cite{Brandenburg2017A}
as somewhat different, because the 80 years Gleissberg cycle
does not fit well on the active branch and, unlike all the
other stars with two cycle periods, which are all younger than
$3.3\Gyr$, the Sun is relatively old.

The early data of \cite{NWV84} did already suggest
\begin{equation}
\omega_{\rm cyc}\propto \Omega^{1.25}
\label{omcyc}
\end{equation}
for the cycle frequency $\omega_{\rm cyc}=2\pi/P_{\rm cyc}$ versus angular
rotation rate $\Omega$.
This dependence is reproduced by considering free dynamo waves
and assuming axisymmetric mean fields, $\meanBB=b\pphi+\nab\times a\pphi$,
with $(a,b)\propto e^{\ii(ky-\omega t)}$ and writing
$-\ii\omega=-\ii\omega_{\rm cyc}+\lambda$, where both
$\omega_{\rm cyc}$ and $\lambda$ are assumed to be real.
The mean-field dynamo equations result in traveling wave solutions
with a dispersion relation of the form
\begin{equation}
\omega_{\rm cyc}=\sqrt{\alpha\Omega' kL/2},\quad
\lambda=\omega_{\rm cyc}-\etaT k^2.
\end{equation}
At least up to moderate rotation rates, it is reasonable to assume that $\alpha$
and $\Omega'$ are proportional to $\Omega$.
The crucial assumption in arriving at an approximation that matches
\Eq{omcyc} is to assume that the relevant wavenumber $k_y$ is selected
not by the condition of marginal excitation, but by the assumption that
$\lambda=\lambda(k)$ is maximized.
Thus, $k$ has to obey $\dd\lambda/\dd k=0$, which yields
$\omega_{\rm cyc}\propto(\alpha\Omega')^{2/3}\propto\Omega^{4/3}$.
By contrast, if the dynamo is quenched to the being marginally excited, then
$\omega_{\rm cyc}\approx\etaT/L^2$, which would be either independent
of $\Omega$, or perhaps even decreasing with $\Omega$, if $\etaT$ decreases
with increasing $\Omega$ due to quenching.

Of course, nonlinear dynamos must always be quenched to reach a steady state.
This led \cite{BST98} to suggest that \Eq{omcyc} could be obeyed if both
$\alpha$ and $\etaT$ are {\em antiquenched} in such a way that $\etaT$
is antiquenched more slowly than $\alpha$, so that $\omega_{\rm cyc}$
would increase with increasing magnetic field strength, and yet the
dynamo would still saturate.
Whether this is the only viable solution to this puzzle remains still
unclear.

\begin{figure}[t]
\begin{center}
\includegraphics[width=0.72\columnwidth]{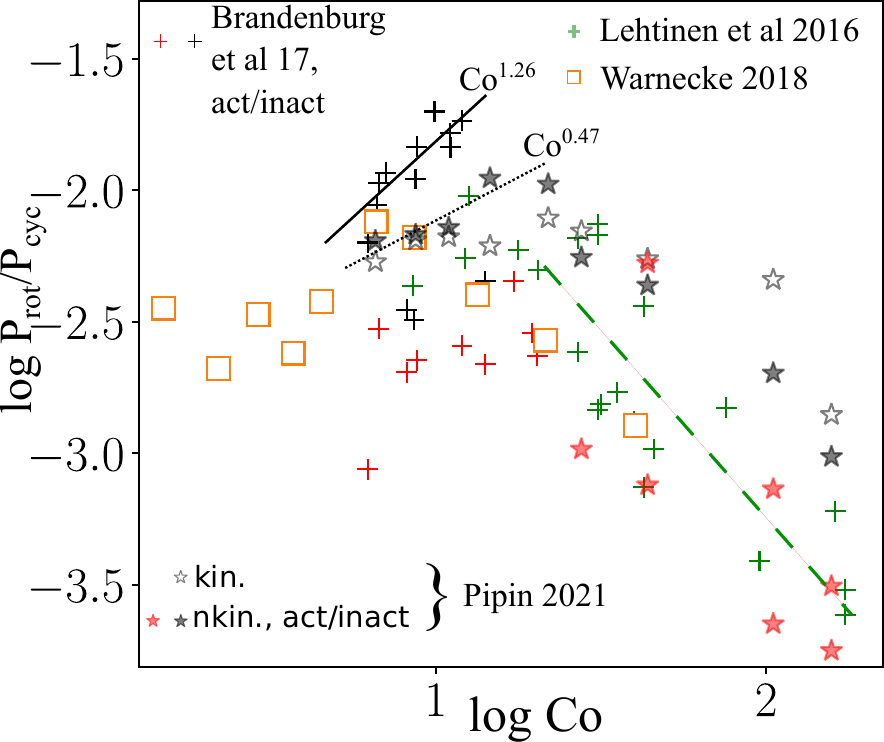}
\end{center}
\caption{\label{corp}Dependence of cycle period on stellar rotation rate.
Red and black crosses show the results of \citet{Brandenburg2017A},
green crosses those of  \citet{Lehtinen2016},
orange squares the models of \citet{warnecke18},
and the asterisks are from the models of \citet{Pipin21c};
act/inact marks the active and inactive branches of activity, respectively;
while kin/nkin stand for kinematic and nonkinematic models, respectively
(adapted by permission from \citealp{Pipin21c}).
}\end{figure}

Recently, a number of numerical dynamo models were applied to investigate
the relation of the cycle period on the stellar rotation rate in solar
analogs \citep{Pipin2015, Strugarek2017, warnecke18, Viviani2018,
Hazra2019, Noraz2022}.
Figure~\ref{corp} shows some of these results including the results
of observations of \citet{Brandenburg2017A} and the survey of
\citet{Lehtinen2016}.
It is interesting that the saturation branch of stellar activity for
the young solar analogs with rotation periods of less than 10 days
is well reproduced in the very different solar-like dynamo models
including various GCD simulations \citep{Strugarek2017, warnecke18,
Viviani2018}, flux transport model of \cite{Hazra2019} and mean-field
model of \cite{Pipin21c}.
In Fig.\ref{corp}, this branch is marked by the green line.

The mean cycle period in this branch is almost independent of stellar
rotation rate.
The nonkinematic nonlinear model of \cite{Pipin21c} shows multiple
periods along this line.
\cite{Pipin21c} found that saturation  of dynamo activity is accompanied
by a depression of latitudinal shear, a concentration of magnetic
activity to the surface, and changes in the meridional circulations from
one cell to multiple cells per hemisphere structure.
According to the conclusions of this paper, it is clear that in the
saturated state, dynamo waves do not follow the Parker--Yoshimura rule.
Their cycle period is determined by turbulent diffusion and meridional
circulation.
This is why the predictions of flux transport and nonkinematic mean-field
dynamo models coincide.
The independence of the cycle period on the rotation rate can be typical
for the dynamo solutions which show concentration of the magnetic activity
toward the boundaries of the dynamo (see \citealp{Pipin2015,Pipin2016b}).

The inactive branch of the nonkinematic mean-field dynamo models shows
a fairly strong positive inclination (see \Fig{corp}), which is absent
in the kinematic models.
We see that the dynamo model can reproduce a power law $\sim\Co^{0.5}$,
avoiding the antiquenching concept of \cite{BST98}.
In fact, the nonkinematic dynamo models show a frequency doubling
phenomenon for models in the range from 10 to 15 days rotation period
(see Figs.~3 and 8 of \citealp{Pipin21c}).
The frequency doubling or the second harmonic generation is known from
nonlinear optics.
This is typical for wave propagation in nonlinear media.
In dynamo waves, second harmonics are generated because of $\BB^2$
effects such as the magnetic feedback on the large-scale flow, magnetic
helicity conservation, and magnetic buoyancy effects.
The second harmonics can be found in the solar activity as well
\citep{Setal20}.
For the solar case, they are subdominant.
However, they can become dominant for fast rotating stars.
This makes the interpretation of magnetic activity cycles difficult
\citep{Stepanov2020MN}.

It is important to note that the GCD simulations of \cite{Viviani2018}
show an increased level of nonaxisymmetry with an increase of
rotation rate and a transition to preferred nonaxisymmetric dynamos
for solar-type stars with a rotation period of less than 14 days.
In simulations, this transition happens when the rotation profile changes
from an antisolar to a solar-like profile with a faster equator and
slow poles.
Stellar Zeeman Doppler Imaging observations of \cite{Donati2009} and
\cite{See2016}, for example, show that the magnetic topology depends on
stellar mass and rotation rate.
In a certain interval of rotation periods (below $14\days$) and
stellar mass, the results of \cite{Viviani2018} are compatible
with the observational findings mentioned above.

Summarizing, we suggest that the Parker--Yoshimura dynamo regime can
work for solar analogs rotating with periods above 14--15 days.
In an interval of stellar rotation periods between 10 and 15 days,
frequency doubling and a transition to the nonaxisymmetric dynamos occurs.
For lower rotation periods, the dynamo transits to a saturation stage.
It can be characterized by high magnetic activity and multiple dynamo
periods which are independent of stellar rotation rate.
This new picture should be further improved by including possible dynamo
effects of surface activity in the form of bipolar magnetic regions and
star spots.

Another interesting observation is that different types of stars, including
partially convective main sequence solar-type stars, fully
convective M-dwarfs, and evolved post-main sequence giant stars show
similar scaling with the Rossby number for the unsaturated regime (see,
e.g., \citealp{WrightDrake2016N,Wright2018W,Lehtinen2020}).
In particular, to derive the Rossby number, these study employ a
simple parameterization of the convective turnover time suggested by
stellar interior models.
It was found that for evolved stars, the Rossby-independent
parameterizations break down in the rotation–activity relation
\citep{Lehtinen2020}.
This constitutes a strong argument in favor of the turbulent dynamo
paradigm suggesting a common role of the turbulent process in dynamos
operating in these stars.

\subsection{Analogy of mean-field $\alpha$ and the chiral magnetic effect}

The $\alpha$ effect in mean-field dynamo theory is an effect that
emerges after averaging over the scale of several turbulent eddies.
We know already that turbulent diffusion is somewhat analogous to
microphysical diffusion, which also emerges after averaging,
but here after averaging over atomic scales.
Interestingly, even for the $\alpha$ effect there can be an effect
on atomic and subatomic scales, because fermions, such as electrons,
are chiral.
The spin of an electron emerging from the decay of a neutron is
anti-aligned with its momentum vector, so their dot product is a
negative pseudo-scalar, called the chirality.
Positrons have positive chirality.
In the presence of an ambient magnetic field, the spins align,
but electrons and positrons move in opposite directions, causing
therefore an electric current.
This constitutes a microscopic $\alpha$ effect \citep{Roga+17,Bran+17},
\begin{equation}
\alpha_{\rm micro}\equiv\mu_5\eta=24\alpha_{\rm fine}
(n_{\rm L}-n_{\rm R})(\hbar c/\kB T)^2,
\end{equation}
where $\mu_5$ is the normalized chiral chemical potential (with units
of inverse length), $\eta$ is the microscopic magnetic diffusivity,
$\alpha_{\rm fine}\approx1/137$ is the fine structure constant (quantifying
the strength of electromagnetic interaction between charged particles),
$n_{\rm L}$ and $n_{\rm R}$ are the number densities of left- and
right-handed fermions, $\hbar\approx10^{-27}\erg\s$ is the reduced Planck
constant, $c\approx3\times10^{10}\cm\s^{-1}$ is the speed of light,
$\kB\approx10^{-16}\erg\K^{-1}$ is the Boltzmann constant, and $T$
is the temperature.

The applications of chiral MHD are manifold and range from condensed matter
systems and heavy ion collisions to neutron stars and the early Universe;
see \cite{Khar14} for a review.
Interestingly, because this microscopic $\alpha$ effect produces
helical magnetic fields, and because the total chirality is conserved
\citep{BFR12}, this effect does not last forever, but is being quenched
in a form analogous to the catastrophic quenching formula, which takes
the form \citep{Roga+17}
\begin{equation}
\frac{\partial\mu_5}{\partial t}=-\lambda\eta
\left(\mu_5\meanBB^2-\etat\mu_0\meanJJ\cdot\meanBB\right)
-\Gamma_{\rm flip}\mu_5,
\end{equation}
where $\lambda$ is a coupling constant which, in the catastrophic quenching
formalism, is related to $2\etat\kf^2/\Beq^2$, and $\Gamma_{\rm flip}$
is a spin-flipping parameter, which is related to $2\eta\kf^2$ in the
catastrophic quenching formalism \citep[see, e.g.,][]{FB02,BB02}.

There is a vast range of recent work in this field, which goes well
beyond the scope of the present paper.
We just mention here the paper of \cite{Masada+18}, who studied
chiral magnetohydrodynamic turbulence in core-collapse supernovae.
They found that the inverse cascade related to the chiral effects impacts
the magnetohydrodynamic evolution in the supernova core toward explosion.

\subsection{Direct Statistical Simulations}

An alternative or extension to mean-field theory in the usual
sense is to solve the time-dependent system of one-point and two-point
correlation functions.
This is what is now known as Direct Statistical Simulations \citep{TMD11,
TM13, TM17} and has primarily been applied to two-dimensional turbulent
shear flow problems.
The dimensionality of the two-point correlation function doubles for
those directions over which homogeneity cannot be assumed.
On the other hand, the dynamics of the low order statistics is usually
slower than that of the original equations.
In addition, it is possible to reduce the complexity of the problem by
employing Proper Orthogonal Decomposition \citep{Allawala+Marston16, ATM20}.
This approach has not yet been applied to magnetohydrodynamics and the
dynamo problem.
Such an approach would be able to incorporate both small-scale and
large-scale dynamos at the same time.
The small-scale dynamo problem would be solved through correlation
functions, as has been done in \cite{BS00}, for example.
But their simulations were isotropic and did not include anisotropic
dependencies on position.
Nevertheless, this approach has the potential of being a strong competitor in
addressing the high Reynolds number dynamics of problems of astrophysical
and geophysical relevance; see \cite{Tobias21} for a recent review
touching on these aspects.

\section{Looking forward}

In this review, we have provided some insight into recent developments
in our understanding of the generation of astrophysical large-scale
magnetic fields.
The current development of mean-field theory allows us to go beyond some of
the original restrictions that were related to the assumption of large
scale-separation and the inappropriate neglect of nonlinear effects
due to higher order correlations in contributions to the mean turbulent
electromotive force.
A big portion of the progress comes from the development in the DNS
of astrophysical turbulence.
Noteworthy, the classical mean-field theory is based on the fundamental
equations of electrodynamics and has well-known limits.
With the new steps forward, we can take into account results of the DNS,
e.g., the spectral kernels, and treat them as experimental facts.
The necessity of some phenomenological additions to classical mean-field
theory are motivated both by DNS and observations of the magnetic activity
in astrophysical systems, such as those in our Sun and other stars.
In this way, mean-field models become a valuable tool to understand
the real and virtual worlds of the dynamo in stars and in DNS.

\begin{acknowledgement}
We thank Hantao Ji for important discussions on the reversed field pinch
in laboratory plasmas and dynamos in magnetically dominated settings.
We are also indebted to the two reviewers for their constructive remarks
that have led to improvements in the presentation.
Support through the grant 2019-04234 from the Swedish Research Council (Vetenskapsr{\aa}det) (AB)
is gratefully acknowledged.
We thank for the allocation of computing resources provided by the
Swedish National Infrastructure for Computing (SNIC)
at the PDC Center for High Performance Computing Stockholm and Link\"oping.
VP acknowledges financial support of the Ministry of Science
and Higher Education (Subsidy No.075-GZ/C3569/278). 
YM has been supported by MEXT/JSPS KAKENHI Grant numbers JP18H01212, JP18K03700, JP21K03612, JP23H01199;
funding from Fukuoka University (Grant No. 171042, 177103 and GR2302).
\end{acknowledgement}

\section*{Declarations}

{\bf Competing Interests}~
The authors declare they have no conflicts of interest.
\\

\noindent
{\bf Open Access}~
This article is licensed under a Creative Commons Attribution 4.0 International License, which
permits use, sharing, adaptation, distribution and reproduction in any medium or format, as long as you give
appropriate credit to the original author(s) and the source, provide a link to the Creative Commons licence,
and indicate if changes were made. The images or other third party material in this article are included in the
article's Creative Commons licence, unless indicated otherwise in a credit line to the material. If material is
not included in the article's Creative Commons licence and your intended use is not permitted by statutory
regulation or exceeds the permitted use, you will need to obtain permission directly from the copyright holder.
To view a copy of this licence, visit \url{http://creativecommons.org/licenses/by/4.0/.}

\bibliographystyle{spbasic}
\bibliography{ref}

\begin{thebibliography}{262}
\providecommand{\natexlab}[1]{#1}
\providecommand{\url}[1]{{#1}}
\providecommand{\urlprefix}{URL }
\expandafter\ifx\csname urlstyle\endcsname\relax
  \providecommand{\doi}[1]{DOI~\discretionary{}{}{}#1}\else
  \providecommand{\doi}{DOI~\discretionary{}{}{}\begingroup
  \urlstyle{rm}\Url}\fi
\providecommand{\eprint}[2][]{\url{#2}}

\bibitem[{{Allawala} and {Marston}(2016)}]{Allawala+Marston16}
{Allawala} A, {Marston} JB (2016) {Statistics of the stochastically forced
  Lorenz attractor by the Fokker-Planck equation and cumulant expansions}. \pre
  94(5):052218, \doi{10.1103/PhysRevE.94.052218}, \eprint{1604.00867}

\bibitem[{{Allawala} et~al(2020){Allawala}, {Tobias}, and {Marston}}]{ATM20}
{Allawala} A, {Tobias} SM, {Marston} JB (2020) {Dimensional reduction of direct
  statistical simulation}. J Fluid Mech 898:A21, \doi{10.1017/jfm.2020.382},
  \eprint{1708.07805}

\bibitem[{{Altrock} et~al(2008){Altrock}, {Howe}, and {Ulrich}}]{Altrock2008}
{Altrock} R, {Howe} R, {Ulrich} R (2008) Solar torsional oscillations and their
  relationship to coronal activity. In: {Howe} R, {Komm} RW, {Balasubramaniam}
  KS, {Petrie} GJD (eds) Subsurface and Atmospheric Influences on Solar
  Activity, Astron. Soc. Pacif. Conf. Ser., vol 383, p 335

\bibitem[{{Augustson} et~al(2015){Augustson}, {Brun}, {Miesch}, and
  {Toomre}}]{augustson+15}
{Augustson} K, {Brun} AS, {Miesch} M, {Toomre} J (2015) {Grand Minima and
  Equatorward Propagation in a Cycling Stellar Convective Dynamo}. \apj
  809(2):149, \doi{10.1088/0004-637X/809/2/149}, \eprint{1410.6547}

\bibitem[{{Bagashvili} et~al(2017){Bagashvili}, {Shergelashvili}, {Japaridze},
  {Chargeishvili}, {Kosovichev}, {Kukhianidze}, {Ramishvili}, {Zaqarashvili},
  {Poedts}, {Khodachenko}, and {De Causmaecker}}]{Bagashvili17}
{Bagashvili} SR, {Shergelashvili} BM, {Japaridze} DR, {Chargeishvili} BB,
  {Kosovichev} AG, {Kukhianidze} V, {Ramishvili} G, {Zaqarashvili} TV, {Poedts}
  S, {Khodachenko} ML, {De Causmaecker} P (2017) {Statistical properties of
  coronal hole rotation rates: Are they linked to the solar interior?} \aap
  603:A134, \doi{10.1051/0004-6361/201630377}, \eprint{1706.04464}

\bibitem[{{Baliunas} et~al(1995){Baliunas}, {Donahue}, {Soon}, {Horne},
  {Frazer}, {Woodard-Eklund}, {Bradford}, {Rao}, {Wilson}, {Zhang}, {Bennett},
  {Briggs}, {Carroll}, {Duncan}, {Figueroa}, {Lanning}, {Misch}, {Mueller},
  {Noyes}, {Poppe}, {Porter}, {Robinson}, {Russell}, {Shelton}, {Soyumer},
  {Vaughan}, and {Whitney}}]{Bal+95}
{Baliunas} SL, {Donahue} RA, {Soon} WH, {Horne} JH, {Frazer} J,
  {Woodard-Eklund} L, {Bradford} M, {Rao} LM, {Wilson} OC, {Zhang} Q, {Bennett}
  W, {Briggs} J, {Carroll} SM, {Duncan} DK, {Figueroa} D, {Lanning} HH, {Misch}
  T, {Mueller} J, {Noyes} RW, {Poppe} D, {Porter} AC, {Robinson} CR, {Russell}
  J, {Shelton} JC, {Soyumer} T, {Vaughan} AH, {Whitney} JH (1995)
  {Chromospheric variations in main-sequence stars}. \apj 438:269--287,
  \doi{10.1086/175072}

\bibitem[{{Baryshnikova} and {Shukurov}(1987)}]{baryshnikova+87}
{Baryshnikova} I, {Shukurov} A (1987) {Oscillatory alpha-squared dynamo -
  Numerical investigation}. Astronomische Nachrichten 308(2):89--100,
  \doi{10.1002/asna.2113080202}

\bibitem[{{Beaudoin} et~al(2013){Beaudoin}, {Charbonneau}, {Racine}, and
  {Smolarkiewicz}}]{Beaudoin2013}
{Beaudoin} P, {Charbonneau} P, {Racine} E, {Smolarkiewicz} PK (2013) Torsional
  oscillations in a global solar dynamo. \solphys 282(2):335--360,
  \doi{10.1007/s11207-012-0150-2}, \eprint{1210.1209}

\bibitem[{{Berdyugina}(2005)}]{Berdyugina2005LRSP}
{Berdyugina} SV (2005) {Starspots: A Key to the Stellar Dynamo}. Living Reviews
  in Solar Physics 2(1):8, \doi{10.12942/lrsp-2005-8}

\bibitem[{{Biskamp} and {M{\"u}ller}(1999)}]{BM99}
{Biskamp} D, {M{\"u}ller} WC (1999) {Decay Laws for Three-Dimensional
  Magnetohydrodynamic Turbulence}. \prl 83(11):2195--2198,
  \doi{10.1103/PhysRevLett.83.2195}, \eprint{physics/9903028}

\bibitem[{{Blackman} and {Brandenburg}(2002)}]{BB02}
{Blackman} EG, {Brandenburg} A (2002) {Dynamic Nonlinearity in Large-Scale
  Dynamos with Shear}. \apj 579(1):359--373, \doi{10.1086/342705},
  \eprint{astro-ph/0204497}

\bibitem[{{Blackman} and {Ji}(2006)}]{BJ06}
{Blackman} EG, {Ji} H (2006) {Laboratory plasma dynamos, astrophysical dynamos
  and magnetic helicity evolution}. \mnras 369(4):1837--1848,
  \doi{10.1111/j.1365-2966.2006.10431.x}, \eprint{astro-ph/0604221}

\bibitem[{{Bonanno} and {Corsaro}(2022)}]{BC22}
{Bonanno} A, {Corsaro} E (2022) {On the Origin of the Dichotomy of Stellar
  Activity Cycles}. \apjl 939(2):L26, \doi{10.3847/2041-8213/ac9c05},
  \eprint{2210.11305}

\bibitem[{{Boro Saikia} et~al(2018){Boro Saikia}, {Marvin}, {Jeffers},
  {Reiners}, {Cameron}, {Marsden}, {Petit}, {Warnecke}, and
  {Yadav}}]{BoroSaikia2018}
{Boro Saikia} S, {Marvin} CJ, {Jeffers} SV, {Reiners} A, {Cameron} R, {Marsden}
  SC, {Petit} P, {Warnecke} J, {Yadav} AP (2018) {Chromospheric activity
  catalogue of 4454 cool stars. Questioning the active branch of stellar
  activity cycles}. \aap 616:A108, \doi{10.1051/0004-6361/201629518},
  \eprint{1803.11123}

\bibitem[{{Boyarsky} et~al(2012){Boyarsky}, {Fr{\"o}hlich}, and
  {Ruchayskiy}}]{BFR12}
{Boyarsky} A, {Fr{\"o}hlich} J, {Ruchayskiy} O (2012) {Self-Consistent
  Evolution of Magnetic Fields and Chiral Asymmetry in the Early Universe}.
  \prl 108(3):031301, \doi{10.1103/PhysRevLett.108.031301}, \eprint{1109.3350}

\bibitem[{{Brandenburg}(1998)}]{Brandenbur1998proc}
{Brandenburg} A (1998) {Disc turbulence and viscosity.} In: {Abramowicz} MA,
  {Bj{\"o}rnsson} G, {Pringle} JE (eds) Theory of Black Hole Accretion Disks,
  pp 61--90

\bibitem[{{Brandenburg}(2001)}]{Bra01}
{Brandenburg} A (2001) {The Inverse Cascade and Nonlinear Alpha-Effect in
  Simulations of Isotropic Helical Hydromagnetic Turbulence}. \apj
  550(2):824--840, \doi{10.1086/319783}, \eprint{astro-ph/0006186}

\bibitem[{{Brandenburg}(2008)}]{Bra08AN}
{Brandenburg} A (2008) {The dual role of shear in large-scale dynamos}. Astron
  Nachr 329(7):725, \doi{10.1002/asna.200811027}, \eprint{0808.0959}

\bibitem[{{Brandenburg}(2016)}]{B16}
{Brandenburg} A (2016) {Stellar Mixing Length Theory with Entropy Rain}. \apj
  832(1):6, \doi{10.3847/0004-637X/832/1/6}, \eprint{1504.03189}

\bibitem[{{Brandenburg}(2017)}]{Bra17}
{Brandenburg} A (2017) {Analytic solution of an oscillatory migratory
  {\ensuremath{\alpha}}$^{2}$ stellar dynamo}. \aap 598:A117,
  \doi{10.1051/0004-6361/201630033}, \eprint{1611.02671}

\bibitem[{{Brandenburg}(2018{\natexlab{a}})}]{Bran18}
{Brandenburg} A (2018{\natexlab{a}}) {Advances in mean-field dynamo theory and
  applications to astrophysical turbulence}. J Plasma Phys 84(4):735840404,
  \doi{10.1017/S0022377818000806}, \eprint{1801.05384}

\bibitem[{{Brandenburg}(2018{\natexlab{b}})}]{Brandenburg2018AN}
{Brandenburg} A (2018{\natexlab{b}}) {Magnetic helicity and fluxes in an
  inhomogeneous {\ensuremath{\alpha}}$^{2}$ dynamo}. Astron Nachr
  339(631):631--640, \doi{10.1002/asna.201913604}

\bibitem[{{Brandenburg} and {Chatterjee}(2018)}]{BC18}
{Brandenburg} A, {Chatterjee} P (2018) {Strong nonlocality variations in a
  spherical mean-field dynamo}. Astron Nachr 339:118--126,
  \doi{10.1002/asna.201813472}, \eprint{1802.04231}

\bibitem[{{Brandenburg} and {Chen}(2020)}]{BC20}
{Brandenburg} A, {Chen} L (2020) {The nature of mean-field generation in three
  classes of optimal dynamos}. J Plasma Phys 86(1):905860110,
  \doi{10.1017/S0022377820000082}, \eprint{1911.01712}

\bibitem[{{Brandenburg} and {Dobler}(2001)}]{BD01}
{Brandenburg} A, {Dobler} W (2001) {Large scale dynamos with helicity loss
  through boundaries}. \aap 369:329--338, \doi{10.1051/0004-6361:20010123},
  \eprint{astro-ph/0012472}

\bibitem[{{Brandenburg} and {Hubbard}(2015)}]{BranHubKap2015}
{Brandenburg} A, {Hubbard} PJ A~K{\"a}pyl{\"a} (2015) {Dynamical quenching with
  non-local {\ensuremath{\alpha}} and downward pumping}. Astronomische
  Nachrichten 336(1):91--96, \doi{10.1002/asna.201412141}, \eprint{1412.0997}

\bibitem[{{Brandenburg} and {K{\"a}pyl{\"a}}(2007)}]{BranKap2007}
{Brandenburg} A, {K{\"a}pyl{\"a}} PJ (2007) {Magnetic helicity effects in
  astrophysical and laboratory dynamos}. New J Phys 9(8):305,
  \doi{10.1088/1367-2630/9/8/305}, \eprint{0705.3507}

\bibitem[{{Brandenburg} and {Larsson}(2023)}]{BL23}
{Brandenburg} A, {Larsson} G (2023) {Turbulence with Magnetic Helicity That Is
  Absent on Average}. Atmosphere 14(6):932, \doi{10.3390/atmos14060932},
  \eprint{2305.08769}

\bibitem[{{Brandenburg} and {Ntormousi}(2023)}]{Bran+Ntor23}
{Brandenburg} A, {Ntormousi} E (2023) {Galactic Dynamos}. Ann Rev Astron
  Astrophys 61(1):561--606, \doi{10.1146/annurev-astro-071221-052807},
  \urlprefix\url{https://doi.org/10.1146/annurev-astro-071221-052807}

\bibitem[{{Brandenburg} and {R{\"a}dler}(2013)}]{BR13}
{Brandenburg} A, {R{\"a}dler} KH (2013) {Yoshizawa's cross-helicity effect and
  its quenching}. Geophysical and Astrophysical Fluid Dynamics
  107(1-2):207--217, \doi{10.1080/03091929.2012.681307}, \eprint{1112.1237}

\bibitem[{{Brandenburg} and {Sokoloff}(2002)}]{BS02}
{Brandenburg} A, {Sokoloff} D (2002) {Local and Nonlocal Magnetic Diffusion and
  Alpha-Effect Tensors in Shear Flow Turbulence}. Geophysical and Astrophysical
  Fluid Dynamics 96(4):319--344, \doi{10.1080/03091920290032974},
  \eprint{astro-ph/0111568}

\bibitem[{{Brandenburg} and {Subramanian}(2000)}]{BS00}
{Brandenburg} A, {Subramanian} K (2000) {Large scale dynamos with ambipolar
  diffusion nonlinearity}. \aap 361:L33--L36,
  \doi{10.48550/arXiv.astro-ph/0007450}, \eprint{astro-ph/0007450}

\bibitem[{{Brandenburg} and {Subramanian}(2005{\natexlab{a}})}]{BS05}
{Brandenburg} A, {Subramanian} K (2005{\natexlab{a}}) Astrophysical magnetic
  fields and nonlinear dynamo theory. \physrep 417:1--209,
  \doi{10.1016/j.physrep.2005.06.005}, \eprint{arXiv:astro-ph/0405052}

\bibitem[{{Brandenburg} and
  {Subramanian}(2005{\natexlab{b}})}]{brandenburg+05b}
{Brandenburg} A, {Subramanian} K (2005{\natexlab{b}}) {Minimal tau
  approximation and simulations of the alpha effect}. \aap 439(3):835--843,
  \doi{10.1051/0004-6361:20053221}, \eprint{astro-ph/0504222}

\bibitem[{{Brandenburg} and {Subramanian}(2005{\natexlab{c}})}]{BS05c}
{Brandenburg} A, {Subramanian} K (2005{\natexlab{c}}) {Strong mean field
  dynamos require supercritical helicity fluxes}. Astron Nachr 326:400--408,
  \doi{10.1002/asna.200510362}, \eprint{astro-ph/0505457}

\bibitem[{{Brandenburg} and {Subramanian}(2007)}]{BS07}
{Brandenburg} A, {Subramanian} K (2007) {Simulations of the anisotropic kinetic
  and magnetic alpha effects}. Astronomische Nachrichten 328(6):507,
  \doi{10.1002/asna.200710772}, \eprint{0705.3508}

\bibitem[{{Brandenburg} and {Tuominen}(1988)}]{brandenburg+88}
{Brandenburg} A, {Tuominen} I (1988) {Variation of magnetic fields and flows
  during the solar cycle}. Adv Spa Res 8(7):185--189,
  \doi{10.1016/0273-1177(88)90190-1}

\bibitem[{{Brandenburg} et~al(1990){Brandenburg}, {Tuominen}, {Nordlund},
  {Pulkkinen}, and {Stein}}]{brandenburg+90}
{Brandenburg} A, {Tuominen} I, {Nordlund} A, {Pulkkinen} P, {Stein} RF (1990)
  {3-D simulation of turbulent cyclonic magneto-convection.} \aap 232:277--291

\bibitem[{{Brandenburg} et~al(1992){Brandenburg}, {Moss}, and
  {Tuominen}}]{Brandenburg1992}
{Brandenburg} A, {Moss} D, {Tuominen} I (1992) Stratification and
  thermodynamics in mean-field dynamos. \aap 265:328--344

\bibitem[{{Brandenburg} et~al(1996){Brandenburg}, {Jennings}, {Nordlund},
  {Rieutord}, {Stein}, and {Tuominen}}]{brandenburg+96}
{Brandenburg} A, {Jennings} RL, {Nordlund} {\r{A}}, {Rieutord} M, {Stein} RF,
  {Tuominen} I (1996) {Magnetic structures in a dynamo simulation}. J Fluid
  Mech 306:325--352, \doi{10.1017/S0022112096001322}

\bibitem[{{Brandenburg} et~al(1998){Brandenburg}, {Saar}, and {Turpin}}]{BST98}
{Brandenburg} A, {Saar} SH, {Turpin} CR (1998) {Time Evolution of the Magnetic
  Activity Cycle Period}. \apjl 498(1):L51--L54, \doi{10.1086/311297}

\bibitem[{{Brandenburg} et~al(2008){Brandenburg}, {R{\"a}dler}, and
  {Schrinner}}]{BRS08}
{Brandenburg} A, {R{\"a}dler} KH, {Schrinner} M (2008) {Scale dependence of
  alpha effect and turbulent diffusivity}. \aap 482(3):739--746,
  \doi{10.1051/0004-6361:200809365}, \eprint{0801.1320}

\bibitem[{{Brandenburg} et~al(2009){Brandenburg}, {Candelaresi}, and
  {Chatterjee}}]{BCC09}
{Brandenburg} A, {Candelaresi} S, {Chatterjee} P (2009) {Small-scale magnetic
  helicity losses from a mean-field dynamo}. \mnras 398(3):1414--1422,
  \doi{10.1111/j.1365-2966.2009.15188.x}, \eprint{0905.0242}

\bibitem[{{Brandenburg} et~al(2010){Brandenburg}, {Kleeorin}, and
  {Rogachevskii}}]{BKR10}
{Brandenburg} A, {Kleeorin} N, {Rogachevskii} I (2010) {Large-scale magnetic
  flux concentrations from turbulent stresses}. Astron Nachr 331(1):5,
  \doi{10.1002/asna.200911311}, \eprint{0910.1835}

\bibitem[{{Brandenburg} et~al(2011){Brandenburg}, {Kemel}, {Kleeorin}, {Mitra},
  and {Rogachevskii}}]{brandenburg+11}
{Brandenburg} A, {Kemel} K, {Kleeorin} N, {Mitra} D, {Rogachevskii} I (2011)
  {Detection of Negative Effective Magnetic Pressure Instability in Turbulence
  Simulations}. \apjl 740(2):L50, \doi{10.1088/2041-8205/740/2/L50},
  \eprint{1109.1270}

\bibitem[{{Brandenburg} et~al(2012){Brandenburg}, {R{\"a}dler}, and
  {Kemel}}]{Brandenburg2012AA}
{Brandenburg} A, {R{\"a}dler} KH, {Kemel} K (2012) {Mean-field transport in
  stratified and/or rotating turbulence}. \aap 539:A35,
  \doi{10.1051/0004-6361/201117871}, \eprint{1108.2264}

\bibitem[{{Brandenburg} et~al(2013){Brandenburg}, {Kleeorin}, and
  {Rogachevskii}}]{BKR13}
{Brandenburg} A, {Kleeorin} N, {Rogachevskii} I (2013) {Self-assembly of
  Shallow Magnetic Spots through Strongly Stratified Turbulence}. \apjl
  776(2):L23, \doi{10.1088/2041-8205/776/2/L23}, \eprint{1306.4915}

\bibitem[{{Brandenburg} et~al(2014){Brandenburg}, {Gressel}, {Jabbari},
  {Kleeorin}, and {Rogachevskii}}]{BGJKR14}
{Brandenburg} A, {Gressel} O, {Jabbari} S, {Kleeorin} N, {Rogachevskii} I
  (2014) {Mean-field and direct numerical simulations of magnetic flux
  concentrations from vertical field}. \aap 562:A53,
  \doi{10.1051/0004-6361/201322681}, \eprint{1309.3547}

\bibitem[{{Brandenburg} et~al(2017{\natexlab{a}}){Brandenburg}, {Ashurova}, and
  {Jabbari}}]{BAJ17}
{Brandenburg} A, {Ashurova} MB, {Jabbari} S (2017{\natexlab{a}}) {Compensating
  Faraday Depolarization by Magnetic Helicity in the Solar Corona}. \apjl
  845(2):L15, \doi{10.3847/2041-8213/aa844b}, \eprint{1706.09540}

\bibitem[{{Brandenburg} et~al(2017{\natexlab{b}}){Brandenburg}, {Mathur}, and
  {Metcalfe}}]{Brandenburg2017A}
{Brandenburg} A, {Mathur} S, {Metcalfe} TS (2017{\natexlab{b}}) {Evolution of
  Co-existing Long and Short Period Stellar Activity Cycles}. \apj 845(1):79,
  \doi{10.3847/1538-4357/aa7cfa}, \eprint{1704.09009}

\bibitem[{{Brandenburg} et~al(2017{\natexlab{c}}){Brandenburg}, {Schober},
  {Rogachevskii}, {Kahniashvili}, {Boyarsky}, {Fr{\"o}hlich}, {Ruchayskiy}, and
  {Kleeorin}}]{Bran+17}
{Brandenburg} A, {Schober} J, {Rogachevskii} I, {Kahniashvili} T, {Boyarsky} A,
  {Fr{\"o}hlich} J, {Ruchayskiy} O, {Kleeorin} N (2017{\natexlab{c}}) {The
  Turbulent Chiral Magnetic Cascade in the Early Universe}. \apjl 845(2):L21,
  \doi{10.3847/2041-8213/aa855d}, \eprint{1707.03385}

\bibitem[{{Browning} et~al(2006){Browning}, {Miesch}, {Brun}, and
  {Toomre}}]{browning+06}
{Browning} MK, {Miesch} MS, {Brun} AS, {Toomre} J (2006) {Dynamo Action in the
  Solar Convection Zone and Tachocline: Pumping and Organization of Toroidal
  Fields}. \apjl 648(2):L157--L160, \doi{10.1086/507869},
  \eprint{astro-ph/0609153}

\bibitem[{{Brummell} et~al(1998){Brummell}, {Hurlburt}, and
  {Toomre}}]{brummell+98}
{Brummell} NH, {Hurlburt} NE, {Toomre} J (1998) {Turbulent Compressible
  Convection with Rotation. II. Mean Flows and Differential Rotation}. \apj
  493(2):955--969, \doi{10.1086/305137}

\bibitem[{{Brummell} et~al(2002){Brummell}, {Clune}, and
  {Toomre}}]{brummell+02}
{Brummell} NH, {Clune} TL, {Toomre} J (2002) {Penetration and Overshooting in
  Turbulent Compressible Convection}. \apj 570(2):825--854,
  \doi{10.1086/339626}

\bibitem[{{Bushby} and {Tobias}(2007)}]{Bushby2007}
{Bushby} PJ, {Tobias} SM (2007) On predicting the solar cycle using mean-field
  models. \apj 661:1289--1296, \doi{10.1086/516628}, \eprint{0704.2345}

\bibitem[{{Bushby} et~al(2018){Bushby}, {K{\"a}pyl{\"a}}, {Masada},
  {Brandenburg}, {Favier}, {Guervilly}, and {K{\"a}pyl{\"a}}}]{bushby+18}
{Bushby} PJ, {K{\"a}pyl{\"a}} PJ, {Masada} Y, {Brandenburg} A, {Favier} B,
  {Guervilly} C, {K{\"a}pyl{\"a}} MJ (2018) {Large-scale dynamos in rapidly
  rotating plane layer convection}. \aap 612:A97,
  \doi{10.1051/0004-6361/201732066}, \eprint{1710.03174}

\bibitem[{{Cameron} and {Sch{\"u}ssler}(2015)}]{CS15}
{Cameron} R, {Sch{\"u}ssler} M (2015) {The crucial role of surface magnetic
  fields for the solar dynamo}. Science 347(6228):1333--1335,
  \doi{10.1126/science.1261470}, \eprint{1503.08469}

\bibitem[{{Candelaresi} and {Brandenburg}(2013)}]{CB13}
{Candelaresi} S, {Brandenburg} A (2013) {Kinetic helicity needed to drive
  large-scale dynamos}. \pre 87:043104, \doi{10.1103/PhysRevE.87.043104},
  \eprint{1208.4529}

\bibitem[{{Cattaneo} and {Hughes}(2006)}]{cattaneo+06}
{Cattaneo} F, {Hughes} DW (2006) {Dynamo action in a rotating convective
  layer}. J Fluid Mech 553:401--418, \doi{10.1017/S0022112006009165}

\bibitem[{{Chabrier} and {K{\"u}ker}(2006)}]{Chabrier2006AA}
{Chabrier} G, {K{\"u}ker} M (2006) {Large-scale {\ensuremath{\alpha}}\^2-dynamo
  in low-mass stars and brown dwarfs}. \aap 446(3):1027--1037,
  \doi{10.1051/0004-6361:20042475}, \eprint{astro-ph/0510075}

\bibitem[{{Charbonneau}(2014)}]{Charbonneau14}
{Charbonneau} P (2014) {Solar Dynamo Theory}. \araa 52:251--290,
  \doi{10.1146/annurev-astro-081913-040012}

\bibitem[{{Charbonneau}(2020)}]{Charbonneau10}
{Charbonneau} P (2020) {Dynamo Models of the Solar Cycle}. Living Reviews in
  Solar Physics 17(1):3, \doi{10.1007/s41116-020-00025-6}

\bibitem[{{Chatterjee} et~al(2010){Chatterjee}, {Brandenburg}, and
  {Guerrero}}]{Chatterjee2010G}
{Chatterjee} P, {Brandenburg} A, {Guerrero} G (2010) {Can catastrophic
  quenching be alleviated by separating shear and {\ensuremath{\alpha}}
  effect?} Geophysical and Astrophysical Fluid Dynamics 104(5):591--599,
  \doi{10.1080/03091929.2010.504185}, \eprint{1005.5708}

\bibitem[{{Chatterjee} et~al(2011){Chatterjee}, {Guerrero}, and
  {Brandenburg}}]{Chatterjee2011}
{Chatterjee} P, {Guerrero} G, {Brandenburg} A (2011) Magnetic helicity fluxes
  in interface and flux transport dynamos. \aap 525:A5,
  \doi{10.1051/0004-6361/201015073}, \eprint{1005.5335}

\bibitem[{{Choudhuri} et~al(2007){Choudhuri}, {Chatterjee}, and
  {Jiang}}]{Choudhuri2007}
{Choudhuri} AR, {Chatterjee} P, {Jiang} J (2007) Predicting solar cycle 24 with
  a solar dynamo model. Phys Rev Lett 98(13):131103,
  \doi{10.1103/PhysRevLett.98.131103}, \eprint{arXiv:astro-ph/0701527}

\bibitem[{{Covas} et~al(2000){Covas}, {Tavakol}, {Moss}, and
  {Tworkowski}}]{Covas2000}
{Covas} E, {Tavakol} R, {Moss} D, {Tworkowski} A (2000) Torsional oscillations
  in the solar convection zone. \aap 360:L21--L24, \eprint{astro-ph/0010323}

\bibitem[{{Cowling}(1933)}]{Cowling1933}
{Cowling} TG (1933) {The magnetic field of sunspots}. \mnras 94:39--48,
  \doi{10.1093/mnras/94.1.39}

\bibitem[{{Davidson}(2000)}]{Davidson00}
{Davidson} PA (2000) {Was Loitsyansky correct? A review of the arguments}. J
  Turbulence 1(1):6, \doi{10.1088/1468-5248/1/1/006}

\bibitem[{{Deardorff}(1972)}]{Deardorff72}
{Deardorff} JW (1972) {Theoretical expression for the countergradient vertical
  heat flux}. J Geophys Res 77(30):5900--5904, \doi{10.1029/JC077i030p05900}

\bibitem[{{Del Sordo} et~al(2013){Del Sordo}, {Guerrero}, and
  {Brandenburg}}]{DSordo2013MN}
{Del Sordo} F, {Guerrero} G, {Brandenburg} A (2013) {Turbulent dynamos with
  advective magnetic helicity flux}. \mnras 429(2):1686--1694,
  \doi{10.1093/mnras/sts398}, \eprint{1205.3502}

\bibitem[{{Donati} and {Landstreet}(2009)}]{Donati2009}
{Donati} JF, {Landstreet} JD (2009) Magnetic fields of nondegenerate stars.
  \araa 47:333--370, \doi{10.1146/annurev-astro-082708-101833},
  \eprint{0904.1938}

\bibitem[{{Egorov} et~al(2004){Egorov}, {R{\"u}diger}, and
  {Ziegler}}]{egorov+04}
{Egorov} P, {R{\"u}diger} G, {Ziegler} U (2004) {Vorticity and helicity of the
  solar supergranulation flow-field}. \aap 425:725--728,
  \doi{10.1051/0004-6361:20040531}

\bibitem[{{Elstner} and {R{\"u}diger}(2007)}]{Elstner2007AN}
{Elstner} D, {R{\"u}diger} G (2007) {How can
  {\ensuremath{\alpha}}$^{2}$-dynamos generate axisymmetric magnetic fields?}
  Astron Nachr 328(10):1130--1132, \doi{10.1002/asna.200710864}

\bibitem[{{Elstner} et~al(2020){Elstner}, {Fournier}, and {Arlt}}]{Elstner20}
{Elstner} D, {Fournier} Y, {Arlt} R (2020) {Various scenarios for the
  equatorward migration of sunspots}. In: {Kosovichev} A, {Strassmeier} S,
  {Jardine} M (eds) Solar and Stellar Magnetic Fields: Origins and
  Manifestations, vol 354, pp 134--137, \doi{10.1017/S1743921319009888},
  \eprint{2003.08131}

\bibitem[{{Fan} and {Fang}(2014)}]{fan+14}
{Fan} Y, {Fang} F (2014) {A Simulation of Convective Dynamo in the Solar
  Convective Envelope: Maintenance of the Solar-like Differential Rotation and
  Emerging Flux}. \apj 789(1):35, \doi{10.1088/0004-637X/789/1/35},
  \eprint{1405.3926}

\bibitem[{{Favier} and {Bushby}(2013)}]{favier+13}
{Favier} B, {Bushby} PJ (2013) {On the problem of large-scale magnetic field
  generation in rotating compressible convection}. J Fluid Mech 723:529--555,
  \doi{10.1017/jfm.2013.132}, \eprint{1302.7243}

\bibitem[{{Ferriere}(1993)}]{Ferriere+93}
{Ferriere} K (1993) {The Full Alpha-Tensor Due to Supernova Explosions and
  Superbubbles in the Galactic Disk}. \apj 404:162, \doi{10.1086/172266}

\bibitem[{{Field} and {Blackman}(2002)}]{FB02}
{Field} GB, {Blackman} EG (2002) {Dynamical Quenching of the
  {\ensuremath{\alpha}}$^{2}$ Dynamo}. \apj 572(1):685--692,
  \doi{10.1086/340233}, \eprint{astro-ph/0111470}

\bibitem[{{Getling} et~al(2021){Getling}, {Kosovichev}, and
  {Zhao}}]{Getling2021}
{Getling} AV, {Kosovichev} AG, {Zhao} J (2021) {Evolution of Subsurface Zonal
  and Meridional Flows in Solar Cycle 24 from Helioseismological Data}. \apjl
  908(2):L50, \doi{10.3847/2041-8213/abe45a}, \eprint{2012.15555}

\bibitem[{{Ghizaru} et~al(2010){Ghizaru}, {Charbonneau}, and
  {Smolarkiewicz}}]{ghizaru+10}
{Ghizaru} M, {Charbonneau} P, {Smolarkiewicz} PK (2010) {Magnetic Cycles in
  Global Large-eddy Simulations of Solar Convection}. \apjl 715(2):L133--L137,
  \doi{10.1088/2041-8205/715/2/L133}

\bibitem[{{Gopalakrishnan} and {Subramanian}(2023)}]{Gopalakr2023}
{Gopalakrishnan} K, {Subramanian} K (2023) {Magnetic Helicity Fluxes from
  Triple Correlators}. \apj 943(1):66, \doi{10.3847/1538-4357/aca808},
  \eprint{2209.14810}

\bibitem[{{Gressel}(2010)}]{Gressel10}
{Gressel} O (2010) {A mean-field approach to the propagation of field patterns
  in stratified magnetorotational turbulence}. \mnras 405(1):41--48,
  \doi{10.1111/j.1365-2966.2010.16440.x}, \eprint{1001.5250}

\bibitem[{{Gruzinov} and {Diamond}(1994)}]{GD94}
{Gruzinov} AV, {Diamond} PH (1994) {Self-consistent theory of mean-field
  electrodynamics}. \prl 72(11):1651--1653, \doi{10.1103/PhysRevLett.72.1651}

\bibitem[{{Guerrero} et~al(2016){Guerrero}, {Smolarkiewicz}, {de Gouveia Dal
  Pino}, {Kosovichev}, and {Mansour}}]{Guerrero2016}
{Guerrero} G, {Smolarkiewicz} PK, {de Gouveia Dal Pino} EM, {Kosovichev} AG,
  {Mansour} NN (2016) Understanding solar torsional oscillations from global
  dynamo models. \apjl 828:L3, \doi{10.3847/2041-8205/828/1/L3},
  \eprint{1608.02278}

\bibitem[{{Hanasoge} et~al(2016){Hanasoge}, {Gizon}, and
  {Sreenivasan}}]{hanasoge+16}
{Hanasoge} S, {Gizon} L, {Sreenivasan} KR (2016) {Seismic Sounding of
  Convection in the Sun}. Annual Review of Fluid Mechanics 48(1):191--217,
  \doi{10.1146/annurev-fluid-122414-034534}, \eprint{1503.07961}

\bibitem[{{Hathaway}(2012)}]{Hathaway2012}
{Hathaway} DH (2012) Supergranules as probes of the sun's meridional
  circulation. \apj 760:84, \doi{10.1088/0004-637X/760/1/84},
  \eprint{1210.3343}

\bibitem[{{Hathaway}(2015)}]{Hathaway2015}
{Hathaway} DH (2015) {The Solar Cycle}. Living Reviews in Solar Physics
  12(1):4, \doi{10.1007/lrsp-2015-4}, \eprint{1502.07020}

\bibitem[{{Hatori}(1984)}]{Hat84}
{Hatori} T (1984) {Kolmogorov-Style Argument for the Decaying Homogeneous MHD
  Turbulence}. J Phys Soc Jpn 53(8):2539, \doi{10.1143/JPSJ.53.2539}

\bibitem[{{Hazra} and {Choudhuri}(2017)}]{Hazra2017}
{Hazra} G, {Choudhuri} AR (2017) A theoretical model of the variation of the
  meridional circulation with the solar cycle. \mnras 472(3):2728--2741,
  \doi{10.1093/mnras/stx2152}, \eprint{1708.05204}

\bibitem[{{Hazra} et~al(2019){Hazra}, {Jiang}, {Karak}, and
  {Kitchatinov}}]{Hazra2019}
{Hazra} G, {Jiang} J, {Karak} BB, {Kitchatinov} L (2019) {Exploring the Cycle
  Period and Parity of Stellar Magnetic Activity with Dynamo Modeling}. \apj
  884(1):35, \doi{10.3847/1538-4357/ab4128}, \eprint{1909.01286}

\bibitem[{{Hazra} et~al(2023){Hazra}, {Nandy}, {Kitchatinov}, and
  {Choudhuri}}]{Hazra2023}
{Hazra} G, {Nandy} D, {Kitchatinov} L, {Choudhuri} AR (2023) {Mean Field Models
  of Flux Transport Dynamo and Meridional Circulation in the Sun and Stars}.
  \ssr 219(5):39, \doi{10.1007/s11214-023-00982-y}, \eprint{2302.09390}

\bibitem[{{Hotta} and {Kusano}(2021)}]{hotta+21}
{Hotta} H, {Kusano} K (2021) {Solar differential rotation reproduced with
  high-resolution simulation}. Nature Astronomy 5:1100--1102,
  \doi{10.1038/s41550-021-01459-0}, \eprint{2109.06280}

\bibitem[{{Hotta} et~al(2016){Hotta}, {Rempel}, and {Yokoyama}}]{hotta+16}
{Hotta} H, {Rempel} M, {Yokoyama} T (2016) {Large-scale magnetic fields at high
  Reynolds numbers in magnetohydrodynamic simulations}. Science
  351(6280):1427--1430, \doi{10.1126/science.aad1893}

\bibitem[{{Howard} and {Labonte}(1980)}]{Howard1980}
{Howard} R, {Labonte} BJ (1980) {The sun is observed to be a torsional
  oscillator with a period of 11 years}. \apjl 239:L33--L36,
  \doi{10.1086/183286}

\bibitem[{{Howe} et~al(2011){Howe}, {Hill}, {Komm}, {Christensen-Dalsgaard},
  {Larson}, {Schou}, {Thompson}, and {Ulrich}}]{Howe2011}
{Howe} R, {Hill} F, {Komm} R, {Christensen-Dalsgaard} J, {Larson} TP, {Schou}
  J, {Thompson} MJ, {Ulrich} R (2011) The torsional oscillation and the new
  solar cycle. J Phys Conf Ser 271(1):012074,
  \doi{10.1088/1742-6596/271/1/012074}

\bibitem[{{Hubbard} and {Brandenburg}(2009)}]{HB09}
{Hubbard} A, {Brandenburg} A (2009) {Memory Effects in Turbulent Transport}.
  \apj 706(1):712--726, \doi{10.1088/0004-637X/706/1/712}, \eprint{0811.2561}

\bibitem[{{Hubbard} and {Brandenburg}(2012)}]{Hubbard2012}
{Hubbard} A, {Brandenburg} A (2012) Catastrophic quenching in
  {$\alpha$}{$\Omega$} dynamos revisited. \apj 748:51,
  \doi{10.1088/0004-637X/748/1/51}, \eprint{1107.0238}

\bibitem[{{Hubbard} et~al(2009){Hubbard}, {Del Sordo}, {K{\"a}pyl{\"a}}, and
  {Brandenburg}}]{Hubbard+09}
{Hubbard} A, {Del Sordo} F, {K{\"a}pyl{\"a}} PJ, {Brandenburg} A (2009) {The
  {\ensuremath{\alpha}} effect with imposed and dynamo-generated magnetic
  fields}. \mnras 398(4):1891--1899, \doi{10.1111/j.1365-2966.2009.15108.x},
  \eprint{0904.2773}

\bibitem[{{Jabbari} et~al(2014){Jabbari}, {Brandenburg}, {Losada}, {Kleeorin},
  and {Rogachevskii}}]{jabbari+14}
{Jabbari} S, {Brandenburg} A, {Losada} IR, {Kleeorin} N, {Rogachevskii} I
  (2014) {Magnetic flux concentrations from dynamo-generated fields}. \aap
  568:A112, \doi{10.1051/0004-6361/201423499}, \eprint{1401.6107}

\bibitem[{{Jabbari} et~al(2016){Jabbari}, {Brandenburg}, {Mitra}, {Kleeorin},
  and {Rogachevskii}}]{jabbari+16}
{Jabbari} S, {Brandenburg} A, {Mitra} D, {Kleeorin} N, {Rogachevskii} I (2016)
  {Turbulent reconnection of magnetic bipoles in stratified turbulence}. \mnras
  459(4):4046--4056, \doi{10.1093/mnras/stw888}, \eprint{1601.08167}

\bibitem[{{Ji}(1999)}]{Ji99}
{Ji} H (1999) {Turbulent Dynamos and Magnetic Helicity}. \prl
  83(16):3198--3201, \doi{10.1103/PhysRevLett.83.3198},
  \eprint{astro-ph/0102321}

\bibitem[{{Ji} and {Prager}(2002)}]{Ji+Prager02}
{Ji} H, {Prager} SC (2002) {The {\ensuremath{\alpha}} dynamo effects in
  laboratory plasmas}. Magnetohydrodynamics 38:191--210,
  \doi{10.22364/mhd.38.1-2.15}, \eprint{astro-ph/0110352}

\bibitem[{{Ji} et~al(1995){Ji}, {Prager}, and {Sarff}}]{Ji+95}
{Ji} H, {Prager} SC, {Sarff} JS (1995) {Conservation of Magnetic Helicity
  during Plasma Relaxation}. \prl 74(15):2945--2948,
  \doi{10.1103/PhysRevLett.74.2945}

\bibitem[{{Kaneda} et~al(2003){Kaneda}, {Ishihara}, {Yokokawa}, {Itakura}, and
  {Uno}}]{kaneda+03}
{Kaneda} Y, {Ishihara} T, {Yokokawa} M, {Itakura} K, {Uno} A (2003) {Energy
  dissipation rate and energy spectrum in high resolution direct numerical
  simulations of turbulence in a periodic box}. Physics of Fluids
  15(2):L21--L24, \doi{10.1063/1.1539855}

\bibitem[{{K{\"a}pyl{\"a}} et~al(2016{\natexlab{a}}){K{\"a}pyl{\"a}},
  {K{\"a}pyl{\"a}}, {Olspert}, {Brandenburg}, {Warnecke}, {Karak}, and
  {Pelt}}]{Kapyla2016}
{K{\"a}pyl{\"a}} MJ, {K{\"a}pyl{\"a}} PJ, {Olspert} N, {Brandenburg} A,
  {Warnecke} J, {Karak} BB, {Pelt} J (2016{\natexlab{a}}) Multiple dynamo modes
  as a mechanism for long-term solar activity variations. \aap 589:A56,
  \doi{10.1051/0004-6361/201527002}, \eprint{1507.05417}

\bibitem[{{K{\"a}pyl{\"a}} et~al(2022){K{\"a}pyl{\"a}}, {Rheinhardt}, and
  {Brandenburg}}]{Maarit2022}
{K{\"a}pyl{\"a}} MJ, {Rheinhardt} M, {Brandenburg} A (2022) {Compressible
  Test-field Method and Its Application to Shear Dynamos}. \apj 932(1):8,
  \doi{10.3847/1538-4357/ac5b78}, \eprint{2106.01107}

\bibitem[{{K{\"a}pyl{\"a}} et~al(2004){K{\"a}pyl{\"a}}, {Korpi}, and
  {Tuominen}}]{kapyla+04}
{K{\"a}pyl{\"a}} PJ, {Korpi} MJ, {Tuominen} I (2004) {Local models of stellar
  convection:. Reynolds stresses and turbulent heat transport}. \aap
  422:793--816, \doi{10.1051/0004-6361:20035874}, \eprint{astro-ph/0312376}

\bibitem[{{K{\"a}pyl{\"a}} et~al(2006{\natexlab{a}}){K{\"a}pyl{\"a}}, {Korpi},
  {Ossendrijver}, and {Stix}}]{kapyla+06a}
{K{\"a}pyl{\"a}} PJ, {Korpi} MJ, {Ossendrijver} M, {Stix} M
  (2006{\natexlab{a}}) {Magnetoconvection and dynamo coefficients. III.
  {\ensuremath{\alpha}}-effect and magnetic pumping in the rapid rotation
  regime}. \aap 455(2):401--412, \doi{10.1051/0004-6361:20064972},
  \eprint{astro-ph/0602111}

\bibitem[{{K{\"a}pyl{\"a}} et~al(2006{\natexlab{b}}){K{\"a}pyl{\"a}}, {Korpi},
  and {Tuominen}}]{kapyla+06b}
{K{\"a}pyl{\"a}} PJ, {Korpi} MJ, {Tuominen} I (2006{\natexlab{b}}) {Solar
  dynamo models with {\ensuremath{\alpha}}-effect and turbulent pumping from
  local 3D convection calculations}. Astronomische Nachrichten 327(9):884,
  \doi{10.1002/asna.200610636}, \eprint{astro-ph/0606089}

\bibitem[{{K{\"a}pyl{\"a}} et~al(2009{\natexlab{a}}){K{\"a}pyl{\"a}}, {Korpi},
  and {Brandenburg}}]{Kaepylae2009AA}
{K{\"a}pyl{\"a}} PJ, {Korpi} MJ, {Brandenburg} A (2009{\natexlab{a}}) {Alpha
  effect and turbulent diffusion from convection}. \aap 500(2):633--646,
  \doi{10.1051/0004-6361/200811498}, \eprint{0812.1792}

\bibitem[{{K{\"a}pyl{\"a}} et~al(2009{\natexlab{b}}){K{\"a}pyl{\"a}}, {Korpi},
  and {Brandenburg}}]{kapyla+09}
{K{\"a}pyl{\"a}} PJ, {Korpi} MJ, {Brandenburg} A (2009{\natexlab{b}})
  {Large-scale Dynamos in Rigidly Rotating Turbulent Convection}. \apj
  697(2):1153--1163, \doi{10.1088/0004-637X/697/2/1153}, \eprint{0812.3958}

\bibitem[{{K{\"a}pyl{\"a}} et~al(2012){K{\"a}pyl{\"a}}, {Mantere}, and
  {Brandenburg}}]{kapyla+12}
{K{\"a}pyl{\"a}} PJ, {Mantere} MJ, {Brandenburg} A (2012) {Cyclic Magnetic
  Activity due to Turbulent Convection in Spherical Wedge Geometry}. \apjl
  755(1):L22, \doi{10.1088/2041-8205/755/1/L22}, \eprint{1205.4719}

\bibitem[{{K{\"a}pyl{\"a}} et~al(2013){K{\"a}pyl{\"a}}, {Mantere}, and
  {Brandenburg}}]{kapyla+13}
{K{\"a}pyl{\"a}} PJ, {Mantere} MJ, {Brandenburg} A (2013) {Oscillatory
  large-scale dynamos from Cartesian convection simulations}. Geophysical and
  Astrophysical Fluid Dynamics 107(1-2):244--257,
  \doi{10.1080/03091929.2012.715158}, \eprint{1111.6894}

\bibitem[{{K{\"a}pyl{\"a}} et~al(2016{\natexlab{b}}){K{\"a}pyl{\"a}},
  {Brandenburg}, {Kleeorin}, {K{\"a}pyl{\"a}}, and {Rogachevskii}}]{KBKKR16}
{K{\"a}pyl{\"a}} PJ, {Brandenburg} A, {Kleeorin} N, {K{\"a}pyl{\"a}} MJ,
  {Rogachevskii} I (2016{\natexlab{b}}) {Magnetic flux concentrations from
  turbulent stratified convection}. \aap 588:A150,
  \doi{10.1051/0004-6361/201527731}, \eprint{1511.03718}

\bibitem[{{K{\"a}pyl{\"a}} et~al(2023){K{\"a}pyl{\"a}}, {Browning}, {Brun},
  {Guerrero}, and {Warnecke}}]{Kapyla+23}
{K{\"a}pyl{\"a}} PJ, {Browning} MK, {Brun} AS, {Guerrero} G, {Warnecke} J
  (2023) {Simulations of solar and stellar dynamos and their theoretical
  interpretation}. arXiv e-prints arXiv:2305.16790,
  \doi{10.48550/arXiv.2305.16790}, \eprint{2305.16790}

\bibitem[{{Karak}(2023)}]{Karak23}
{Karak} BB (2023) {Models for the long-term variations of solar activity}.
  Living Reviews in Solar Physics 20(1):3, \doi{10.1007/s41116-023-00037-y},
  \eprint{2305.17188}

\bibitem[{{Karak} et~al(2014){Karak}, {Rheinhardt}, {Brandenburg},
  {K{\"a}pyl{\"a}}, and {K{\"a}pyl{\"a}}}]{Karak+14}
{Karak} BB, {Rheinhardt} M, {Brandenburg} A, {K{\"a}pyl{\"a}} PJ,
  {K{\"a}pyl{\"a}} MJ (2014) {Quenching and Anisotropy of Hydromagnetic
  Turbulent Transport}. \apj 795(1):16, \doi{10.1088/0004-637X/795/1/16},
  \eprint{1406.4521}

\bibitem[{{Katsova} et~al(2021){Katsova}, {Obridko}, {Sokoloff}, and
  {Livshits}}]{Katsova2021G}
{Katsova} MM, {Obridko} VN, {Sokoloff} DD, {Livshits} IM (2021) {Estimating the
  Energy of Solar and Stellar Superflares}. Geomagnetism and Aeronomy
  61(7):1063--1068, \doi{10.1134/S0016793221070094}

\bibitem[{{Kemel} et~al(2011){Kemel}, {Brandenburg}, and {Ji}}]{Kemel+11}
{Kemel} K, {Brandenburg} A, {Ji} H (2011) {Model of driven and decaying
  magnetic turbulence in a cylinder}. \pre 84(5):056407,
  \doi{10.1103/PhysRevE.84.056407}, \eprint{1106.1129}

\bibitem[{{Kharzeev}(2014)}]{Khar14}
{Kharzeev} DE (2014) {The Chiral Magnetic Effect and anomaly-induced
  transport}. Progress in Particle and Nuclear Physics 75:133--151,
  \doi{10.1016/j.ppnp.2014.01.002}, \eprint{1312.3348}

\bibitem[{{Kitchatinov}(2013)}]{Kitchatinov2013}
{Kitchatinov} LL (2013) Theory of differential rotation and meridional
  circulation. In: {Kosovichev} AG, {de Gouveia Dal Pino} E, {Yan} Y (eds) IAU
  Symposium, IAU Symposium, vol 294, pp 399--410,
  \doi{10.1017/S1743921313002834}, \eprint{1210.7041}

\bibitem[{{Kitchatinov} and {Olemskoy}(2011)}]{KitOle2011AN}
{Kitchatinov} LL, {Olemskoy} SV (2011) {Alleviation of catastrophic quenching
  in solar dynamo model with nonlocal alpha-effect}. Astronomische Nachrichten
  332(5):496, \doi{10.1002/asna.201011549}, \eprint{1101.3115}

\bibitem[{{Kitchatinov} et~al(1994){Kitchatinov}, {Pipin}, and
  {R\"udiger}}]{Kitchatinov1994}
{Kitchatinov} LL, {Pipin} VV, {R\"udiger} G (1994) Turbulent viscosity,
  magnetic diffusivity, and heat conductivity under the influence of rotation
  and magnetic field. Astron Nachr 315:157--170

\bibitem[{Kitchatinov et~al(1994)Kitchatinov, R\"udiger, and
  K\"uker}]{Kitchatinov1994a}
Kitchatinov LL, R\"udiger G, K\"uker M (1994) Lambda-quenching as the
  nonlinearity in stellar-turbulence dynamos. \aap 292:125--132

\bibitem[{Kleeorin and Rogachevskii(1999)}]{Kleeorin1999}
Kleeorin N, Rogachevskii I (1999) Magnetic helicity tensor for an anisotropic
  turbulence. Phys RevE 59:6724--6729

\bibitem[{{Kleeorin} and {Rogachevskii}(2022)}]{Kleeorin2022}
{Kleeorin} N, {Rogachevskii} I (2022) {Turbulent magnetic helicity fluxes in
  solar convective zone}. \mnras 515(4):5437--5448,
  \doi{10.1093/mnras/stac2141}, \eprint{2206.14152}

\bibitem[{{Kleeorin} et~al(1996){Kleeorin}, {Mond}, and
  {Rogachevskii}}]{Kleeorin1996}
{Kleeorin} N, {Mond} M, {Rogachevskii} I (1996) Magnetohydrodynamic turbulence
  in the solar convective zone as a source of oscillations and sunspots
  formation. \aap 307:293--309

\bibitem[{{Kleeorin} et~al(2000){Kleeorin}, {Moss}, {Rogachevskii}, and
  {Sokoloff}}]{Kleeorin2000}
{Kleeorin} N, {Moss} D, {Rogachevskii} I, {Sokoloff} D (2000) Helicity balance
  and steady-state strength of the dynamo generated galactic magnetic field.
  \aap 361:L5--L8, \eprint{arXiv:astro-ph/0205266}

\bibitem[{{Kleeorin} et~al(2020){Kleeorin}, {Safiullin}, {Kuzanyan},
  {Rogachevskii}, {Tlatov}, and {Porshnev}}]{Kleeorin2020m}
{Kleeorin} N, {Safiullin} N, {Kuzanyan} K, {Rogachevskii} I, {Tlatov} A,
  {Porshnev} S (2020) {The mean tilt of sunspot bipolar regions: theory,
  simulations and comparison with observations}. \mnras 495(1):238--248,
  \doi{10.1093/mnras/staa1047}, \eprint{2001.01932}

\bibitem[{Kleeorin and Ruzmaikin(1982)}]{Kleeorin1982}
Kleeorin NI, Ruzmaikin AA (1982) Dynamics of the average turbulent helicity in
  a magnetic field. Magnetohydrodynamics 18:116--122

\bibitem[{{Kleeorin} and {Ruzmaikin}(1991)}]{Kleeorin1991}
{Kleeorin} NI, {Ruzmaikin} AA (1991) {Large-scale flows excited by magnetic
  fields in the solar convective zone}. \solphys 131(2):211--230,
  \doi{10.1007/BF00151634}

\bibitem[{{Kleeorin} et~al(1989){Kleeorin}, {Rogachevskii}, and
  {Ruzmaikin}}]{KRR89}
{Kleeorin} NI, {Rogachevskii} IV, {Ruzmaikin} AA (1989) {Negative Magnetic
  Pressure as a Trigger of Largescale Magnetic Instability in the Solar
  Convective Zone}. Sov Astron Lett 15:274

\bibitem[{{Kochukhov}(2021)}]{Kochukhov2021AAr}
{Kochukhov} O (2021) {Magnetic fields of M dwarfs}. \aapr 29(1):1,
  \doi{10.1007/s00159-020-00130-3}, \eprint{2011.01781}

\bibitem[{{K{\"o}hler}(1973)}]{Koehler1973}
{K{\"o}hler} H (1973) {The Solar Dynamo and Estimate of the Magnetic
  Diffusivity and the {\ensuremath{\alpha}}-effect}. \aap 25:467

\bibitem[{{Kosovichev} and {Pipin}(2019)}]{Kosovichev2019}
{Kosovichev} AG, {Pipin} VV (2019) {Dynamo Wave Patterns inside of the Sun
  Revealed by Torsional Oscillations}. \apjl 871(2):L20,
  \doi{10.3847/2041-8213/aafe82}

\bibitem[{{Kosovichev} et~al(1997){Kosovichev}, {Schou}, {Scherrer}, {Bogart},
  {Bush}, {Hoeksema}, {Aloise}, {Bacon}, {Burnette}, {de Forest}, {Giles},
  {Leibrand}, {Nigam}, {Rubin}, {Scott}, {Williams}, {Basu},
  {Christensen-Dalsgaard}, {Dappen}, {Rhodes}, {Duvall}, {Howe}, {Thompson},
  {Gough}, {Sekii}, {Toomre}, {Tarbell}, {Title}, {Mathur}, {Morrison}, {Saba},
  {Wolfson}, {Zayer}, and {Milford}}]{Kosovichev1997}
{Kosovichev} AG, {Schou} J, {Scherrer} PH, {Bogart} RS, {Bush} RI, {Hoeksema}
  JT, {Aloise} J, {Bacon} L, {Burnette} A, {de Forest} C, {Giles} PM,
  {Leibrand} K, {Nigam} R, {Rubin} M, {Scott} K, {Williams} SD, {Basu} S,
  {Christensen-Dalsgaard} J, {Dappen} W, {Rhodes} EJ Jr, {Duvall} TL Jr, {Howe}
  R, {Thompson} MJ, {Gough} DO, {Sekii} T, {Toomre} J, {Tarbell} TD, {Title}
  AM, {Mathur} D, {Morrison} M, {Saba} JLR, {Wolfson} CJ, {Zayer} I, {Milford}
  PN (1997) Structure and rotation of the solar interior: Initial results from
  the mdi medium-l program. \solphys 170:43--61, \doi{10.1023/A:1004949311268}

\bibitem[{{Krause} and {R{\"a}dler}(1980)}]{KR80}
{Krause} F, {R{\"a}dler} KH (1980) {Mean-Field Magnetohydrodynamics and Dynamo
  Theory}. Pergamon Press (also Akademie-Verlag: Berlin), Oxford

\bibitem[{{K\"uker} et~al(1996){K\"uker}, {R\"udiger}, and
  {Pipin}}]{Kueker1996}
{K\"uker} M, {R\"udiger} G, {Pipin} VV (1996) Solar torsional oscillations due
  to the magnetic quenching of the reynolds stress. \aap 312:615--623

\bibitem[{{Kulsrud} and {Zweibel}(2008)}]{KZ08}
{Kulsrud} RM, {Zweibel} EG (2008) {On the origin of cosmic magnetic fields}.
  Reports on Progress in Physics 71(4):046901,
  \doi{10.1088/0034-4885/71/4/046901}, \eprint{0707.2783}

\bibitem[{{Lehtinen} et~al(2016){Lehtinen}, {Jetsu}, {Hackman}, {Kajatkari},
  and {Henry}}]{Lehtinen2016}
{Lehtinen} J, {Jetsu} L, {Hackman} T, {Kajatkari} P, {Henry} GW (2016)
  {Activity trends in young solar-type stars}. \aap 588:A38,
  \doi{10.1051/0004-6361/201527420}, \eprint{1509.06606}

\bibitem[{{Lehtinen} et~al(2020){Lehtinen}, {Spada}, {K{\"a}pyl{\"a}},
  {Olspert}, and {K{\"a}pyl{\"a}}}]{Lehtinen2020}
{Lehtinen} JJ, {Spada} F, {K{\"a}pyl{\"a}} MJ, {Olspert} N, {K{\"a}pyl{\"a}} PJ
  (2020) {Common dynamo scaling in slowly rotating young and evolved stars}.
  Nature Astronomy 4:658--662, \doi{10.1038/s41550-020-1039-x},
  \eprint{2003.08997}

\bibitem[{{Leighton}(1969)}]{Leighton1969}
{Leighton} RB (1969) {A Magneto-Kinematic Model of the Solar Cycle}. \apj
  156:1, \doi{10.1086/149943}

\bibitem[{{Losada} et~al(2012){Losada}, {Brandenburg}, {Kleeorin}, {Mitra}, and
  {Rogachevskii}}]{losada+12}
{Losada} IR, {Brandenburg} A, {Kleeorin} N, {Mitra} D, {Rogachevskii} I (2012)
  {Rotational effects on the negative magnetic pressure instability}. \aap
  548:A49, \doi{10.1051/0004-6361/201220078}, \eprint{1207.5392}

\bibitem[{{Malkus} and {Proctor}(1975)}]{Malkus1975}
{Malkus} WVR, {Proctor} MRE (1975) The macrodynamics of alpha-effect dynamos in
  rotating fluids. J Fluid Mech 67:417--443, \doi{10.1017/S0022112075000390}

\bibitem[{{Masada} and {Sano}(2014{\natexlab{a}})}]{masada14b}
{Masada} Y, {Sano} T (2014{\natexlab{a}}) {Long-term evolution of large-scale
  magnetic fields in rotating stratified convection}. \pasj 66:S2,
  \doi{10.1093/pasj/psu081}, \eprint{1403.6221}

\bibitem[{{Masada} and {Sano}(2014{\natexlab{b}})}]{masada14a}
{Masada} Y, {Sano} T (2014{\natexlab{b}}) {Mean-Field Modeling of an
  {\ensuremath{\alpha}}$^{2}$ Dynamo Coupled with Direct Numerical Simulations
  of Rigidly Rotating Convection}. \apjl 794(1):L6,
  \doi{10.1088/2041-8205/794/1/L6}, \eprint{1409.3256}

\bibitem[{{Masada} and {Sano}(2016)}]{masada+16}
{Masada} Y, {Sano} T (2016) {Spontaneous Formation of Surface Magnetic
  Structure from Large-scale Dynamo in Strongly Stratified Convection}. \apjl
  822(2):L22, \doi{10.3847/2041-8205/822/2/L22}, \eprint{1604.05374}

\bibitem[{{Masada} and {Sano}(2022)}]{masada+22}
{Masada} Y, {Sano} T (2022) {Rotational Dependence of Large-scale Dynamo in
  Strongly-stratified Convection: What Causes It?} arXiv e-prints
  arXiv:2206.06566, \eprint{2206.06566}

\bibitem[{{Masada} et~al(2013){Masada}, {Yamada}, and {Kageyama}}]{masada+13}
{Masada} Y, {Yamada} K, {Kageyama} A (2013) {Effects of Penetrative Convection
  on Solar Dynamo}. \apj 778(1):11, \doi{10.1088/0004-637X/778/1/11},
  \eprint{1304.1252}

\bibitem[{{Masada} et~al(2018){Masada}, {Kotake}, {Takiwaki}, and
  {Yamamoto}}]{Masada+18}
{Masada} Y, {Kotake} K, {Takiwaki} T, {Yamamoto} N (2018) {Chiral
  magnetohydrodynamic turbulence in core-collapse supernovae}. \prd
  98(8):083018, \doi{10.1103/PhysRevD.98.083018}, \eprint{1805.10419}

\bibitem[{{Matthaeus} et~al(2008){Matthaeus}, {Pouquet}, {Mininni}, {Dmitruk},
  and {Breech}}]{Matthaeus2008}
{Matthaeus} WH, {Pouquet} A, {Mininni} PD, {Dmitruk} P, {Breech} B (2008)
  {Rapid Alignment of Velocity and Magnetic Field in Magnetohydrodynamic
  Turbulence}. \prl 100(8):085003, \doi{10.1103/PhysRevLett.100.085003},
  \eprint{0708.0801}

\bibitem[{{Miesch} and {Toomre}(2009)}]{MT09}
{Miesch} MS, {Toomre} J (2009) {Turbulence, Magnetism, and Shear in Stellar
  Interiors}. Annual Review of Fluid Mechanics 41(1):317--345,
  \doi{10.1146/annurev.fluid.010908.165215}

\bibitem[{{Mitra} et~al(2010{\natexlab{a}}){Mitra}, {Candelaresi},
  {Chatterjee}, {Tavakol}, and {Brandenburg}}]{Mitra2010}
{Mitra} D, {Candelaresi} S, {Chatterjee} P, {Tavakol} R, {Brandenburg} A
  (2010{\natexlab{a}}) Equatorial magnetic helicity flux in simulations with
  different gauges. Astron Nachr 331:130, \doi{10.1002/asna.200911308},
  \eprint{0911.0969}

\bibitem[{{Mitra} et~al(2010{\natexlab{b}}){Mitra}, {Tavakol},
  {K{\"a}pyl{\"a}}, and {Brandenburg}}]{mitra+10}
{Mitra} D, {Tavakol} R, {K{\"a}pyl{\"a}} PJ, {Brandenburg} A
  (2010{\natexlab{b}}) {Oscillatory Migrating Magnetic Fields in Helical
  Turbulence in Spherical Domains}. \apjl 719(1):L1--L4,
  \doi{10.1088/2041-8205/719/1/L1}, \eprint{0901.2364}

\bibitem[{{Mitra} et~al(2014){Mitra}, {Brandenburg}, {Kleeorin}, and
  {Rogachevskii}}]{mitra+14}
{Mitra} D, {Brandenburg} A, {Kleeorin} N, {Rogachevskii} I (2014) {Intense
  bipolar structures from stratified helical dynamos}. \mnras 445(1):761--769,
  \doi{10.1093/mnras/stu1755}, \eprint{1404.3194}

\bibitem[{{Moffatt}(1978)}]{Mof78}
{Moffatt} HK (1978) {Magnetic Field Generation in Electrically Conducting
  Fluids}. Cambridge University Press, Cambridge

\bibitem[{{Moss} and {Brandenburg}(1992)}]{Moss1992}
{Moss} D, {Brandenburg} A (1992) The influence of boundary conditions on the
  excitation of disk dynamo modes. \aap 256:371--374

\bibitem[{{Noraz} et~al(2022){Noraz}, {Brun}, {Strugarek}, and
  {Depambour}}]{Noraz2022}
{Noraz} Q, {Brun} AS, {Strugarek} A, {Depambour} G (2022) {Impact of anti-solar
  differential rotation in mean-field solar-type dynamos. Exploring possible
  magnetic cycles in slowly rotating stars}. \aap 658:A144,
  \doi{10.1051/0004-6361/202141946}, \eprint{2111.12722}

\bibitem[{{Nordlund} et~al(1992){Nordlund}, {Brandenburg}, {Jennings},
  {Rieutord}, {Ruokolainen}, {Stein}, and {Tuominen}}]{nordlund+92}
{Nordlund} A, {Brandenburg} A, {Jennings} RL, {Rieutord} M, {Ruokolainen} J,
  {Stein} RF, {Tuominen} I (1992) {Dynamo Action in Stratified Convection with
  Overshoot}. \apj 392:647, \doi{10.1086/171465}

\bibitem[{{Noyes} et~al(1984){Noyes}, {Weiss}, and {Vaughan}}]{NWV84}
{Noyes} RW, {Weiss} NO, {Vaughan} AH (1984) {The relation between stellar
  rotation rate and activity cycle periods}. \apj 287:769--773,
  \doi{10.1086/162735}

\bibitem[{{Obridko} et~al(2021){Obridko}, {Pipin}, {Sokoloff}, and
  {Shibalova}}]{Obridko21}
{Obridko} VN, {Pipin} VV, {Sokoloff} D, {Shibalova} AS (2021) {Solar
  large-scale magnetic field and cycle patterns in solar dynamo}. \mnras
  504(4):4990--5000, \doi{10.1093/mnras/stab1062}, \eprint{2104.06808}

\bibitem[{{Olspert} et~al(2018){Olspert}, {Lehtinen}, {K{\"a}pyl{\"a}}, {Pelt},
  and {Grigorievskiy}}]{Olspert2018}
{Olspert} N, {Lehtinen} JJ, {K{\"a}pyl{\"a}} MJ, {Pelt} J, {Grigorievskiy} A
  (2018) {Estimating activity cycles with probabilistic methods. II. The Mount
  Wilson Ca H\&K data}. \aap 619:A6, \doi{10.1051/0004-6361/201732525},
  \eprint{1712.08240}

\bibitem[{{Ossendrijver}(2003)}]{ossendrijver03}
{Ossendrijver} M (2003) {The solar dynamo}. \aapr 11(4):287--367,
  \doi{10.1007/s00159-003-0019-3}

\bibitem[{{Ossendrijver} et~al(2001){Ossendrijver}, {Stix}, and
  {Brandenburg}}]{ossendrijver+01}
{Ossendrijver} M, {Stix} M, {Brandenburg} A (2001) {Magnetoconvection and
  dynamo coefficients:. Dependence of the alpha effect on rotation and magnetic
  field}. \aap 376:713--726, \doi{10.1051/0004-6361:20011041},
  \eprint{astro-ph/0108274}

\bibitem[{Parker(1955)}]{Parker1955}
Parker E (1955) Hydromagnetic dynamo models. Astrophys J 122:293--314

\bibitem[{{Parker}(1967)}]{Parker67}
{Parker} EN (1967) {The Dynamical State of the Interstellar Gas and Field. III.
  Turbulence and Enhanced Diffusion}. \apj 149:535, \doi{10.1086/149283}

\bibitem[{{Parker}(1979)}]{Par79}
{Parker} EN (1979) {Cosmical Magnetic Fields: Their Origin and Their Activity}.
  Clarendon Press, Oxford

\bibitem[{{Parker}(1993)}]{Par93}
{Parker} EN (1993) {A Solar Dynamo Surface Wave at the Interface between
  Convection and Nonuniform Rotation}. \apj 408:707, \doi{10.1086/172631}

\bibitem[{{Paxton} et~al(2015){Paxton}, {Marchant}, {Schwab}, {Bauer},
  {Bildsten}, {Cantiello}, {Dessart}, {Farmer}, {Hu}, {Langer}, {Townsend},
  {Townsley}, and {Timmes}}]{Paxton2015}
{Paxton} B, {Marchant} P, {Schwab} J, {Bauer} EB, {Bildsten} L, {Cantiello} M,
  {Dessart} L, {Farmer} R, {Hu} H, {Langer} N, {Townsend} RHD, {Townsley} DM,
  {Timmes} FX (2015) Modules for experiments in stellar astrophysics (mesa):
  Binaries, pulsations, and explosions. \apjs 220:15,
  \doi{10.1088/0067-0049/220/1/15}, \eprint{1506.03146}

\bibitem[{{Pipin}(2008)}]{Pipin2008a}
{Pipin} VV (2008) The mean electro-motive force and current helicity under the
  influence of rotation, magnetic field and shear. Geophysical and
  Astrophysical Fluid Dynamics 102:21--49, \eprint{arXiv:astro-ph/0606265}

\bibitem[{{Pipin}(2015)}]{Pipin2015}
{Pipin} VV (2015) Dependence of magnetic cycle parameters on period of rotation
  in non-linear solar-type dynamos. \mnras 451:1528--1539,
  \doi{10.1093/mnras/stv1026}, \eprint{1412.5284}

\bibitem[{{Pipin}(2018)}]{Pipin2018b}
{Pipin} VV (2018) Nonkinematic solar dynamo models with double-cell meridional
  circulation. J Atmosph Solar-Terr Phys 179:185--201,
  \doi{10.1016/j.jastp.2018.07.010}, \eprint{1803.09459}

\bibitem[{{Pipin}(2021)}]{Pipin21c}
{Pipin} VV (2021) {Solar dynamo cycle variations with a rotational period}.
  \mnras 502(2):2565--2581, \doi{10.1093/mnras/stab033}, \eprint{2008.05083}

\bibitem[{{Pipin}(2022)}]{Pipin2022}
{Pipin} VV (2022) {On the effect of surface bipolar magnetic regions on the
  convection zone dynamo}. \mnras 514(1):1522--1534,
  \doi{10.1093/mnras/stac1434}, \eprint{2112.09460}

\bibitem[{{Pipin}(2023)}]{Pipin23}
{Pipin} VV (2023) {Spatio-temporal non-localities in a solar-like mean-field
  dynamo}. \mnras 522(2):2919--2927, \doi{10.1093/mnras/stad1150},
  \eprint{2302.11176}

\bibitem[{{Pipin} and {Kosovichev}(2011)}]{Pipin2011a}
{Pipin} VV, {Kosovichev} AG (2011) The subsurface-shear-shaped solar
  {$\alpha$}{$\Omega$} dynamo. ApJL 727:L45--L48,
  \doi{10.1088/2041-8205/727/2/L45}, \eprint{1011.4276}

\bibitem[{{Pipin} and {Kosovichev}(2016)}]{Pipin2016b}
{Pipin} VV, {Kosovichev} AG (2016) Dependence of stellar magnetic activity
  cycles on rotational period in a nonlinear solar-type dynamo. \apj 823:133,
  \doi{10.3847/0004-637X/823/2/133}, \eprint{1602.07815}

\bibitem[{{Pipin} and {Kosovichev}(2018)}]{Pipin2018c}
{Pipin} VV, {Kosovichev} AG (2018) On the origin of the double-cell meridional
  circulation in the solar convection zone. \apj 854:67,
  \doi{10.3847/1538-4357/aaa759}, \eprint{1708.03073}

\bibitem[{{Pipin} and {Kosovichev}(2019)}]{Pipin2019c}
{Pipin} VV, {Kosovichev} AG (2019) {On the Origin of Solar Torsional
  Oscillations and Extended Solar Cycle}. \apj 887(2):215,
  \doi{10.3847/1538-4357/ab5952}

\bibitem[{{Pipin} and {Seehafer}(2009)}]{Pipin2009}
{Pipin} VV, {Seehafer} N (2009) Stellar dynamos with {$\Omega$} {$\times$} j
  effect. A\&A 493:819--828, \doi{10.1051/0004-6361:200810766},
  \eprint{0811.4225}

\bibitem[{{Pipin} and {Yokoi}(2018)}]{Yokoi2018}
{Pipin} VV, {Yokoi} N (2018) {Generation of a Large-scale Magnetic Field in a
  Convective Full-sphere Cross-helicity Dynamo}. \apj 859(1):18,
  \doi{10.3847/1538-4357/aabae6}, \eprint{1712.01527}

\bibitem[{{Pipin} et~al(2013){Pipin}, {Sokoloff}, {Zhang}, and
  {Kuzanyan}}]{Pipin2013c}
{Pipin} VV, {Sokoloff} DD, {Zhang} H, {Kuzanyan} KM (2013) Helicity
  conservation in nonlinear mean-field solar dynamo. \apj 768:46,
  \doi{10.1088/0004-637X/768/1/46}, \eprint{1211.2420}

\bibitem[{{Pipin} et~al(2023){Pipin}, {Kosovichev}, and {Tomin}}]{Pipin2023a}
{Pipin} VV, {Kosovichev} AG, {Tomin} VE (2023) {Effects of Emerging Bipolar
  Magnetic Regions in Mean-field Dynamo Model of Solar Cycles 23 and 24}. \apj
  949(1):7, \doi{10.3847/1538-4357/acaf69}, \eprint{2210.08764}

\bibitem[{{Pouquet} et~al(1976){Pouquet}, {Frisch}, and {Leorat}}]{pouquet+76}
{Pouquet} A, {Frisch} U, {Leorat} J (1976) {Strong MHD helical turbulence and
  the nonlinear dynamo effect}. J Fluid Mech 77:321--354,
  \doi{10.1017/S0022112076002140}

\bibitem[{{Racine} et~al(2011){Racine}, {Charbonneau}, {Ghizaru}, {Bouchat},
  and {Smolarkiewicz}}]{racine+11}
{Racine} {\'E}, {Charbonneau} P, {Ghizaru} M, {Bouchat} A, {Smolarkiewicz} PK
  (2011) {On the Mode of Dynamo Action in a Global Large-eddy Simulation of
  Solar Convection}. \apj 735(1):46, \doi{10.1088/0004-637X/735/1/46}

\bibitem[{R\"adler(1969)}]{Raedler1969}
R\"adler KH (1969) On the electrodynamics of turbulent fields under the
  influence of corilois forces. Monats Dt Akad Wiss 11:194--201

\bibitem[{R\"adler et~al(2003)R\"adler, Kleeorin, and
  Rogachevskii}]{Raedler2003a}
R\"adler KH, Kleeorin N, Rogachevskii I (2003) The mean electromotive force for
  mhd turbulence: the case of a weak mean magnetic field and slow rotation.
  Geophys Astrophys Fluid Dyn 97:249--269

\bibitem[{{R{\"a}dler} et~al(2011){R{\"a}dler}, {Brandenburg}, {Del Sordo}, and
  {Rheinhardt}}]{Radl+11}
{R{\"a}dler} KH, {Brandenburg} A, {Del Sordo} F, {Rheinhardt} M (2011)
  {Mean-field diffusivities in passive scalar and magnetic transport in
  irrotational flows}. \pre 84(4):046321, \doi{10.1103/PhysRevE.84.046321},
  \eprint{1104.1613}

\bibitem[{{Rempel}(2007)}]{Rempel2007ApJ}
{Rempel} M (2007) {Origin of Solar Torsional Oscillations}. \apj
  655(1):651--659, \doi{10.1086/509866}, \eprint{astro-ph/0610221}

\bibitem[{{Rempel} et~al(2023){Rempel}, {Bhatia}, {Bellot Rubio}, and
  {Korpi-Lagg}}]{Rempel2023a}
{Rempel} M, {Bhatia} T, {Bellot Rubio} L, {Korpi-Lagg} MJ (2023) {Small-Scale
  Dynamos: From Idealized Models to Solar and Stellar Applications}. \ssr
  219(5):36, \doi{10.1007/s11214-023-00981-z}, \eprint{2305.02787}

\bibitem[{{Rheinhardt} and {Brandenburg}(2010)}]{RB10}
{Rheinhardt} M, {Brandenburg} A (2010) {Test-field method for mean-field
  coefficients with MHD background}. \aap 520:A28,
  \doi{10.1051/0004-6361/201014700}, \eprint{1004.0689}

\bibitem[{{Rheinhardt} and {Brandenburg}(2012)}]{RB12}
{Rheinhardt} M, {Brandenburg} A (2012) {Modeling spatio-temporal nonlocality in
  mean-field dynamos}. Astron Nachr 333:71--77, \doi{10.1002/asna.201111625},
  \eprint{1110.2891}

\bibitem[{{Rheinhardt} et~al(2014){Rheinhardt}, {Devlen}, {R{\"a}dler}, and
  {Brandenburg}}]{Rhei+14}
{Rheinhardt} M, {Devlen} E, {R{\"a}dler} KH, {Brandenburg} A (2014) {Mean-field
  dynamo action from delayed transport}. \mnras 441:116--126,
  \doi{10.1093/mnras/stu438}, \eprint{1401.5026}

\bibitem[{{Rincon}(2021)}]{Rincon21}
{Rincon} F (2021) {Helical turbulent nonlinear dynamo at large magnetic
  Reynolds numbers}. Physical Review Fluids 6(12):L121701,
  \doi{10.1103/PhysRevFluids.6.L121701}, \eprint{2108.12037}

\bibitem[{{Roberts}(1972)}]{Rob72}
{Roberts} GO (1972) {Dynamo Action of Fluid Motions with Two-Dimensional
  Periodicity}. Phil Trans Roy Soc Lond Ser A 271(1216):411--454,
  \doi{10.1098/rsta.1972.0015}

\bibitem[{{Rogachevskii} et~al(2017){Rogachevskii}, {Ruchayskiy}, {Boyarsky},
  {Fr{\"o}hlich}, {Kleeorin}, {Brandenburg}, and {Schober}}]{Roga+17}
{Rogachevskii} I, {Ruchayskiy} O, {Boyarsky} A, {Fr{\"o}hlich} J, {Kleeorin} N,
  {Brandenburg} A, {Schober} J (2017) {Laminar and Turbulent Dynamos in Chiral
  Magnetohydrodynamics. I. Theory}. \apj 846(2):153,
  \doi{10.3847/1538-4357/aa886b}, \eprint{1705.00378}

\bibitem[{{R\"udiger} and {Kitchatinov}(1990)}]{Ruediger1990}
{R\"udiger} G, {Kitchatinov} LL (1990) {The turbulent stresses in the theory of
  the solar torsional oscillations}. \aap 236(2):503--508

\bibitem[{{R\"udiger} and {Kitchatinov}(1993)}]{Ruediger1993AA}
{R\"udiger} G, {Kitchatinov} LL (1993) {Alpha-effect and alpha-quenching}. \aap
  269(1-2):581--588

\bibitem[{{R{\"u}diger} and {Pipin}(2000)}]{Ruediger2000AA}
{R{\"u}diger} G, {Pipin} VV (2000) {Viscosity-alpha and dynamo-alpha for
  magnetically driven compressible turbulence in Kepler disks}. \aap
  362:756--761

\bibitem[{{R{\"u}diger} et~al(2011){R{\"u}diger}, {Kitchatinov}, and
  {Brandenburg}}]{Ruediger2011s}
{R{\"u}diger} G, {Kitchatinov} LL, {Brandenburg} A (2011) {Cross Helicity and
  Turbulent Magnetic Diffusivity in the Solar Convection Zone}. \solphys
  269(1):3--12, \doi{10.1007/s11207-010-9683-4}, \eprint{1004.4881}

\bibitem[{{R{\"u}diger} et~al(2012){R{\"u}diger}, {Kitchatinov}, and
  {Schultz}}]{Ruediger2012}
{R{\"u}diger} G, {Kitchatinov} LL, {Schultz} M (2012) {Suppression of the
  large-scale Lorentz force by turbulence}. Astron Nachr 333(1):84--91,
  \doi{10.1002/asna.201111635}, \eprint{1109.3345}

\bibitem[{{Ruzmaikin}(1981)}]{Ruzmaikin1981}
{Ruzmaikin} AA (1981) {The solar cycle as a strange attractor}. Comments on
  Astrophysics 9(2):85--93

\bibitem[{{Schrinner}(2011)}]{Schrinner2011}
{Schrinner} M (2011) Global dynamo models from direct numerical simulations and
  their mean-field counterparts. \aap 533:A108,
  \doi{10.1051/0004-6361/201116642}, \eprint{1105.2912}

\bibitem[{{Schrinner} et~al(2005){Schrinner}, {R{\"a}dler}, {Schmitt},
  {Rheinhardt}, and {Christensen}}]{schrinner+05}
{Schrinner} M, {R{\"a}dler} KH, {Schmitt} D, {Rheinhardt} M, {Christensen} U
  (2005) {Mean-field view on rotating magnetoconvection and a geodynamo model}.
  Astron Nachr 326(3):245--249, \doi{10.1002/asna.200410384}

\bibitem[{{Schrinner} et~al(2007){Schrinner}, {R{\"a}dler}, {Schmitt},
  {Rheinhardt}, and {Christensen}}]{schrinner+07}
{Schrinner} M, {R{\"a}dler} KH, {Schmitt} D, {Rheinhardt} M, {Christensen} UR
  (2007) {Mean-field concept and direct numerical simulations of rotating
  magnetoconvection and the geodynamo}. Geophys Astrophys Fluid Dyn
  101(2):81--116, \doi{10.1080/03091920701345707}, \eprint{astro-ph/0609752}

\bibitem[{{Schrinner} et~al(2011){Schrinner}, {Petitdemange}, and
  {Dormy}}]{Schrinner2011a}
{Schrinner} M, {Petitdemange} L, {Dormy} E (2011) Oscillatory dynamos and their
  induction mechanisms. \aap 530:A140, \doi{10.1051/0004-6361/201016372},
  \eprint{1101.1837}

\bibitem[{{Schuessler}(1981)}]{Schuessler1981}
{Schuessler} M (1981) {The solar torsional oscillation and dynamo models of the
  solar cycle}. \aap 94(2):L17

\bibitem[{{Schuessler}(1983)}]{Schuessler1983I}
{Schuessler} M (1983) {Stellar dynamo theory}. In: {Stenflo} JO (ed) Solar and
  Stellar Magnetic Fields: Origins and Coronal Effects, vol 102, pp 213--234

\bibitem[{{See} et~al(2016){See}, {Jardine}, {Vidotto}, {Donati}, {Boro
  Saikia}, {Bouvier}, {Fares}, {Folsom}, {Gregory}, {Hussain}, {Jeffers},
  {Marsden}, {Morin}, {Moutou}, {do Nascimento}, {Petit}, and
  {Waite}}]{See2016}
{See} V, {Jardine} M, {Vidotto} AA, {Donati} JF, {Boro Saikia} S, {Bouvier} J,
  {Fares} R, {Folsom} CP, {Gregory} SG, {Hussain} G, {Jeffers} SV, {Marsden}
  SC, {Morin} J, {Moutou} C, {do Nascimento} JD, {Petit} P, {Waite} IA (2016)
  The connection between stellar activity cycles and magnetic field topology.
  \mnras 462:4442--4450, \doi{10.1093/mnras/stw2010}, \eprint{1610.03737}

\bibitem[{{Seehafer} and {Pipin}(2009)}]{Seehafer2009}
{Seehafer} N, {Pipin} VV (2009) An advective solar-type dynamo without the
  {$\alpha$} effect. \aap 508:9--16, \doi{10.1051/0004-6361/200912614},
  \eprint{0910.2614}

\bibitem[{{Shimada} et~al(2022){Shimada}, {Hotta}, and {Yokoyama}}]{shimada+22}
{Shimada} R, {Hotta} H, {Yokoyama} T (2022) {Mean-field Analysis on Large-scale
  Magnetic Fields at High Reynolds Numbers}. \apj 935(1):55,
  \doi{10.3847/1538-4357/ac7e43}, \eprint{2207.01639}

\bibitem[{{Simard} et~al(2013){Simard}, {Charbonneau}, and
  {Bouchat}}]{simard+13}
{Simard} C, {Charbonneau} P, {Bouchat} A (2013) {Magnetohydrodynamic
  Simulation-driven Kinematic Mean Field Model of the Solar Cycle}. \apj
  768(1):16, \doi{10.1088/0004-637X/768/1/16}

\bibitem[{{Simard} et~al(2016){Simard}, {Charbonneau}, and
  {Dub{\'e}}}]{simard+16}
{Simard} C, {Charbonneau} P, {Dub{\'e}} C (2016) {Characterisation of the
  turbulent electromotive force and its magnetically-mediated quenching in a
  global EULAG-MHD simulation of solar convection}. Adv Spa Res
  58(8):1522--1537, \doi{10.1016/j.asr.2016.03.041}, \eprint{1604.01533}

\bibitem[{{Snodgrass} and {Howard}(1985)}]{Snodgrass1985}
{Snodgrass} HB, {Howard} R (1985) Torsional oscillations of low mode. \solphys
  95:221--228, \doi{10.1007/BF00152399}

\bibitem[{{Sokoloff} et~al(2020){Sokoloff}, {Shibalova}, {Obridko}, and
  {Pipin}}]{Setal20}
{Sokoloff} DD, {Shibalova} AS, {Obridko} VN, {Pipin} VV (2020) {Shape of solar
  cycles and mid-term solar activity oscillations}. \mnras 497(4):4376--4383,
  \doi{10.1093/mnras/staa2279}, \eprint{2007.14779}

\bibitem[{{Spruit}(2003)}]{Spruit2003}
{Spruit} HC (2003) Origin of the torsional oscillation pattern of solar
  rotation. \solphys 213:1--21, \doi{10.1023/A:1023202605379},
  \eprint{astro-ph/0209146}

\bibitem[{{Steenbeck} and {Krause}(1969)}]{StKr1969}
{Steenbeck} M, {Krause} F (1969) {On the Dynamo Theory of Stellar and Planetary
  Magnetic Fields. I. AC Dynamos of Solar Type}. Astron Nachr 291:49--84,
  \doi{10.1002/asna.19692910201}

\bibitem[{{Steenbeck} et~al(1966){Steenbeck}, {Krause}, and
  {R{\"a}dler}}]{SKR1966}
{Steenbeck} M, {Krause} F, {R{\"a}dler} KH (1966) {Berechnung der mittleren
  Lorentz-Feldst{\"a}rke f{\"u}r ein elektrisch leitendes Medium in
  turbulenter, durch Coriolis-Kr{\"a}fte beeinflu{\ss}ter Bewegung}.
  Zeitschrift Naturforschung Teil A 21:369, \doi{10.1515/zna-1966-0401}

\bibitem[{{Stein} and {Nordlund}(2012)}]{SN12}
{Stein} RF, {Nordlund} {\r{A}} (2012) {On the Formation of Active Regions}.
  \apjl 753(1):L13, \doi{10.1088/2041-8205/753/1/L13}, \eprint{1207.4248}

\bibitem[{{Stejko} et~al(2021){Stejko}, {Kosovichev}, and {Pipin}}]{Stej2021}
{Stejko} AM, {Kosovichev} AG, {Pipin} VV (2021) {Forward Modeling Helioseismic
  Signatures of One- and Two-cell Meridional Circulation}. \apj 911(2):90,
  \doi{10.3847/1538-4357/abec70}, \eprint{2101.01220}

\bibitem[{{Stenflo}(1992)}]{Sten1992}
{Stenflo} JO (1992) {Comments on the Concept of an ``Extended Solar Cycle''}.
  In: {Harvey} KL (ed) The Solar Cycle, Astron. Soc. Pacif. Conf. Ser., vol~27,
  p 421

\bibitem[{{Stepanov} et~al(2020){Stepanov}, {Bondar'}, {Katsova}, {Sokoloff},
  and {Frick}}]{Stepanov2020MN}
{Stepanov} R, {Bondar'} NI, {Katsova} MM, {Sokoloff} D, {Frick} P (2020)
  {Wavelet analysis of the long-term activity of V833 Tau}. \mnras
  495(4):3788--3794, \doi{10.1093/mnras/staa1458}, \eprint{2005.11136}

\bibitem[{{Stix}(1974)}]{Stix1974}
{Stix} M (1974) {Comments on the solar dynamo.} \aap 37(1):121--133

\bibitem[{Stix(1976)}]{Stix1976}
Stix M (1976) Differential rotation and the solar dynamo. Astron Astrophys
  47:243--254

\bibitem[{{Stix}(1977)}]{Stix1977}
{Stix} M (1977) Coronal holes and the large-scale solar magnetic field. \aap
  59:73--78

\bibitem[{{Strugarek} et~al(2017){Strugarek}, {Beaudoin}, {Charbonneau},
  {Brun}, and {do Nascimento}}]{Strugarek2017}
{Strugarek} A, {Beaudoin} P, {Charbonneau} P, {Brun} AS, {do Nascimento} JD
  (2017) Reconciling solar and stellar magnetic cycles with nonlinear dynamo
  simulations. Science 357:185--187, \doi{10.1126/science.aal3999},
  \eprint{1707.04335}

\bibitem[{{Sur} and {Brandenburg}(2009)}]{Sur2009MN}
{Sur} S, {Brandenburg} A (2009) {The role of the Yoshizawa effect in the
  Archontis dynamo}. \mnras 399(1):273--280,
  \doi{10.1111/j.1365-2966.2009.15254.x}, \eprint{0902.2394}

\bibitem[{{Taylor}(1974)}]{Taylor74}
{Taylor} JB (1974) {Relaxation of Toroidal Plasma and Generation of Reverse
  Magnetic Fields}. \prl 33(19):1139--1141, \doi{10.1103/PhysRevLett.33.1139}

\bibitem[{{Taylor}(1986)}]{Taylor86}
{Taylor} JB (1986) {Relaxation and magnetic reconnection in plasmas}. Reviews
  of Modern Physics 58(3):741--763, \doi{10.1103/RevModPhys.58.741}

\bibitem[{{Timothy} et~al(1975){Timothy}, {Krieger}, and {Vaiana}}]{Timothy75}
{Timothy} AF, {Krieger} AS, {Vaiana} GS (1975) {The Structure and Evolution of
  Coronal Holes}. \solphys 42(1):135--156, \doi{10.1007/BF00153291}

\bibitem[{{Tobias} and {Weiss}(2007)}]{TobiasWeiss2007}
{Tobias} S, {Weiss} N (2007) {The solar dynamo and the tachocline}. In:
  {Hughes} DW, {Rosner} R, {Weiss} NO (eds) The Solar Tachocline, p 319

\bibitem[{{Tobias}(1996)}]{Tobias1996}
{Tobias} SM (1996) Grand minimia in nonlinear dynamos. \aap 307:L21

\bibitem[{{Tobias}(2021)}]{Tobias21}
{Tobias} SM (2021) {The turbulent dynamo}. J Fluid Mech 912:P1,
  \doi{10.1017/jfm.2020.1055}, \eprint{1907.03685}

\bibitem[{{Tobias} and {Marston}(2013)}]{TM13}
{Tobias} SM, {Marston} JB (2013) {Direct Statistical Simulation of
  Out-of-Equilibrium Jets}. \prl 110(10):104502,
  \doi{10.1103/PhysRevLett.110.104502}, \eprint{1209.3862}

\bibitem[{{Tobias} and {Marston}(2017)}]{TM17}
{Tobias} SM, {Marston} JB (2017) {Three-dimensional rotating Couette flow via
  the generalised quasilinear approximation}. J Fluid Mech 810:412--428,
  \doi{10.1017/jfm.2016.727}, \eprint{1605.07410}

\bibitem[{{Tobias} et~al(2008){Tobias}, {Cattaneo}, and {Brummell}}]{tobias+08}
{Tobias} SM, {Cattaneo} F, {Brummell} NH (2008) {Convective Dynamos with
  Penetration, Rotation, and Shear}. \apj 685(1):596--605, \doi{10.1086/590422}

\bibitem[{{Tobias} et~al(2011){Tobias}, {Dagon}, and {Marston}}]{TMD11}
{Tobias} SM, {Dagon} K, {Marston} JB (2011) {Astrophysical Fluid Dynamics via
  Direct Statistical Simulation}. \apj 727(2):127,
  \doi{10.1088/0004-637X/727/2/127}, \eprint{1009.2684}

\bibitem[{{Ulrich}(2001)}]{Ulrich2001}
{Ulrich} RK (2001) Very long lived wave patterns detected in the solar surface
  velocity signal. \apj 560:466--475, \doi{10.1086/322524}

\bibitem[{{Vainshtein} and {Cattaneo}(1992)}]{VC92}
{Vainshtein} SI, {Cattaneo} F (1992) {Nonlinear Restrictions on Dynamo Action}.
  \apj 393:165, \doi{10.1086/171494}

\bibitem[{{Vainshtein} et~al(1980){Vainshtein}, {Zeldovich}, and
  {Ruzmaikin}}]{VZR1980}
{Vainshtein} SI, {Zeldovich} IB, {Ruzmaikin} AA (1980) {The turbulent dynamo in
  astrophysics}. Moscow Izdatel Nauka

\bibitem[{{Vidotto} et~al(2018){Vidotto}, {Lehmann}, {Jardine}, and
  {Pevtsov}}]{Vidotto2018}
{Vidotto} AA, {Lehmann} LT, {Jardine} M, {Pevtsov} AA (2018) The magnetic field
  vector of the sun-as-a-star - ii. evolution of the large-scale vector field
  through activity cycle 24. \mnras 480:477--487, \doi{10.1093/mnras/sty1926},
  \eprint{1807.06334}

\bibitem[{Vishniac and Cho(2001)}]{Vishniac2001}
Vishniac ET, Cho J (2001) Magnetic helicity conservation and astrophysical
  dynamos. Astrophys J 550:752--760

\bibitem[{{Viviani} et~al(2018){Viviani}, {Warnecke}, {K{\"a}pyl{\"a}},
  {K{\"a}pyl{\"a}}, {Olspert}, {Cole-Kodikara}, {Lehtinen}, and
  {Brandenburg}}]{Viviani2018}
{Viviani} M, {Warnecke} J, {K{\"a}pyl{\"a}} MJ, {K{\"a}pyl{\"a}} PJ, {Olspert}
  N, {Cole-Kodikara} EM, {Lehtinen} JJ, {Brandenburg} A (2018) Transition from
  axi- to nonaxisymmetric dynamo modes in spherical convection models of
  solar-like stars. \aap 616:A160, \doi{10.1051/0004-6361/201732191},
  \eprint{1710.10222}

\bibitem[{{Warnecke}(2018)}]{warnecke18}
{Warnecke} J (2018) {Dynamo cycles in global convection simulations of
  solar-like stars}. \aap 616:A72, \doi{10.1051/0004-6361/201732413},
  \eprint{1712.01248}

\bibitem[{{Warnecke} et~al(2011){Warnecke}, {Brandenburg}, and
  {Mitra}}]{War+11}
{Warnecke} J, {Brandenburg} A, {Mitra} D (2011) {Dynamo-driven plasmoid
  ejections above a spherical surface}. \aap 534:A11,
  \doi{10.1051/0004-6361/201117023}, \eprint{1104.0664}

\bibitem[{{Warnecke} et~al(2012){Warnecke}, {Brandenburg}, and
  {Mitra}}]{War+12}
{Warnecke} J, {Brandenburg} A, {Mitra} D (2012) {Magnetic twist: a source and
  property of space weather}. J Spa Weather Spa Climate 2:A11,
  \doi{10.1051/swsc/2012011}, \eprint{1203.0959}

\bibitem[{{Warnecke} et~al(2013){Warnecke}, {Losada}, {Brandenburg},
  {Kleeorin}, and {Rogachevskii}}]{warnecke+13}
{Warnecke} J, {Losada} IR, {Brandenburg} A, {Kleeorin} N, {Rogachevskii} I
  (2013) {Bipolar Magnetic Structures Driven by Stratified Turbulence with a
  Coronal Envelope}. \apjl 777(2):L37, \doi{10.1088/2041-8205/777/2/L37},
  \eprint{1308.1080}

\bibitem[{{Warnecke} et~al(2016){Warnecke}, {K{\"a}pyl{\"a}}, {K{\"a}pyl{\"a}},
  and {Brandenburg}}]{Warnecke16}
{Warnecke} J, {K{\"a}pyl{\"a}} PJ, {K{\"a}pyl{\"a}} MJ, {Brandenburg} A (2016)
  {Influence of a coronal envelope as a free boundary to global convective
  dynamo simulations}. \aap 596:A115, \doi{10.1051/0004-6361/201526131},
  \eprint{1503.05251}

\bibitem[{{Warnecke} et~al(2018){Warnecke}, {Rheinhardt}, {Tuomisto},
  {K{\"a}pyl{\"a}}, {K{\"a}pyl{\"a}}, and {Brandenburg}}]{Warnecke2018a}
{Warnecke} J, {Rheinhardt} M, {Tuomisto} S, {K{\"a}pyl{\"a}} PJ,
  {K{\"a}pyl{\"a}} MJ, {Brandenburg} A (2018) {Turbulent transport coefficients
  in spherical wedge dynamo simulations of solar-like stars}. \aap 609:A51,
  \doi{10.1051/0004-6361/201628136}, \eprint{1601.03730}

\bibitem[{{Warnecke} et~al(2021){Warnecke}, {Rheinhardt}, {Viviani}, {Gent},
  {Tuomisto}, and {K{\"a}pyl{\"a}}}]{War+21}
{Warnecke} J, {Rheinhardt} M, {Viviani} M, {Gent} FA, {Tuomisto} S,
  {K{\"a}pyl{\"a}} MJ (2021) {Investigating Global Convective Dynamos with
  Mean-field Models: Full Spectrum of Turbulent Effects Required}. \apjl
  919(2):L13, \doi{10.3847/2041-8213/ac1db5}, \eprint{2105.07708}

\bibitem[{{Willis}(2012)}]{Willis12}
{Willis} AP (2012) {Optimization of the Magnetic Dynamo}. \prl 109(25):251101,
  \doi{10.1103/PhysRevLett.109.251101}, \eprint{1209.1559}

\bibitem[{{Wilson} et~al(1988){Wilson}, {Altrocki}, {Harvey}, {Martin}, and
  {Snodgrass}}]{Wilson1988}
{Wilson} PR, {Altrocki} RC, {Harvey} KL, {Martin} SF, {Snodgrass} HB (1988) The
  extended solar activity cycle. \nat 333:748--750, \doi{10.1038/333748a0}

\bibitem[{{Wright} and {Drake}(2016)}]{WrightDrake2016N}
{Wright} NJ, {Drake} JJ (2016) {Solar-type dynamo behaviour in fully convective
  stars without a tachocline}. \nat 535(7613):526--528,
  \doi{10.1038/nature18638}, \eprint{1607.07870}

\bibitem[{{Wright} et~al(2018){Wright}, {Newton}, {Williams}, {Drake}, and
  {Yadav}}]{Wright2018W}
{Wright} NJ, {Newton} ER, {Williams} PKG, {Drake} JJ, {Yadav} RK (2018) {The
  stellar rotation-activity relationship in fully convective M dwarfs}. \mnras
  479(2):2351--2360, \doi{10.1093/mnras/sty1670}, \eprint{1807.03304}

\bibitem[{{Yokoi}(2013)}]{Yokoi2013}
{Yokoi} N (2013) Cross helicity and related dynamo. Geophysical and
  Astrophysical Fluid Dynamics 107:114--184,
  \doi{10.1080/03091929.2012.754022}, \eprint{1306.6348}

\bibitem[{{Yokoi} et~al(2016){Yokoi}, {Schmitt}, {Pipin}, and
  {Hamba}}]{Yokoi2016}
{Yokoi} N, {Schmitt} D, {Pipin} V, {Hamba} F (2016) {A New Simple Dynamo Model
  for Stellar Activity Cycle}. \apj 824(2):67,
  \doi{10.3847/0004-637X/824/2/67}, \eprint{1601.06348}

\bibitem[{{Yoshimura}(1978)}]{Yoshimura1978}
{Yoshimura} H (1978) Nonlinear astrophysical dynamos - multiple-period dynamo
  wave oscillations and long-term modulations of the 22 year solar cycle. \apj
  226:706--719, \doi{10.1086/156653}

\bibitem[{{Yoshimura}(1981)}]{Yoshimura1981}
{Yoshimura} H (1981) Solar cycle lorentz force waves and the torsional
  oscillations of the sun. \apj 247:1102--1112, \doi{10.1086/159120}

\bibitem[{{Yoshizawa} and {Yokoi}(1993)}]{Yoshizawa1993}
{Yoshizawa} A, {Yokoi} N (1993) {Turbulent Magnetohydrodynamic Dynamo for
  Accretion Disks Using the Cross-Helicity Effect}. \apj 407:540,
  \doi{10.1086/172535}

\bibitem[{{Yoshizawa} et~al(2000){Yoshizawa}, {Kato}, and
  {Yokoi}}]{Yoshizawa2000}
{Yoshizawa} A, {Kato} H, {Yokoi} N (2000) {Mean Field Theory Interpretation of
  Solar Polarity Reversal}. \apj 537(2):1039--1053, \doi{10.1086/309057}

\bibitem[{{Zeldovich} et~al(1983){Zeldovich}, {Ruzmaikin}, and
  {Sokoloff}}]{ZRS83}
{Zeldovich} {\relax Ya}B, {Ruzmaikin} AA, {Sokoloff} DD (1983) {Magnetic Fields
  in Astrophysics}. Gordon and Breach, New York

\bibitem[{{Zhou} et~al(2018){Zhou}, {Blackman}, and {Chamandy}}]{ZBC18}
{Zhou} H, {Blackman} EG, {Chamandy} L (2018) {Derivation and precision of mean
  field electrodynamics with mesoscale fluctuations}. J Plasma Phys
  84(3):735840302, \doi{10.1017/S0022377818000375}, \eprint{1710.04064}

\end{thebibliography}
\end{document}